\documentclass[useAMS,usenatbib]{mn2e}

\let\jnlstyle=\rm
\def\refjnl#1{{\jnlstyle#1}}

\def\apjl{\refjnl{ApJ}}                
\def\apjs{\refjnl{ApJS}}               
\def\ao{\refjnl{Appl.~Opt.}}           
\def\aap{\refjnl{A\&A}}                

\input{psfig.sty}

\usepackage{graphicx}
\usepackage{epsfig}
\usepackage{amssymb}
\usepackage{amsmath}
\usepackage{amsfonts}
\usepackage{txfonts}
\usepackage{multirow}
\usepackage{afterpage}
\usepackage{xcolor}

\definecolor{mypink1}{rgb}{0.858, 0.188, 0.478}
\definecolor{mygreen1}{rgb}{0.258, 0.788, 0.878}
\definecolor{myorange1}{rgb}{0.5, 0.2, 0.2}

\newcommand{\mycomment}[1]{}

\defcitealias{FORGE}{A21}
\defcitealias{BRIDGE}{CL22}

\title[MGLenS]{\textsc{MGLenS}: Modified gravity weak lensing simulations for emulation-based cosmological inference}

\author[J. Harnois-D\'{e}raps et al.]{Joachim Harnois-D\'{e}raps$^{1}$\thanks{e-mail: joachim.harnois-deraps@ncl.ac.uk}, Cesar Hernandez-Aguayo$^{2,3}$, Carolina Cuesta-Lazaro$^{4,5,6,7}$, \newauthor Christian Arnold$^{7}$, Baojiu Li$^{7}$, Christopher T. Davies$^{8}$ \& Yan-Chuan Cai$^9$
\\
$^{1}$School of Mathematics, Statistics and Physics, Newcastle University, Herschel Building, NE1 7RU, Newcastle-upon-Tyne, UK\\
$^{2}$Max-Planck-Institut f\"ur Astrophysik, Karl-Schwarzschild-Str. 1, D-85748, Garching, Germany\\%
$^{3}$Excellence Cluster ORIGINS, Boltzmannstrasse 2, D-85748 Garching, Germany\\%
$^{4}$Center for Astrophysics | Harvard \& Smithsonian, 60 Garden St, Cambridge, MA 02138, USA\\
$^{5}$The NSF AI Institute for Artificial Intelligence and Fundamental Interactions \\
$^{6}$Department of Physics, Massachusetts Institute of Technology, Cambridge, MA 02139, USA \\
$^{7}$ Institute for Computational Cosmology, Department of Physics, Durham University, South Road, DH1 3LE, Durham, UK\\
$^{8}$ Faculty of Physics, Ludwig-Maximilians-Universit\"{a}t, Scheinerstr. 1, 81679 Munich, Germany\\
$^9$ Institute for Astronomy, University of Edinburgh, Blackford Hill, Edinburgh, EH9 3HJ, UK\\
}
\date{Accepted XXX. Received YYY; in original form ZZZ}

\pubyear{2022}

\begin{document}
\label{firstpage}
\pagerange{\pageref{firstpage}--\pageref{lastpage}}
\maketitle

\begin{abstract}
 We present \textsc{MGLenS}, a large series of  modified gravity lensing simulations tailored for cosmic shear data analyses and forecasts in which cosmological and modified gravity parameters are varied simultaneously. Based on the {\sc forge} and {\sc bridge} $N$-body simulation suites presented in companion papers, we construct 500,000 deg$^2$ of mock Stage-IV lensing data, sampling a pair of 4-dimensional volumes designed for the training of emulators. We validate the accuracy of \textsc{MGLenS} with inference analyses based on the lensing power spectrum exploiting our implementation of $f(R)$ and nDGP theoretical predictions within the {\sc cosmoSIS} cosmological inference package. A Fisher analysis reveals that the vast majority of the constraining power from such a survey comes from the highest redshift galaxies alone. We further find from a full likelihood sampling that cosmic shear can achieve 95\% CL constraints on the modified gravity parameters  of log$_{10}\left[ f_{R_0}\right] < -5.24$ and log$_{10}\left[ H_0 r_c\right] > -0.05$, after marginalising over intrinsic alignments of galaxies and including scales up to $\ell=5000$. Such a survey setup could in fact detect with more than $3\sigma$ confidence  $f(R)$ values larger than $3 \times 10^{-6}$ and $H_0 r_c$ smaller than 1.0. 
Scale cuts at $\ell=3000$ reduce the degeneracy breaking between $S_8$ and the modified gravity parameters, while photometric redshift uncertainty seem to play a subdominant role in our error budget. We finally explore the consequences of analysing data with the wrong gravity model, and report the catastrophic biases for a number of possible scenarios. The Stage-IV \textsc{MGLenS} simulations, the {\sc forge} and {\sc bridge} emulators and the {\sc cosmoSIS} interface modules will be made publicly available upon journal acceptance.
\end{abstract}




\section{Introduction}
\label{sec:intro}

Recent measurements from dedicated cosmic shear surveys such as the Kilo Degree Survey\footnote{KiDS:kids.strw.leidenuniv.nl} \citep{KiDS1000_Asgari, KiDS1000_vdBusch}, the Dark Energy Survey\footnote{DES:www.darkenergysurvey.org} \citep{DESY3_Amon, DESY3_Secco} and the HyperSuprime Camera Survey\footnote{HSC:www.naoj.org/Projects/HSC} \citep{HSCY1_2PCF, HSCY1_Cell} have established weak gravitational lensing as one of the most competitive probe of the dark sector of our Universe. In addition to constraining key parameters such as the total matter abundance $\Omega_{\rm m}$, the clustering amplitude $\sigma_8$ and the dark-energy equation of state $w_0$, lensing data has also been used to
test 
the gravitational sector. Indeed, the matter density field could be affected by deviations from the theory of General Relativity (GR) on cosmic scales, where the presence of a fifth force would increase the clustering in a manner detectable by lensing \citep{WL_MGReview}. In most viable models, a screening mechanism is invoked to suppress the impact of modified gravity (MG hereafter) on small scales or high-density regions, as required to satisfy the tight  Solar System constraints on GR \citep{Hu:2007nk}. Screening can be achieved in a number of ways, including: 1- {\it Chameleon mechanism} \citep{Khoury:2003rn}, in which the range of the fifth force is decreased in regions of high space-time curvature, thus, effectively hiding the additional force; 2- {\it Symmetron} \citep{Symmetron,Symmetron_Cosmo}, in which the coupling of the scalar field mediating the fifth force is density dependent; 3-
{\it Vainshtein screening} \citep{Vainshtein:1972plb},  in which the screening effect is sourced by the second derivative of the field value and happens mostly on small scales; 4- {\it k-mouflage} screening \citep{k-mouflage}.
We refer to reader to \citet{MGReview} for a review on modified theories of gravitation. 

In any case, a clear detection of  the resulting excess clustering in galaxy surveys is made difficult by the large uncertainty on the galaxy bias, especially on small non-linear scales. Weak gravitational lensing, however, naturally emerges as a  potentially cleaner probe of MG, being unaffected by this severe limitation \citep{WL_MGReview}. While travelling through the foreground large scale structure on its way to our telescopes, the light emitted by distant galaxies acquires coherent distortions, which we measure in cosmic shear surveys. Recently, \citet{2015MNRAS.454.2722H} constrained a series of MG models from the cosmic shear analysis of the Canada-France-Hawaii Telescope lensing survey in a pathfinder analysis. Upgraded investigations including a number of systematics inherent to cosmic shear data have since been carried out with the KiDS and DES data  \citep{2017MNRAS.471.1259J,DESY1_MG, KiDS1000_Troester, DESY3_Ferte}, however the constraining power on MG remains weak and model-dependent. As discussed in the above
references, exploring multiple MG hypotheses is  essential in light of the current $S_8\equiv\sigma_8 \sqrt{\Omega_{\rm m}/0.3}$ tension between low- and high-redshift cosmological data analyses \citep[e.g.][]{KiDS1000_Heymans}, although it likely will not be the sole solution since MG moves $S_8$ upwards in both weak lensing and CMB data \citep{KiDS1000_Troester}, preserving the tension.

In these previous analyses, the constraints on MG parameters are derived from measurements of lensing two-point statistics, either the two-point correlation functions or the lensing power spectrum.  These choices of summary statistics are largely motivated by the simplicity of their modelling, which involves  tractable modifications to the matter power spectrum that are well measured from $N$-body simulations. Recent computational efforts led to public power spectrum {\it emulators}, which  predict the enhancement of clustering for a variety of MG models, over a wide range of cosmological parameters\footnote{{\sc MGEmu}:github.com/LSSTDESC/mgemu}$^,$\footnote{{\sc MGcamb}:github.com/sfu-cosmo/MGCAMB}$^,$\footnote{{\sc forge}:bitbucket.org/arnoldcn/forge\_emulator}$^,$\footnote{{\sc HMCode}:github.com/alexander-mead/HMcode}$^,$\footnote{{\sc ReACT}:github.com/nebblu/ReACT}.

 It is expected that two-point statistics are not optimal 
for constraining MG, largely due to the fact that the screening mechanism is typically density-dependent. Instead, statistics that are more sensitive to low-density regions, for example those measuring signals around under-dense regions \citep[e.g.][]{BarreiraWLthroughs, LensingVoids} or up-weighting these with marked correlation functions \citep{Austin2018, Armijo2018,Hernandez-Aguayo:2018yrp}, have been shown to better constrain the parameters that describe a fifth force. The main difficulty with these alternative measurement methods is the absence of  theoretical models to describe this signal, forcing one to rely on emulators trained of a large number of accurate weak lensing simulations to facilitate their interpretation. 

Searching for modifications to GR is a complicated enterprise, since different theories predict sometimes radically different effects on the formation of large-scale structures, making this a model-dependent search. Moreover, among all existing MG simulations, only a few have been designed to enable the extraction of weak lensing statistics at the field level, including for example the DUSTGRAIN Pathfinder \citep{Dustgrain}, in which MG models were used to co-evolve dark matter and massive neutrinos. These simulations have shown again that non-Gaussian statistics are better suited to break down the known degeneracy between the increase in structure formation caused by the fifth force, and the decrease caused by neutrino free-streaming. Other simulation efforts studying weak lensing statistics include that of \citet{Higushi2016}, \citet{BarreiraWLthroughs},  \citet{Shirasaki2017} and \citet{Li_Shirasaki2018}, which examine various non-Gaussian statistics in light-cones produced by the ECOSMOG  modified-gravity $N$-body solver \citep{ECOSMOG}.  Fast approximate $N$-body methods such as MG-COLA \citep{MG-COLA} are generally not accurate enough to model the small scales physics probed by lensing, however speed-up of the MG sector as in  \citet{MG-GLAM} might prove helpful to reduce the computational cost overhead in the future.

We present in this work the first suite of MG weak lensing simulations designed for the analysis of current cosmic shear surveys. Based on the \textsc{forge} (F Of R Gravity Emulator) simulations described in \citet[][hereafter A21]{FORGE} and the \textsc{bridge} (BRaneworld-Inspired DGP Gravity Emulator) simulations presented in \citet[][hereafter CL22]{BRIDGE}, the Modified Gravity Lensing Simulations (\textsc{MGLenS}) consist of two suites of lensing maps in which  three cosmological and one modified gravity parameters are varied on a Latin hypercube over a volume that encloses most of the 2$\sigma$ posterior allowed by current lensing surveys.  The two MG scenarios are modelled separately, and their respective parameters capture the strength of the deviations from GR in the widely studied $f(R)$ \citep{Hu:2007nk} and the normal branch of the DGP \citep[nDGP hereafter, see][]{Dvali:2000hr} gravity models respectively. With its 2$\times$50 nodes, \textsc{MGLenS} has enough sampling points to emulate with better than 2.5\% accuracy most lensing statistics. Combined with Gaussian Process Regression (GPR) or Neural  Network (NN) emulators, the summary statistics measured from these simulations can be modelled and passed to a likelihood analysis code, which can then serve to constrain cosmology and gravity models from existing lensing data. Additionally, they can serve to forecast the performance of upcoming surveys when analysed with alternative measurement methods. 

The first part of this paper summarises the gravitational physics that are captured by the \textsc{forge} and \textsc{bridge} simulation suites (Secs. \ref{subsec:fR} and \ref{subsec:DGP}), while Sec. \ref{subsec:Nbody} includes a brief overview of their numerical implementation within the high-performance $N$-body code {\sc Arepo-MG} \citep{Arnold:2019vpg,Hernandez-Aguayo:2020kgq}. After reviewing  our emulator in Sec. \ref{subsec:CNN} and the modelling aspects of weak lensing two-point statistics in Sec. \ref{subsec:wl-th}, we describe our weak lensing simulations in Sec. \ref{sec:sims}. We present in Sec. \ref{sec:MCMC} the results from a series of likelihood analyses where we  first validate both our cosmology inference pipeline based on these emulators and the {\sc MGLenS} simulations themselves. We then investigate the detection potential from measurements of the weak lensing power spectrum in a Stage-IV survey such as those of to be probed by the  Vera Rubin\footnote{Rubin:www.lsst.org}, {\it Euclid}\footnote{{\it Euclid}:euclid-ec.org} or Nancy Grace\footnote{Grace:wfirst.gsfc.nasa.gov} telescopes.  Finally, we explicitly demonstrate the model-dependence of such searches by running cosmological analyses on MG data assuming the wrong gravity model, recording extreme biases both on the gravity and cosmology sectors. 

Throughout this paper we assume a flat $\Lambda$CDM universe. Optimal studies involving beyond-two-point statistics will be presented in companion papers.

\section{Background}
\label{sec:theory}

Although GR is well tested on small scales in laboratory experiments and in the Solar system \citep[e.g.][]{Will:2005va,Will:2014kxa}, possible deviations are at the moment largely unconstrained on cosmological scales (Mpc and above). To quantify such deviations in a self-consistent way, it is useful to develop an array of simple representative models to be used as templates for making predictions, which can be compared
to observational data. There is a large (probably infinite) number of currently viable MG models, 
and this paper focuses on two of the most widely-studied examples, namely the Hu-Sawicky $f(R)$ and the nDGP gravity models, which we introduce in this section.  Note that although these do not support self-acceleration and therefore require dark energy as well, they are two viable, representative MG models that can guide our search.

\subsection{$f(R)$ gravity}
\label{subsec:fR}

The modified Einstein equations in $f(R)$ gravity can be obtained from a modified Einstein-Hilbert action in which the standard Ricci scalar $R$ is supplemented by an algebraic function, $f(R)$ (hence its name):
\begin{equation}\label{eq:S-f(R)}
S = \frac{1}{16\pi G} \int \mathrm{d}^4x\sqrt{-g} (R + f(R)) + S_{\rm m}(g_{\mu\nu},\psi_i)\, .
\end{equation} 
In this expression $G$ is the gravitational constant, $g_{\mu\nu}$ is the metric, $g\equiv\det\left(g_{\mu\nu}\right)$ is its determinant and  $S_{\rm m}$ the action of the matter field, which depends on the metric and the different matter fluids $\psi_i$.
Varying  $S$  with respect to $g_{\mu\nu}$, we obtain:
\begin{equation}\label{eq:M-EQ}
G_{\mu\nu} + f_R R_{\mu\nu} - g_{\mu\nu} \left(\frac{1}{2}f(R) - \Box f_R \right) - \nabla_\mu \nabla_\nu f_R = 8\pi G T^{\rm m}_{\mu\nu}\,,
\end{equation}
where $R_{\mu\nu}$ and $G_{\mu\nu}$ are respectively the Ricci and Einstein tensors, $\nabla_\mu$ is the covariant derivative compatible with the spacetime metric $g_{\mu\nu}$ (i.e. $\nabla_{\lambda}g_{\mu\nu}=0$), $\Box \equiv \nabla^\mu\nabla_\mu = g^{\mu\nu}\nabla_\mu\nabla_\nu$ is the d'Alembert operator in the 4-dimensional spacetime, $f_R \equiv {\rm d} f(R) / {\rm d} R$ and $T^{\rm m}_{\mu \nu}$ is the energy-momentum tensor for matter. 

Despite the small modification to the standard Einstein-Hilbert action, Eq.~\eqref{eq:M-EQ} differs from the usual Einstein equation in that it contains up to fourth-order, rather than second-order, derivatives of the metric, as a result of the terms $\Box{f}_R$ and $\nabla_\mu\nabla_\nu{f}_R$. However, both terms are second derivatives of a scalar quantity $f_R$, which indicates that the fourth-order differential equation \eqref{eq:M-EQ} can be written as a second-order Einstein equation if $f_R$ is treated as a (new) scalar degree of freedom (the {\it scalaron} field), which has its own evolution equation obtained by taking the trace of Eq.~\eqref{eq:M-EQ}. Namely: 
\begin{equation}\label{eq:f_R}
\Box f_R = \frac{1}{3} \left[R - f_R R + 2f(R) + 8\pi G\rho_{\rm m}\right] \equiv \frac{{\rm d}V_{\rm eff}\left(f_R\right)}{{\rm d}{f}_R}\,,
\end{equation}
where $\rho_{\rm m}$ is the non-relativistic matter density of the Universe -- this terms originates from the trace of the energy momentum tensor, and thus relativistic species do not contribute directly (i.e., through direct coupling) to the dynamics of the scalar field. In the second equality above we have defined an effective potential, $V_{\rm eff}\left(f_R\right)$, of the scalaron field. 

Cosmological structure formation can be well described by the quasi-static and weak-field approximations \citep[see, e.g.,][for some quantitative analyses of beyond-Newtonian effects in cosmological settings]{Barrera-Hinojosa:2020gnx}. The former approximation applies in the limit of slow, non-relativistic, motions of matter, where the time derivatives of the metric potentials can be neglected; the latter assumes that the potentials created by large-scale structure are shallow so that their higher-order products can also be ignored. In the presence of a scalar field as in the case of $f(R)$ gravity, these approximations also apply to the scalaron itself since, as we will show shortly, the latter can be considered as the potential of the modified gravitational force. Note that in general the quasi-static approximation only means that the perturbations of the scalaron have negligible time derivatives compared to spatial derivatives\footnote{See, e.g., \citet{Oyaizu:2008sr,Bose:2014zba} for some results showing the goodness of the quasi-static approximation in $f(R)$ gravity.}, though in the case of $f(R)$ models with a viable chameleon screening mechanism, this can apply to the full scalar field $f_R$. Under these approximations, the modified Einstein's equation \eqref{eq:M-EQ} and the scalaron equation of motion \eqref{eq:f_R} become:
\begin{equation}\label{eq:Phi}
\boldsymbol{\nabla}^2 \Phi = \frac{16\pi G}{3} a^2 (\rho_{\rm m} - \bar{\rho}_{\rm m}) + \frac{1}{6} a^2 (R(f_R) - \bar{R})\,,
\end{equation}
\begin{equation}\label{eq:fR}
\boldsymbol{\nabla}^2 f_R = -\frac{a^2}{3} [R(f_R) - \bar{R} + 8\pi G(\rho_{\rm m} - \bar{\rho}_{\rm m})]\,,
\end{equation}
where $\Phi$ is the gravitational potential, $\boldsymbol{\nabla}$ is the gradient operator in 3-dimensional space, and $a$ is the scale factor. Over-bars denote the cosmic mean, or background, value of the quantity. Note that the modified Poisson equation \eqref{eq:Phi} can be rewritten as
\begin{equation}\label{eq:Phi2}
\boldsymbol{\nabla}^2 \Phi = {4\pi G} a^2 \delta \rho_{\rm m} - \frac{1}{2} \boldsymbol{\nabla}^2f_R\,,
\end{equation}
by using Eq.~\eqref{eq:fR}, with $\delta \rho_{\rm m} \equiv \rho_{\rm m} - \bar{\rho}_{\rm m}$. This shows that $-f_R/2$ can be considered as the potential of the modified gravity force.  

In this work we consider the \citet{Hu:2007nk} $f(R)$ model, for which the functional form of $f(R)$ is given by
\begin{equation}\label{eq:f(R)}
f(R) = -m^2 \frac{c_1}{c_2} \frac{(-R/m^2)^n}{(-R/m^2)^n + 1}\,,
\end{equation}
where $m^2 \equiv \Omega_{\rm m} H_0^2$ with $H_0$ and $\Omega_{\rm m}$ respectively the values of the Hubble parameter and the matter density parameter today, while $c_1, c_2$ and $n$ are free dimensionless model parameters, with $n$ a non-negative integer. In the limit $|\bar{R}|\gg{m}^2$ (which holds for the entire cosmic history up to today in the models to be studied), the scalaron field  can then be expressed as
\begin{equation}\label{eq:sc1}
   f_R \simeq -\left| \bar{f}_{R_0} \right|\left(\frac{\bar{R}_0}{R}\right)^{n+1},
\end{equation}
where $\bar{R}_0$, $\bar{f}_{R_0}$ are, respectively, the present-day values of the background Ricci scalar and $\bar{f}_R$. We fix the value of the power-law index to $n=1$ for simplicity \citep[other values of $n$, such as $n=0$ and $2$, lead to qualitatively similar behaviours of the model, see, e.g.,][]{Ruan:2021wup} and we vary $\bar{f}_{R_0}$ in the range $\left[10^{-4.5} ;10^{-7.0}\right]$, where larger values lead to larger deviations from GR. See \citet{FORGE} and Table \ref{table:forge_nodes} below for a complete list of the exact $\bar{f}_{R_0}$ values included in this paper, along with other cosmological parameters used in our simulations. Note that hereafter we use  $f_{R_0}$ instead of $\bar{f}_{R_0}$ to improve notation.

It is well established that viable $f(R)$ models for the late-time Universe must invoke the chameleon screening mechanism \citep[][]{Khoury:2003aq,Khoury:2003rn,Mota:2006ed,Brax:2004px,Brax:2008hh}, an intrinsically non-linear behaviour originating from the functional form of $f(R)$. The $R\left(f_R\right)$ term in the scalaron equation of motion, Eq.~\eqref{eq:fR}, can be considered as a description of the non-linear self-interaction of the scalaron and, along with its interaction with matter, this determines how $f_R$ varies in space. If appropriate parameter values are adopted, for dense spherical objects -- such as dark matter haloes in this toy example -- inside a homogeneous medium of matter, $f_R$ will transitions from the background value $\bar{f}_R$ far from the object to nearly zero at its centre, and the transition takes place in a thin shell at the boundary of the object, which means that $f_R$ stays constant in all but a thin shell. Because $f_R$ is the potential of the modified gravity force, this means that this force vanishes, or is efficiently ``screened'', for most parts inside and outside the object. Another way to see how this screening mechanism works is by looking at Eq.~\eqref{eq:Phi2}, which shows that inside the object where $f_R$ is nearly identically zero, the modified gravity force vanishes. 

Large scale structures offer
a variety of environments, from  high-density regions such as the cores of clusters and galaxies, to low-density regions in cosmic voids. As a result, these are
ideal places for investigating signatures of chameleon screening and constraining $f(R)$ gravity. However, it also poses a computational challenge as the non-linear nature of the chameleon mechanism can only be accurately predicted with  high-resolution simulations such as \textsc{forge}.

\subsection{nDGP gravity}
\label{subsec:DGP}

In the gravitational model of \citeauthor*{Dvali:2000hr},  all particle species are assumed to be confined to 
a four-dimensional hypersurface or `brane', while gravitons can propagate along a fourth spatial dimension and leak into the five-dimensional `bulk' spacetime.
The action of this braneworld model is given by
\begin{equation}
S = \int_{\rm brane} {\rm d}^4x \sqrt{-g} \frac{R}{16\pi G} + \int_{\rm bulk} {\rm d}^5x \sqrt{-g^{(5)}} \frac{R^{(5)}}{16\pi G^{(5)}}, \label{eq:S_dgp}
\end{equation}
where $g$, $R$, $G$ are the values on the brane and have the same meaning as before, while the counterpart bulk quantities are denoted by $g^{(5)}$, $R^{(5)}$ and $G^{(5)}$. 

A new parameter can be introduced from the ratio between $G^{(5)}$ and $G$, known as the {\it crossover scale} and denoted by $r_c$:
\begin{equation}\label{eq:rc}
r_c \equiv \frac{1}{2} \frac{G^{(5)}}{G}.
\end{equation}
This can be considered as a critical scale above (below) which gravity is well described by the 5D (4D) part of the action. Since $r_c$ is a dimensional quantity,  its value is often quoted via $H_0r_c/c$, which can be considered as the ratio between the crossover scale and the horizon size $c/H_0$ (the speed of light $c$ is dropped out hereafter  since $c=1$ in natural units). 

The DGP model has two distinct branches of solutions. The first is a self-accelerating branch (sDGP), which supports an accelerated late-time cosmic expansion without the need for exotic dark energy species. The sDGP model, however, is not deemed as a viable alternative to standard $\Lambda$CDM due both to theoretical difficulties such as ghost instabilities \citep[e.g.,][]{Luty:2003vm,Charmousis:2006pn} and to tensions between its predictions and observational datasets \citep[e.g.,][]{Fairbairn:2005ue,Maartens:2006yt,Fang:2008kc,Lombriser:2009xg}. In this paper, we work with the normal branch  \citep[][nDGP]{Schmidt:2009sv} model, for which the modified Friedmann equation is given by
\begin{equation}\label{eq:H_ndgp}
\frac{H(a)}{H_0} = \sqrt{\Omega_{\rm m}a^{-3} + \Omega_{\rm DE}(a) + \Omega_{\rm rc}} - \sqrt{\Omega_{\rm rc}}\,,
\end{equation}
in which $\Omega_{\rm rc} \equiv 1/(4H^2_0r^2_c)$.  Similarly to the Hu-Sawicky $f(R)$ model, the nDGP model does not support self-acceleration, and as a result some additional dark energy component has to be added in order to explain the late-time cosmic acceleration. This naturally makes it less appealing as an alternative to $\Lambda$CDM, but it is nevertheless widely-used in studies of modified gravity as a representative  model featuring the Vainshtein screening mechanism \citep[][]{Vainshtein:1972plb,Babichev:2013usa} and other interesting phenomenology. In this study, we take advantage of this flexibility by  tuning the additional dark energy component $\Omega_{\rm DE}(a)$ such that it counteracts the effect on the background expansion and gives rise to an expansion history identical to that of $\Lambda$CDM: the motivation for this is to enforce expansion histories that are identical between nDGP and $\Lambda$CDM, so that the two models only differ in terms of structure formation. Therefore, departures from GR are quantified exclusively by the parameter $H_0r_c$. As we can see from Eq.~\eqref{eq:H_ndgp}, if $H_0r_c \rightarrow \infty$ then the expansion of the Universe reduces  to $\Lambda$CDM, with the additional dark energy, whose density parameter is denoted by $\Omega_{\rm DE}(a)$ in Eq.~\eqref{eq:H_ndgp}, closer to a cosmological constant $\Lambda$. 

Cosmological structure formation in the nDGP model is again governed by a modified Poisson equation:
\begin{equation}\label{eq:poisson_nDGP}
\boldsymbol{\nabla}^2 \Phi = 4\pi G a^2 \delta \rho_{\rm m} + \frac{1}{2}\boldsymbol{\nabla}^2\varphi\,,
\end{equation}
and an equation of motion for the scalar field ($\varphi$)  \citep{Koyama:2007ih}:
\begin{equation}\label{eq:phi_dgp}
\boldsymbol{\nabla}^2 \varphi + \frac{r_c^2}{3\beta\,a^2c^2} \left[ (\boldsymbol{\nabla}^2\varphi)^2
- (\boldsymbol{\nabla}_i\boldsymbol{\nabla}_j\varphi)^2 \right] = \frac{8\pi\,G\,a^2}{3\beta} \delta\rho_{\rm m}\,,
\end{equation}
where 
\begin{equation}\label{eq:beta_dgp}
\beta(a) \equiv 1 + 2 H\, r_c \left ( 1 + \frac{\dot H}{3 H^2} \right ) = 1 + \frac{\Omega_{\rm m}a^{-3} + 2\Omega_\Lambda}{2\sqrt{\Omega_{\rm rc}(\Omega_{\rm m}a^{-3} + \Omega_{\Lambda})}}\,,
\end{equation}
is a time-dependent function, with $\Omega_\Lambda \equiv 1 - \Omega_{\rm m}$. In the nDGP model we consider here, $\beta$ decreases over time is always positive. The field $\varphi$ is called the `brane-bending mode', a scalar quantity describing the position of the 4D brane along the fourth spatial dimension. 

Again, from Eq.~\eqref{eq:poisson_nDGP}, we can observe that the brane-bending scalaron field acts as the potential of a fifth force. We can deduce from Eq.~\eqref{eq:phi_dgp} that its solutions have very different behaviours in two opposite limits: (\textit{a}) low-density regions, where $\boldsymbol{\nabla}^2\varphi$ is small and so the $\left(\boldsymbol{\nabla}^2\varphi\right)^2$ and $(\boldsymbol{\nabla}_i\boldsymbol{\nabla}_j\varphi)^2$ terms in Eq.~\eqref{eq:phi_dgp} are subdominant -- in this case we have $\boldsymbol{\nabla}^2\varphi\sim8\pi{G}a^2\delta\rho_{\rm m}/(3\beta)$, and so the strength of the fifth force is proportional to that of the standard Newtonian force, leading to an enhancement of Newton's constant from $G$ to $(1+1/{3\beta})G$; (\textit{b}) high-density regions, where $\boldsymbol{\nabla}^2\varphi$ is large, but the quadratic terms in Eq.~\eqref{eq:phi_dgp} become even larger, so that $\boldsymbol{\nabla}^2\varphi\ll8\pi{G}a^2\delta\rho_{\rm m}/(3\beta)$ -- in this case the fifth force term in Eq.~\eqref{eq:poisson_nDGP} is much smaller than the standard Poisson term. This is essentially the Vainshtein screening mechanism at work.

The \textsc{bridge}  simulations used in this work cover nDGP models with $H_0r_c$ values between $0.25$ and $10$ \citepalias[see Table \ref{table:forge_nodes} for further details, and][]{BRIDGE}. These simulations share the same cosmological parameter values and initial conditions as the \textsc{forge} simulations, and differ only in the gravity model. Moreover, we matched the order in the strength of the MG parameters,  such that models close to GR in \textsc{forge} are also close to GR in \textsc{bridge}.

\subsection{$N$-body simulations}
\label{subsec:Nbody}

To date, cosmological simulations are the only known tool for making accurate predictions of physical quantities and observables of the large-scale structure down to the small non-linear scales where  perturbation theory fails. The need for simulations in the study of modified gravity models is even stronger because of the additional non-linear behaviours caused by the fifth force. Over the past decade, various simulation techniques and codes have been developed for such models \citep[see, e.g.,][and references therein, for some reviews and code comparison results]{Llinares:2018maz,Li:2018book,Winther:2015wla}. 

The simulated lensing data described in this paper are based on the \textsc{forge} and \textsc{bridge} simulation suites described respectively in \citetalias{FORGE} and \citetalias{BRIDGE}. Four parameters are varied simultaneously, namely the matter density parameter $\Omega_{\rm m}$, the structure growth parameter $S_8\equiv \sigma_8\sqrt{\Omega_{\rm m}/0.3}$ where $\sigma_8$ is the usual root-mean-squared of the density fluctuations smoothed on $8\,h^{-1}\textrm{Mpc}$ scales, the reduced Hubble parameter $h$ and either $f_{R_0}$ or $H_0 r_c$ for the \textsc{forge} or the \textsc{bridge} suite, respectively. These two four-dimensional parameter spaces are each sampled at 50 nodes organised in a Latin Hyper cube, as detailed in Table \ref{table:forge_nodes}. Details of the $N$-body calculations are provided in the references mentioned above, but we provide here a brief summary of the numerical methods used.

For the \textsc{forge} simulations, the non-linear evolution of the particle distribution is obtained by the {\sc Arepo} Poisson solver \citep{Springel:2010mn,Weinberger:2020apjs}, which is used to compute the standard Newtonian force. This is augmented by a multigrid relaxation solver for Eq.~\eqref{eq:fR} based on a second-order-accurate finite difference scheme, which computes the fifth force arising from $f(R)$ gravity on the local grid elements. Adaptive mesh refinement (AMR) is adopted, in which grid elements where the matter density exceeds some threshold are refined (split) into eight child cells with doubled spatial resolution: this ensures that higher resolution is used in regions where a higher accuracy is needed in the scalar field solver. The additional force is then interpolated onto the positions of particles and used to update their velocities using the standard leapfrog scheme, achieving second-order accuracy in the time integral. The relaxation algorithm described in \citet[][]{Bose:2016wms} and extended by \citet[][]{Ruan:2021wup} has been implemented, improving the numerical stability and  convergence rate;  complete details on  \textsc{Arepo-MG} can be found in \citet[][]{Arnold:2019vpg}.

The \textsc{bridge} simulations are also carried out with \textsc{Arepo} and using multigrid relaxation with the same code structure, except that we are instead solving the differential equation governing the dynamics of the brane-bending mode $\varphi$ given by Eq.~\eqref{eq:phi_dgp}. Since this equation differs in type from Eq.~\eqref{eq:fR}, the algorithm introduced in \citet[][]{Li:2013tda,Li:2013nua} is used instead to ensure numerical stability. To further improve the efficiency of the code, the scheme described in \citet[][]{Barreira:2015xvp} is used, where, instead of solving the scalar field equation on all levels of mesh refinements (labelled by $l$), it is only solved on the lowest few levels; in other words, the scalaron solver is truncated at some level $l=l_{\rm trunc}$, and the solutions of $\varphi$ on level $l_{\rm trunc}$ is interpolated to all higher levels. \citet[][]{Barreira:2015xvp} show that this truncation, while an approximation, leads to negligible errors in the quantities of interest in cosmology. This is because the Vainshtein screening mechanism is very efficient at suppressing the fifth force in high-density regions, which happen to be the highly refined regions of the simulation grid; while the truncation and interpolation causes certain errors in the calculated fifth force in such regions, these are small errors on a small quantity, which have a small overall impact on the simulation results. For further details of the implementation in \textsc{Arepo-MG}, see \citet[][]{Hernandez-Aguayo:2020kgq}.

Each of the \textsc{forge} and \textsc{bridge} simulation suites consists of a total of 200 collisionless, dark-matter-only runs covering the 50 $f(R)$ and nDGP models mentioned above. For each node we have run two independent realisations with initial conditions chosen such that the sampling variance is greatly reduced in the mean matter power spectrum \citepalias[see][for more details]{FORGE}, at two different resolutions. The \textit{high-resolution} simulations employ $1024^3$ particles in a $500\,h^{-1}\textrm{Mpc}$ simulation box, at a mass resolution of $m_\mathrm{p} \simeq 9.1 \times 10^9h^{-1}M_\odot$\footnote{This number is for the fiducial $\Lambda$CDM model, or Node 0. The actual mass resolution varies in the 50 nodes due to their different cosmological parameter values.}, while  the \textit{low-resolution} simulations evolve $512^3$ particles in a simulation box size of $1500\,h^{-1}\textrm{Mpc}$, with a mass resolution of $m_\mathrm{p} \simeq 1.5 \times 10^{12} h^{-1}M_\odot$ (the values of $m_\mathrm{p}$ quoted here are for the fiducial $\Lambda$CDM node). The gravitational softening lengths are respectively $15\, h^{-1}\mathrm{kpc}$ and $75\, h^{-1} \mathrm{kpc}$ for the high- and low-resolution runs. For all simulations, we have fixed the power index of the primordial power spectrum, the present-day baryonic density parameter and the dark energy equation of state to $n_\mathrm{s} = 0.9652$, $\Omega_{\rm b} = 0.049199$ and $w=-1$.  Note that the lensing maps described in this paper only use the high-resolution runs, and that corresponding GR-$\Lambda$CDM simulations exist for all 50 nodes.

All simulations start at $z_{\rm ini}=127$, with initial conditions (ICs) generated using the 2\textsc{lptic} \citep[][]{Crocce:2006mn} code, an IC generator based on \textsc{n-genic} \citep[][]{Springel:2005nat} that uses second-order Lagrangian perturbation theory to compute  more accurately the initial particle displacements for a given matter power spectrum. The linear matter power spectra for all models are generated with the public Boltzmann code \textsc{camb} \citep[][]{Lewis:1999bs}, with the cosmological parameters specified in Table \ref{table:forge_nodes}. Note that for all $f(R)$ and nDGP models, we assume that the linear power spectra are identical to their $\Lambda$CDM \textit{counterparts}, i.e., the $\Lambda$CDM models with the same cosmological parameters -- this is a good approximation since at the initial time ($z=127$) any effect of modified gravity is negligible for the models considered here. In other words, they share the same primordial amplitude $A_{\rm s}$. 
Finally, for each cosmological model, we precompute the redshifts $z$ at which particle data\footnote{Dark matter haloes are also extracted and will be used in companion papers.} have to be written to disk such that the consecutive projections of half simulation boxes can be used to construct contiguous light-cones up to $z = 3.0$. We describe the construction of our  mass shells in Sec. \ref{subsec:WL_sims}.

It is important to emphasise here that the $\sigma_8$ and $S_8$ quantities reported in this work correspond to  the input truth values at which the GR-$\Lambda$CDM $N$-body simulations are run. When turning on MG however, the non-linear excess structure generated by the fifth force increases the late time $\sigma_8$ values by an amount difficult to predict, hence our choice of labelling the simulations by their GR-$\Lambda$CDM quantities\footnote{To avoid any possible confusion, we could label the structure growth parameters as $\sigma_8^{\rm GR}$ and $S_8^{\rm GR}$ but decided against to declutter  notation.}.

Although this paper focuses on two-point statistics, it serves the additional purpose of presenting the infrastructure necessary for companion papers based on lensing statistics beyond two-point. One of the key ingredients for  such measurements is the covariance matrix, for which analytical solutions generally do not exist. We therefore use the public SLICS simulations\footnote{SLICS: slics.roe.ac.uk} for this, a suite of  954 fully independent $N$-body runs that evolve 1536$^3$ particles in a box size of 505 $h^{-1}$Mpc on the side. These are all produced at a fixed cosmology\footnote{GR-$\Lambda$CDM SLICS cosmology: $\Omega_{\rm m} = 0.2905$, $\sigma_8 = 0.826$, $h=0.6898$, $n_{\rm s} = 0.969$.} and vary only in their initial conditions, therefore providing an ideal tool for estimating sample covariance. We refer the reader to \citet{SLICS_1} for full details on the SLICS $N$-body ensemble.

 The SLICS, \textsc{forge} and \textsc{bridge} simulations are post-processed uniformly, creating mock survey light-cones suitable for cosmological inference. Details on the post-processing involved are presented in Sec. \ref{subsec:WL_sims}. Beforehand, we first introduce the basic ingredients that enter our theoretical predictions.

\subsection{Modified gravity emulators}
\label{subsec:CNN}

The information content of the large scale structure is largely encapsulated in the matter power spectrum, $P_{\delta}(k;z)$, a two-point statistics that is directly measurable from the matter density fields $\delta$ in simulations and that can be inferred from galaxy surveys via clustering or cosmic shear measurements.  For example, the $N$-body simulations described in \citetalias[][]{FORGE} are used to construct the public $P_{\delta}$ \textsc{forge} emulator, obtained by training a Gaussian Process Regressor (GPR) on the measurements obtained from the 50 \textsc{forge} nodes; the emulator provides predictions that are accurate to better than 2.5 percent up to $k=10.0 \,h {\rm Mpc}^{-1}$ over the majority of the parameter volume. 

As an alternative, we use here the same measurements to train instead fully connected neural networks (FCNN), which are especially powerful at high-dimensional interpolation (as in CL22). We train in this work a neural network with the same characteristics on both \textsc{forge} and \textsc{bridge} data, as a function of redshift. The neural network is defined by an input layer, composed of the four cosmological parameters ($\Omega_{\rm m}$, $h$, $\sigma_8$ and the modified gravity parameter, either $\bar{f}_{R0}$ for $f(R)$, or $H_0r_c$ for nDGP gravity) and the redshift $z$, two hidden layers of $256$ units each, and an output layer that returns the power spectrum evaluated at the different $k$-bins. In between hidden layers, we use a Gaussian error Linear Unit (GeLU) activation function \citep{2016arXiv160608415H} to add a differentiable non-linearity to the relation between inputs and outputs.

To find the optimal parameters that reproduce the statistics measured in the $N$-body simulations, we minimise the $\mathcal{L}_1$ loss function, defined as:
\begin{equation}
    \mathcal{L}_1 = \frac{1}{N} \sum_{i=0}^N |y^i_\mathrm{true} - y^i_\mathrm{predicted}|
\end{equation}
using the Adam optimiser \citep{2014arXiv1412.6980K}. In the above expression, the $y^i$ are the true and predicted matter power spectra for each of the simulations and each of the snapshots in the simulation suite, and $N$ is the batch size used in the training.

Moreover, we avoid fine-tuning the learning rate with a scheduler that reduces the learning rate by a factor of $10$ when the validation loss does not improve after $30$ epochs. We also stop training the model when the validation loss does not improve after $100$ epochs.
An in-depth description of the emulator and its validation are presented in \citetalias[][]{BRIDGE}, together with the emulator's code. 

More precisely, the emulator outputs the modified gravity {\it enhancement factor},  $B(k,z)$, which is defined as:
\begin{eqnarray}
B(k;z) = P_{\delta, \rm MG}(k;z)/P_{\delta, \rm HaloFIT}(k;z) \, .
\label{eq:Bk}
\end{eqnarray}
Here $P_{\delta, \rm MG}(k;z)$ is the measurement for a modified gravity model from either the \textsc{forge} or {\sc bridge} simulations, and $P_{\delta, \rm HaloFIT}(k;z)$ is the prediction by \textsc{halofit} \citep{Takahashi2012} for the $\Lambda$CDM counterpart of that model (we refer the reader to \citetalias[][]{FORGE} for a more complete description of how this is achieved in practice). The MG enhancement can be as high as 40 per cent depending on the model, for the \textsc{forge} nodes. We find that the FCNN slightly outperforms the GPR at modelling the enhancement factor and is therefore our method of choice, for all gravity models.

Finally, we notice that in the weak $f_{R_0}$ limit the emulator does not converge exactly to the GR case: residual deviations of a few percent are observed at all scales. These same residuals are also present in the  power spectrum training set, as reported in Figure 5 of A21. Although generally small, some segments of our analysis require a smooth convergence to GR, hence we linearly interpolate the emulated $B(k)$ in the range  log$_{10}\left[f_{R_0} \right] = [-7, -6.0]$, enforcing $B(k)=1.0$ at the lower end and for any values smaller than $-7$. The weak nDGP limit does not show such residuals and hence interpolation is not necessary in that case.

\subsection{Cosmic shear two-point functions}
\label{subsec:wl-th}

Two-point functions are the primary cosmic shear measurement methods and exists in different flavours, including two-point correlation functions, angular power spectra, band powers or COSEBIs \citep[see][for a comparison between some of these]{KiDS1000_Asgari}. One of the key advantage of these cosmic shear statistics is that their modelling can be directly linked to the matter power spectrum, $P_{\delta}(k;z)$. Thanks to an increased precision in the estimation of the redshift distributions, the lensing catalogues  are now routinely split into different redshift bins, allowing for tomographic analyses of the data that better measure those parameters impacting the growth rate of large scale structure. Specifically, predictions for the cosmic shear power spectrum $C_{\ell}^{\kappa,ij}$ can be obtained from a Limber integration over the matter power spectrum via \citep[see][for a review]{Kilbinger17}:
 \begin{eqnarray}
C_{\ell}^{\kappa, ij} = \int_0^{\chi_{\rm H}}  \frac{q^i(\chi) \,q^j(\chi)}{\chi^2} \, P_{\delta}\, \bigg(\frac{\ell+1/2}{\chi};z(\chi)\bigg) \ {\rm d}\chi,
\label{eq:C_ell}
\end{eqnarray}
where $\chi_{\rm H}$ is the comoving distance to the horizon, and ($i, j$) label the different tomographic bins. The lensing kernels $q^i(\chi)$ are computed as:
\begin{eqnarray}
q^i(\chi) = \frac{3}{2}\Omega_{\rm m} \, \bigg(\frac{H_0}{c} \bigg)^2 \frac{\chi}{a(\chi)} \int_{\chi}^{\chi_{\rm H}} n^i(\chi')\frac{{\rm d}z}{{\rm d} \chi'}\frac{\chi' - \chi}{\chi'}{\rm d}\chi',
\label{eq:q_lensing}
\end{eqnarray}
where $n^i(z)$ is the redshift distribution of the source galaxies in tomographic bin $i$.

The matter power spectrum entering Eq. (\ref{eq:C_ell}) can be obtained from an array of public codes such as  {\sc HaloFIT} \citep{Takahashi2012}, HMcode \citep{HMcode2020}, {\sc CosmicEmu} \citep{Coyote3},  {\sc BaccoEmulator} \citep{BACCO} or the  {\sc EuclidEmulator} \citep{EuclidEmulator}. Whereas these codes provide highly accurate predictions tools for many cosmological models, their gravity model is restricted to that of GR only. We therefore generate MG lensing predictions  by multiplying the {\sc HaloFIT} predictions by $B(k;z)$  as in Eq. (\ref{eq:Bk}), and then inserting the results into Eq.~(\ref{eq:C_ell}). The Limber integral is carried out by {\sc cosmoSIS}\footnote{{\sc cosmoSIS}: cosmosis.readthedocs.io/en/latest/index.html} cosmology package \citep{cosmosis}, which we also use for parameter inference (see Sec. \ref{sec:MCMC}).


\begin{figure}
\begin{center}
\includegraphics[width=3.1in]{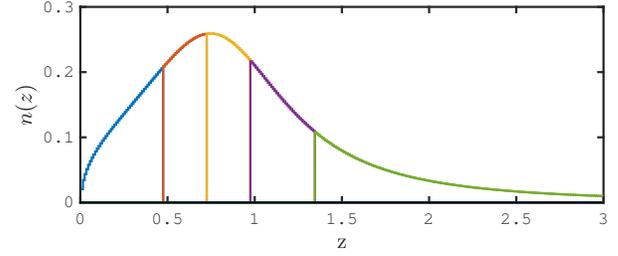}
\caption{Normalised redshift distribution of the five tomographic bins considered in our mock survey. }
\label{fig:nz}
\end{center}
\end{figure}

\begin{table}
   \centering
   \caption{Properties of our Stage-IV survey.   The specifications closely follow those presented in \citet{Martinet20}, with $n_{\rm eff}$=6.0 gal/arcmin$^2$ per tomographic bin and $\sigma_\epsilon$=0.27 per component.}
   \tabcolsep=0.11cm
      \begin{tabular}{@{} lcccc|cc @{}} 
                  
      \hline
      \hline
      tomo     &$z$ range&  $\langle z \rangle$  \\
       \hline
       bin1    & $0.0 - 0.4676$&$0.286$ \\
       bin2    & $0.4676 - 0.7194$& $0.600$ \\
       bin3    & $0.7194 - 0.9625$& $0.841$\\
       bin4    & $0.9625 - 1.3319$ &$1.134$\\
       bin5    & $1.3319 - 3.0$& $1.852$\\
    \hline 
    \hline
    \end{tabular}
    \label{table:survey}
\end{table}

Our mock Stage-IV lensing survey is designed to investigate some of the conditions that would allow MG to be detected by upcoming two-point statistics analyses.
We opted for a source redshift distribution described by:
\begin{eqnarray}
n(z) =A \frac{z^a + z^{ab}}{z^b + c} \,\, , 
\label{eq:nz_euclid}
\end{eqnarray}
with $A=1.7865$, $a=0.4710$, $b=5.1843$, $c=0.7259$. This $n(z)$ is  shown in Fig. \ref{fig:nz} and is taken from \citet{Martinet21, Martinet20} and \citet{2021MNRAS.tmp.2970H}. This sample is further split into five tomographic bins, each with a galaxy density of $n_{\rm gal}=6.0$ gal arcmin$^{-2}$ and shape noise of $\sigma_{\epsilon} = 0.27$ per component.  A summary of the mock survey properties is presented in Table \ref{table:survey}. We assume a survey area of 5000 deg$^2$, which corresponds to the total area sampled by our flat-sky simulations at each cosmological nodes (see Sec. \ref{sec:sims}).

\begin{figure}
\begin{center}
\includegraphics[width=3.4in]{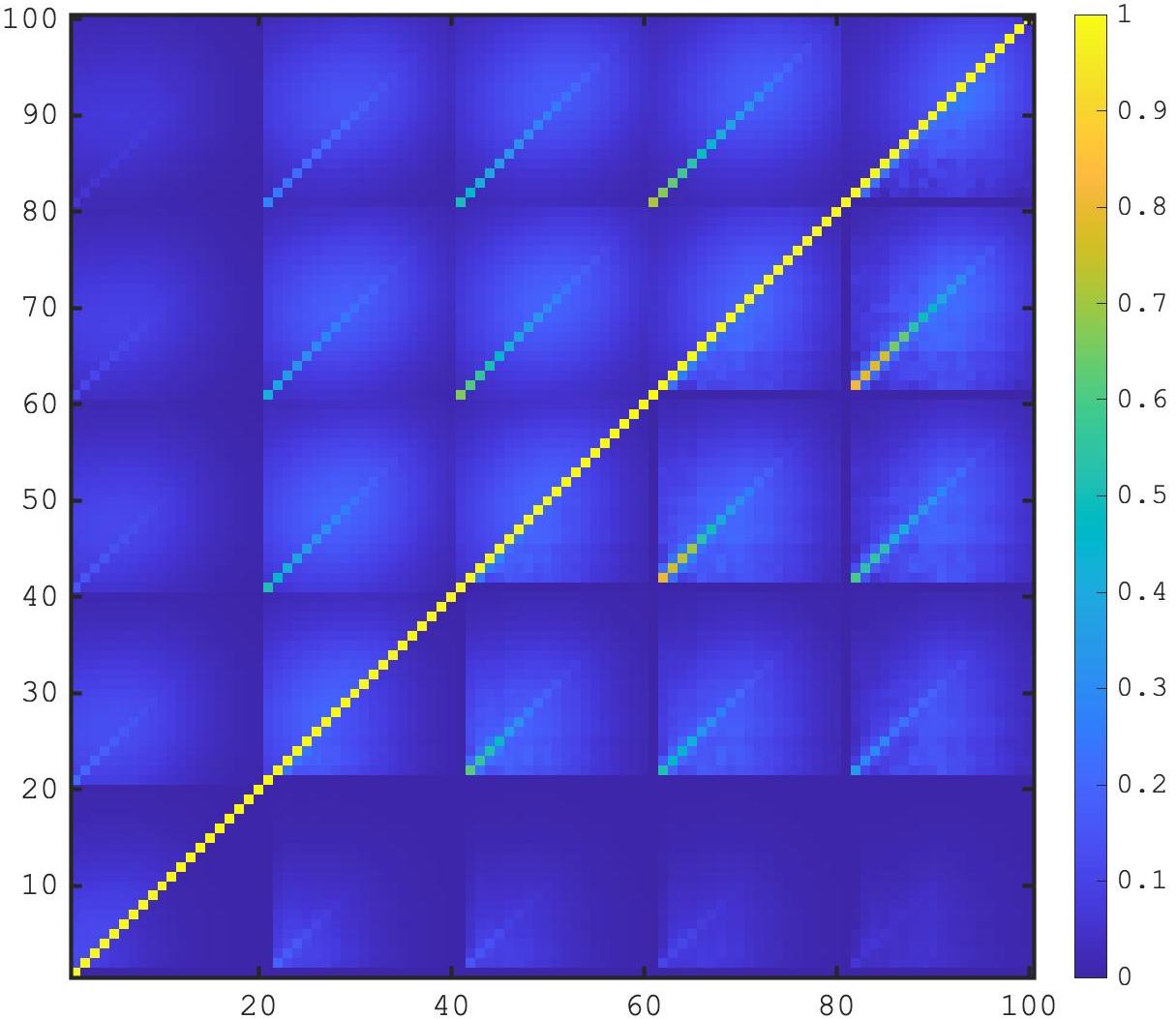}
\caption{Cross-correlation coefficient matrix of our lensing power spectrum data vector, defined as $r_{ij} = C_{ij}/\sqrt{C_{ii}C_{jj}}$. The upper-left triangle shows the analytical calculations, while the lower-right part is estimated from the SLICS simulations (Sec. \ref{subsec:WL_sims}). The redshift bins increase towards the upper-right corner.}
\label{fig:CorrMatrix}
\end{center}
\end{figure}


Sect. \ref{sec:MCMC} details 
our likelihood sampling analysis, which takes as input  a data vector, a covariance matrix  and a theoretical model in which  cosmology, gravity and nuisance parameters are varied simultaneously. As our baseline we use an analytical covariance matrix that describes the mode correlations, the shape noise and the sampling covariance expected for the different elements of our data vector. The calculations  are fully detailed and validated in \citet{cosmoSLICS} and \citet{KiDS1000_Joachimi} and we refer the reader to these for more information. In short they include the Gaussian, non-Gaussian and Super-Sample Covariance terms given a cosmology, a tomographic  redshift distribution, a survey area and binning specifications for the angular multipoles. The non-Gaussian term requires an expensive trispectrum evaluation, while the SSC term assumes a circular survey geometry of 5000 deg$^2$. We show in Fig. \ref{fig:CorrMatrix} the cross-correlation coefficient matrix obtained with our survey specifications\footnote{We use the SLICS cosmology in the analytical covariance matrix calculations.}, and compare our results to an estimate obtained from the SLICS, which we describe in Sec. \ref{subsec:WL_sims}. Aside some residual noise patterns, both methods completely agree. We will quantify the impact of switching between these two later on, but basically the effect is completely subdominant given our statistical precision. This comparison validates both the theoretical approach and the SLICS maps, which will be used in companion non-Gaussian statistics studies.

\section{Weak lensing simulations}
\label{sec:sims}

The \textsc{MGLenS} weak lensing simulations are constructed by ray-tracing\footnote{Ray-tracing in this paper assumes the Born approximation.} through series of mass shells obtained by collapsing the particle data either along one of the Cartesian axes (flat-sky method) or along the radial direction (curved-sky). Both methods have their pros and cons;  we focus mainly on the flat-sky results in this paper for their ability to probe deeper in the small, non-linear regime, and discuss the curved-sky method in Appendix \ref{sec:WL_sims_curved}. In either case, the mass sheets have a comoving thickness equal to exactly half the simulation box size (i.e. 250 $h^{-1}$Mpc), and between 15 and 23 shells are needed to continuously fill the light-cones up to $z=3$, depending on cosmology. We finally produce convergence maps for the five tomographic redshift bins shown in Fig. \ref{fig:nz}.  In this paper we do not train our emulator on statistics measured from these maps and instead aim for their validation, however this logical extension will be presented in companion papers.

\subsection{Weak lensing maps and power spectra}  
\label{subsec:WL_sims}

\begin{figure*}
\begin{center}
\includegraphics[width=6.6in]{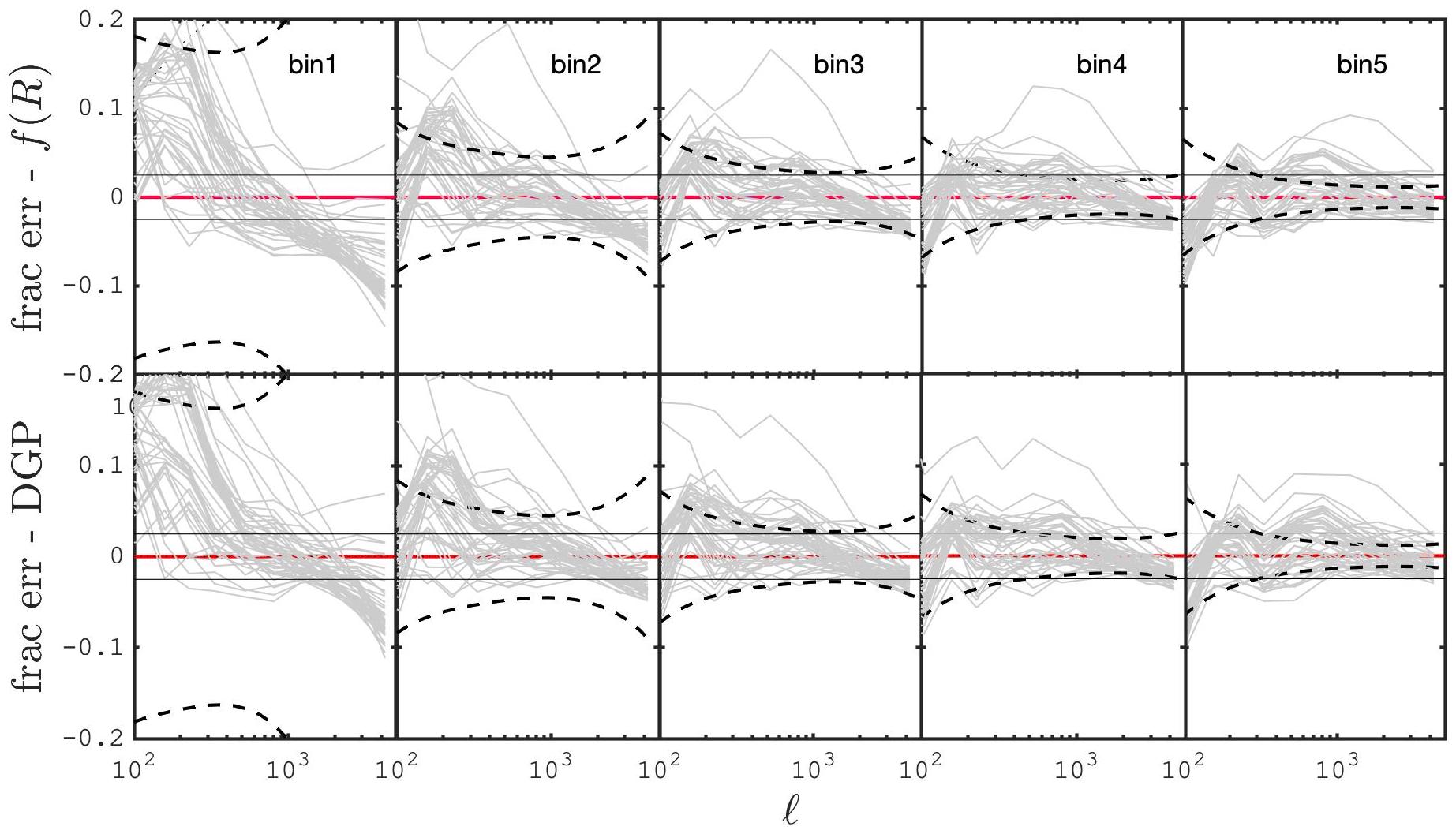}
\caption{Fractional error between the emulated lensing power spectrum and that measured from the \textsc{forge} (upper) and \textsc{bridge} (lower) simulations. The grey lines are obtained for all 50 nodes, each time averaged over the 50  light-cones (two per initial conditions). The black dashed lines indicate the $1\sigma$ statistical error expected from our mock survey. Redshift increases from left to right, and the thin horizontal lines mark the 2.5\% precision target. }
\label{fig:frac_err_all}
\end{center}
\end{figure*}


Our flat-sky method heavily builds from the {\sc SimulLens} algorithm, the multiple-plane technique described in \citet{SLICS_1}: at each preselected redshift
,  the particles from half the simulation volume are projected along the shorter direction and assigned onto a 12,288$^2$ grid. This process is repeated with the other half-volume, and for the other two Cartesian axes, such that 6 density planes are extracted per snapshot. 

Light-cone mass maps, $\delta_{2D}(\boldsymbol \theta, z_{\rm lens})$, are extracted from the density planes with an opening angle of 10 deg$^2$ and $7745^2$ pixels. At each redshift, one of the 6 aforementioned planes is randomly selected and a random origin offset is added.
This means that correlations between different mass shells are broken, but it was shown in \citet{ZorrillaMatilla} that this has a subdominant effect on weak lensing statistics due to the line-of-sight projection. Closely following  \citet{cosmoSLICS}, we repeat this whole ray-tracing procedure in order to create 25 {\it pseudo}-independent light-cones $\delta_{2D}(\boldsymbol \theta, z_{\rm lens})$ maps from  each $N$-body run\footnote{We change the random seeds between the 25 cones at a given cosmology, but use the same 25 seeds for every cosmology node, thereby keeping to a minimum the sampling variance across cosmological models.}. Periodic boundary conditions are used  wherever the area of the light-cone becomes larger than the simulation box itself.

In the multiple-plane approximation, each of these mass shells acts as a discrete gravitational lens, distorting the light as it passes through it. Within the Born approximation, the convergence $\kappa$ experienced by photons propagating along the direction $\boldsymbol \theta$ and originating from a source redshift distribution $n(z)$ can be computed as:
\begin{eqnarray}
\kappa^{i}({\boldsymbol \theta}) = \sum_{{\rm lens}} \delta_{2D}(\boldsymbol \theta, z_{\rm lens}) \,\, q^{i}\left(\chi(z_{\rm lens})\right)\, ,
\end{eqnarray}
where $q^{i}(\chi)$ is the tomographic lensing kernel given by Eq. (\ref{eq:q_lensing}), and  the index `{\rm lens}' runs over all foreground lens planes in the light-cone. 

The cosmic shear power spectra are estimated from the square of the Fourier-transformed convergence map, first averaged in annuli of thickness $\Delta \ell=35$ centred on $\ell \in [35 - 5000]$:
\begin{eqnarray}
C_{\ell}^{\kappa, ij} = \langle \kappa^{i}({\boldsymbol \ell})  \kappa^{j}({\boldsymbol \ell}) \rangle_{d\Omega} \, ,
\end{eqnarray}
with $\langle ... \rangle_{d\Omega}$ denoting an angular average over the solid angle of the annulus.  Our measurements are then rebinned into 25 logarithmically-spaced bands over the same $\ell$-range to further reduced the sampling noise. 
We refer the reader to \citet{SLICS_1} for more details on the numerical implementation of our lensing power spectrum estimation method, which includes a mass-assignment de-biasing step; we have also checked that our measurements are consistent with those using the public code {\sc LensTools}\footnote{lenstools.readthedocs.io/en/latest/} \citep{LensTools}.
Our fiducial cosmological inference excludes $\ell<150$ modes as these are not well measured on our 10$\times$10 deg$^2$ patches, and are  affected by the finite lens thickness. The high-$\ell$ limit is an optimistic scenario, since in the real Universe these multipoles are plagued with systematic effects such as baryonic feedback, which are difficult to model and largely uncertain \citep{HorizonAGN}.   We therefore consider as well a more conservative scenario that further excludes the $\ell>3000$ modes. Note that we only extract the auto-angular power spectra in this work, however it is straight-forward to extend this to include cross-redshift terms as well. 

\subsection{Validation} 
\label{subsec:results}

\begin{figure*}
\begin{center}
\includegraphics[width=5.5in]{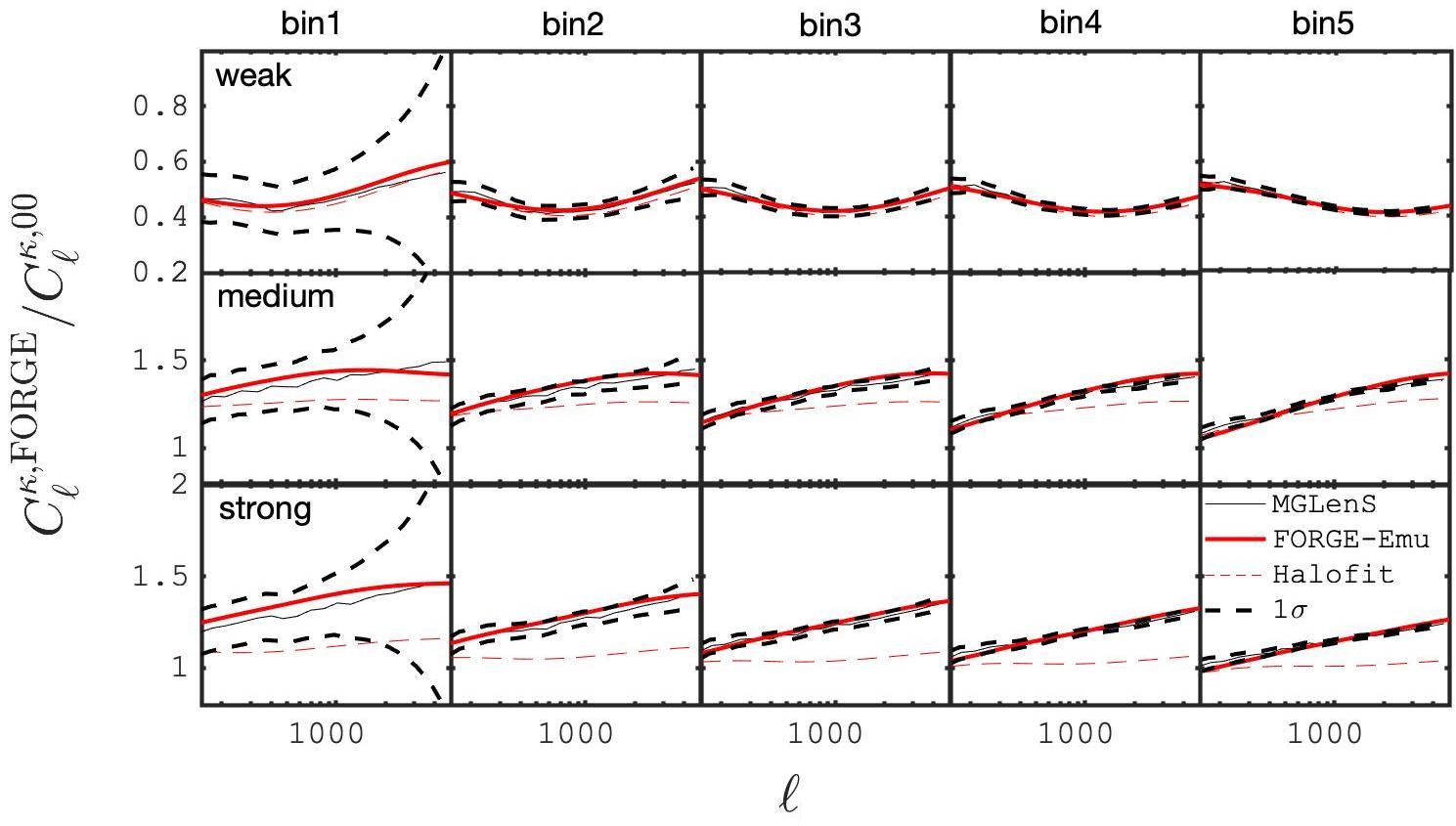}
\caption{Ratio between the tomographic weak lensing power spectrum of different \textsc{forge} models and that of model-00. Redshift increases from left to right, as indicated above the upper panels. Measurements from \textsc{MGLenS} are shown by the thin black lines, predictions from the \textsc{forge} emulator are indicated by the solid red lines, while the pair of thick dashed lines indicate the $\pm1\sigma$ statistical uncertainty expected from our mock Stage-IV lensing survey. We also plot with the thin dashed red lines the GR predictions from \textsc{Halofit} at these cosmologies. The \textsc{bridge} simulations and predictions reach a similar level of agreement.}
\label{fig:Cell_kappa_flat}
\end{center}
\end{figure*}

As a first validation test, we examine the fractional error between the $C_{\ell}$ measured  from the \textsc{forge} and \textsc{bridge} simulations and their respective  emulator predictions. We can see in Fig. \ref{fig:frac_err_all} that the agreement is generally at the few percent level except for the lowest redshift bin, where the deviations are much larger. These are caused by reduced accuracy in the multiple lens approximation, combined with flat-sky projection effects and broken correlations, yielding tilted residuals in the left-most panel. Note that however large this might seem, the precision of lensing surveys is massively reduced at low redshifts, as seen by the black dashed lines, such that these differences should not lead to noticeable biases at the inference stage. On small scales (large $\ell$-modes) most of the measurements scatter inside the  2.5\% region, consistent with the advertised 2.5\% accuracy on the power spectrum emulator reported in \citetalias[][]{FORGE}. The intermediate scales (300 $<\ell<1000$) exhibit a larger scatter reaching $\sim5\%$ at times, caused by limits in the emulator predictions combined with a small amount of residual sampling variance. From this we can expect that emulation of weak lensing statistics from these simulations  should also reach 2-3\% absolute accuracy.



\begin{figure*}
\begin{center}
\includegraphics[width=5.4in]{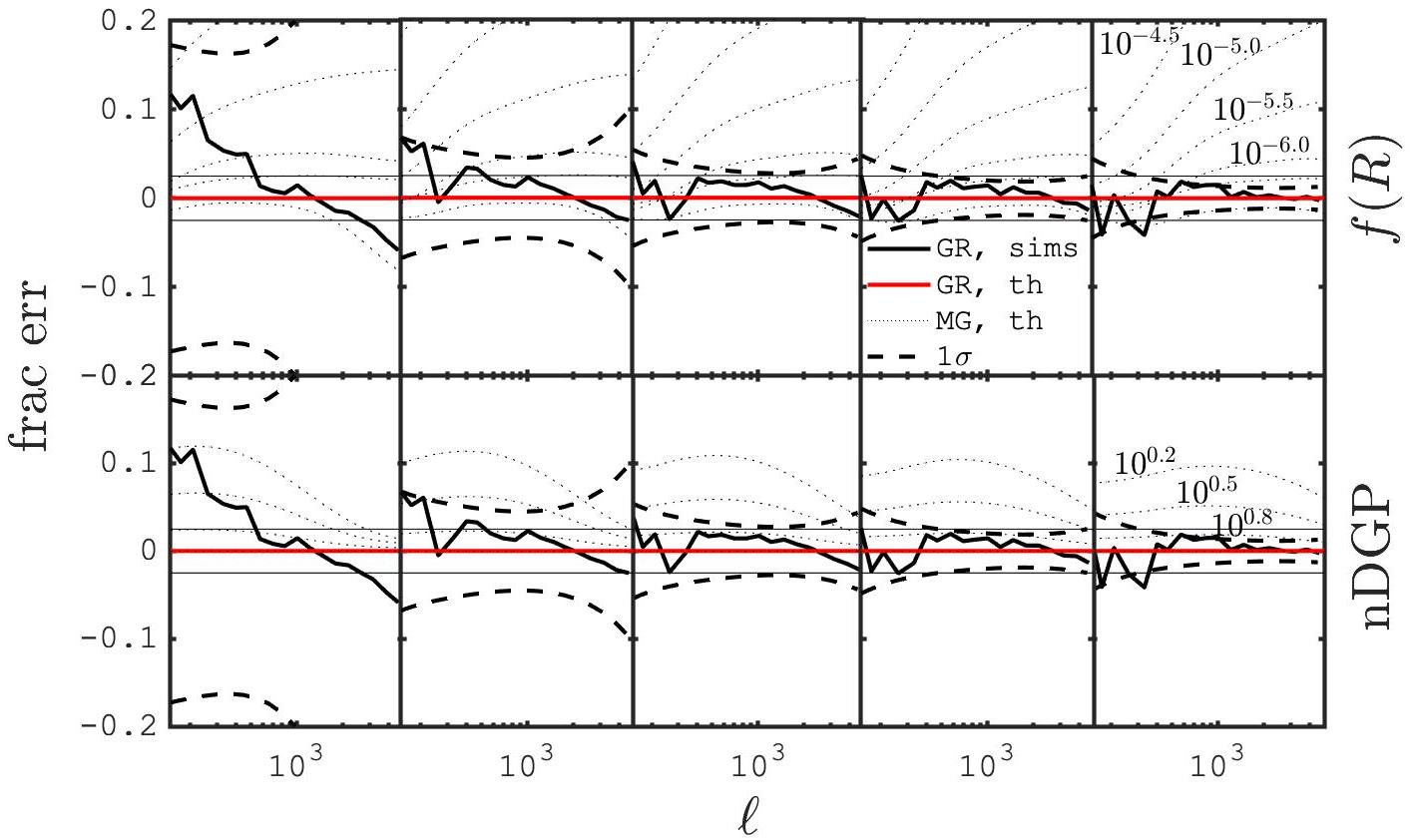}
\caption{{top:} Comparison between the lensing measurements on model-00 (black solid) relative to GR-theory (red solid), along with the expected error from the Stage-IV survey (dashed black), augmented with predictions for many $f(R)$ models (thin dotted). As before, redshift bins increases from left to right. (bottom:) Same as top panels, but now the dotted lines show nDGP model with different values of $H_0 r_c$.}
\label{fig:model_00_fr}
\end{center}
\end{figure*}

We next examine the cosmological dependence of the measurements in comparison with the emulators, shown in  Fig. \ref{fig:Cell_kappa_flat}  for a representative sample of  \textsc{forge}  models.  These are labelled as {\it strong} (model-13), {\it medium} (model-18) and {\it weak} (model-04), referring to the strength of their departure from GR. The match is excellent here and for most other cases; discrepancies occur only for a handful of nodes with exceptionally low $\Omega_{\rm m}$, which are poorly modelled by {\sc HaloFIT} and by the \textsc{forge} emulator. This is well documented in \citetalias[][]{FORGE} and is not expected to affect our cosmology and gravity inference results, which are all centred on larger values of the matter density. 
The emulator predictions (in red solid) is generally within the statistical precision of our mock survey (shown with the dashed black lines) for $\ell < 1000$, beyond which it occasionally deviates by a few percent. This is caused by residual inaccuracies in the \textsc{forge} emulator itself, which was reported in A21 (see their figure 5) to emulate the simulated matter power spectrum only to a few percent precision. Similar agreements are found for all other {\sc forge} and {\sc bridge} models, which validates both the cosmology dependence of the light-cones and the {\sc cosmoSIS} implementation of the two MG emulators.

Also shown in  Fig. \ref{fig:Cell_kappa_flat} are the predictions for the pure GR case (see the thin red-dashed curve), obtained by setting  $B(k,z)=1.0$ while keeping the cosmology unchanged. The  difference with respect to the solid red line is solely due to the absence of the fifth force, and falls well outside the statistical error for most models. In other words, in absence of observational and astrophysical systematics that are not included in this figure, deviations from GR would likely be observed to a high significance in our survey, if the Universe followed either the medium or strong {\sc forge} models.  This raises a key question: given our mock survey, how weak could be detectable deviations from GR, if they exist? 
The first step in answering this is to understand what redshift and angular scales mostly contribute towards such a measurement, an exercise that we carry out next with a Fisher analysis.
 
\subsection{Fisher information}
\label{sec:fisher}

\begin{figure}
\begin{center}
\includegraphics[width=3.1in]{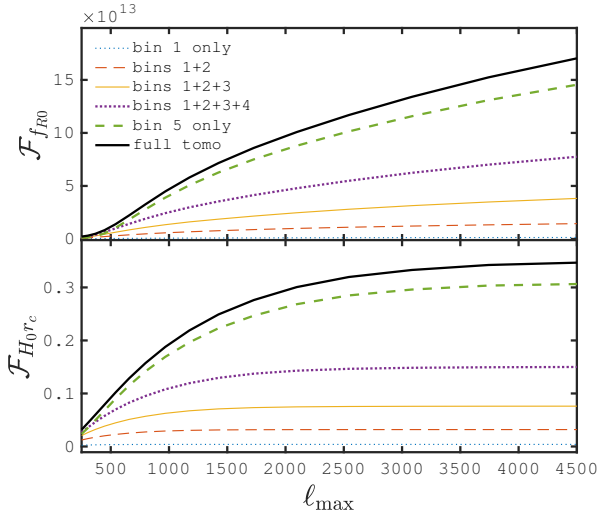}
\caption{Fisher information as a function of $\ell_{\rm max}$, the highest mode included in the data vector, shown here for different selections of tomographic bins. The top and bottom panels are for $f_{R_0}$ and $H_0 r_c$, respectively.}
\label{fig:fisher}
\end{center}
\end{figure}
 
The origin of the constraining potential on $f_{R_0}$ and $H_0 r_c$ from measurements of the lensing power spectrum is best understood by first fixing the cosmology in the emulators and varying only the modified gravity parameter. This is shown in Fig.~\ref{fig:model_00_fr} for cosmology otherwise identical to model-00, where we  compare the measurements from the flat-sky GR-$\Lambda$CDM simulations (solid black) to the {\sc forge} and {\sc bridge} predictions with different values of their MG parameters (the thin dotted lines). Also shown are the expected statistical uncertainty.  This figure suggests that the information about the $f_{R0}$ parameter mostly comes from the high redshift and high-$\ell$ modes, where the deviations with respect to GR are amplified and the statistical error bars vastly reduced.  In comparison, the constraints on $H_0 r_c$ arise from larger scales as well, again with the strongest detection potential coming from the highest redshift bins. This difference is driven by the type of fifth forces and screening mechanisms. In this section we dissect these signals and shine light on the data elements that better contribute to their constraints.

We carry out this investigation with a Fisher analysis \citep[see, e.g.][for a similar Fisher matrix calculation]{Takada2009a}, which is cheaper to run than a full MCMC while providing exactly the information we are seeking. Given measurements of the lensing power spectrum, the Fisher information about a parameter $\pi$ is obtain from 
\begin{eqnarray}
\mathcal{F}_\pi = \left[\frac{{\rm d}C_{\ell}}{{\rm d}\pi} \right] {\rm Cov}^{-1} \left[\frac{{\rm d}C_{\ell}}{{\rm d}\pi}  \right]^{\rm T} \, ,
\label{eq:fisher}
\end{eqnarray}
where Cov is the covariance matrix shown in Fig. \ref{fig:CorrMatrix}, which we assume to be cosmology independent in our calculation. A matrix product is taken between the three terms, resulting in a single scalar quantity per parameter $\pi$. In short, the contribution to the information is large for elements of the data vector that are highly sensitive to changes in $\pi$ (i.e. their derivative is large) and for which the covariance is small (the inverse is large).  The inverse of $\mathcal{F}$ provides an optimistic estimate of  the covariance about $\pi$, which in our one-dimensional case gives $\sigma_{f_{R0}}=\sqrt{\mathcal{F}_{f_{R0}}^{-1}}$ and $\sigma_{H_0 r_c}=\sqrt{\mathcal{F}_{H_0 r_c}^{-1}}$. 

Starting our dissection, we compute the Fisher information for different selections of the full data vector, first allowing variations in the maximal multipole included in our survey, $\ell_{\rm max}$.  The results are shown in Fig. \ref{fig:fisher} with the solid black line, where for $f_{R_0}$ we observe that the  increase in information remains significant for all scales included here. We notice a slight transition  past $\ell_{\rm max}=1500$ where the slope becomes shallower, due to the non-linear coupling between the different Fourier modes \citep{Takada2009a}.  The flattening of the slope is more pronounced for $H_0 r_c$, where a full information saturation is observed beyond $\ell=3000$, similar to that found by \citet[][see their figure 3]{Takada2009a}.

We next explore the impact of adding each of the tomographic bins one at a time. The second line from the top shows the information contained solely in the  highest tomographic bin, while the other lines correspond to different combinations of the lower redshift bins. It is clear from this that most of the information is contained in  bin 5, the other four bins providing only a modest additional gain.       

Using all scales and all tomographic bins, we could expect a  detection of at least $3\sigma$ if $f_{R0}> 2.3\times 10^{-7}$ or if $H_0r_c<5.1$, in absence of systematics and assuming that the cosmology is perfectly known from external data. We could include variations with cosmology and marginalisation over systematics in an upgraded Fisher calculation, however we choose instead to run full MCMC on mock data, yielding the most accurate picture of the inference capabilities provided by the \textsc{MGLenS} simulations. 
\section{Cosmology Inference}
\label{sec:MCMC}

\begin{table}
   \centering
   \caption{Priors used in our cosmological inference. Except for $\delta z^i$, all parameters are sampled with a uniform (i.e. flat) prior; a Gaussian prior of width 0.01 is applied on the redshift parameters, reflecting a realistic precision we should have on the redshift distributions.}
   \tabcolsep=0.11cm
      \begin{tabular}{@{} lcccc|cc @{}} 
                  
      \hline
      \hline
      parameter     & range \\
       \hline
       $\Omega_{\rm m}$    & 0.1 -- 0.55 \\
       $S_8$   & 0.6  -- 0.9 \\
       $h$    & 0.6 -- 0.82 \\
       log$_{10}[f_{R_0}]$    & -8.0 -- -4.5   \\
       log$_{10}[H_0 r_c]$   &  -0.6 -- 1.0  \\
       \hline
       $A_{\rm IA}$ &   -5.0 -- 5.0  \\
       $\delta z^i$ & -0.1 -- 0.1  \\
    \hline 
    \hline
    \end{tabular}
    \label{table:priors}
\end{table}

This section presents the inference method with which we quantify our ability to distinguish  cosmological and gravitational  parameters in different scenarios.  After validating our inference pipeline, we run a sensitivity test on both MG models, this time varying both cosmological and gravity parameters. We next validate the measurements from the \textsc{MGLenS} simulations, 
then investigate the catastrophic impact of analysing mock MG data with the wrong gravity model, thereby demonstrating the strong model-dependence of this approach. We finally study the impact of various systematics effects on these measurements.
Our data vector consists once again of the auto-spectra measured from the weak, medium and strong   \textsc{forge}/\textsc{bridge} models in all five tomographic bins. 

In all cases our likelihood assumes a standard multivariate Gaussian functional form with a fixed covariance matrix (see Section \ref{subsec:wl-th}).  The predictions are computed at arbitrary cosmologies using Eq. \ref{eq:C_ell} augmented with the $B(k,z)$ emulators, with a flat prior on the four parameters ($\Omega_{\rm m}$, $S_8$, $h$ and either log$_{10}[f_{R_0}]$ or log$_{10}[H_0 r_c]$) that spans  the range for which the emulators are supported, listed in Table \ref{table:inference}. One exception to this is the lower bound on  log$_{10}[f_{R_0}]$ which we set to $-8$ in order to  reduce prior effects in the weak MG limit. Otherwise the inference pipeline could wrongly reject log$_{10}[f_{R_0}]\sim -7$ simply because it is poorly sampled. As explained before, we set $B(k,z)$ to 1.0 whenever   log$_{10}[f_{R_0}]\in [-8,-7]$.  The other cosmological parameters are held fixed to the values used in the  $N$-body runs.


We carry out our cosmology inferences with the likelihood sampler {\sc Multinest} \citep{Multinest}, which is run within {\sc cosmoSIS}. This sampling method has been used and validated in a number of previous works, notably in the cosmic shear analysis of the KiDS-1000 data \citep{KiDS1000_Asgari} and of the DES-Year 3 data \citep{DESY3_Secco}.  It has been reported in \citet{LemosSampling} that the projected contours could be slightly over-constraining in some cases compared to alternative samplers, however we opted for {\sc Multinest} as it is much faster and its accuracy is sufficient to support the scientific goals of this paper. The chains all ran in 5000 steps and are analysed with  {\sc getdist}\footnote{{\sc getdist}:getdist.readthedocs.io/en/latest/.}. 

\subsection{Likelihood-based forecasts on $f_{R_0}$ and $H_0 r_c$}
\label{subsec:forecast}

\begin{figure}
\begin{center}
\includegraphics[width=3.4in]{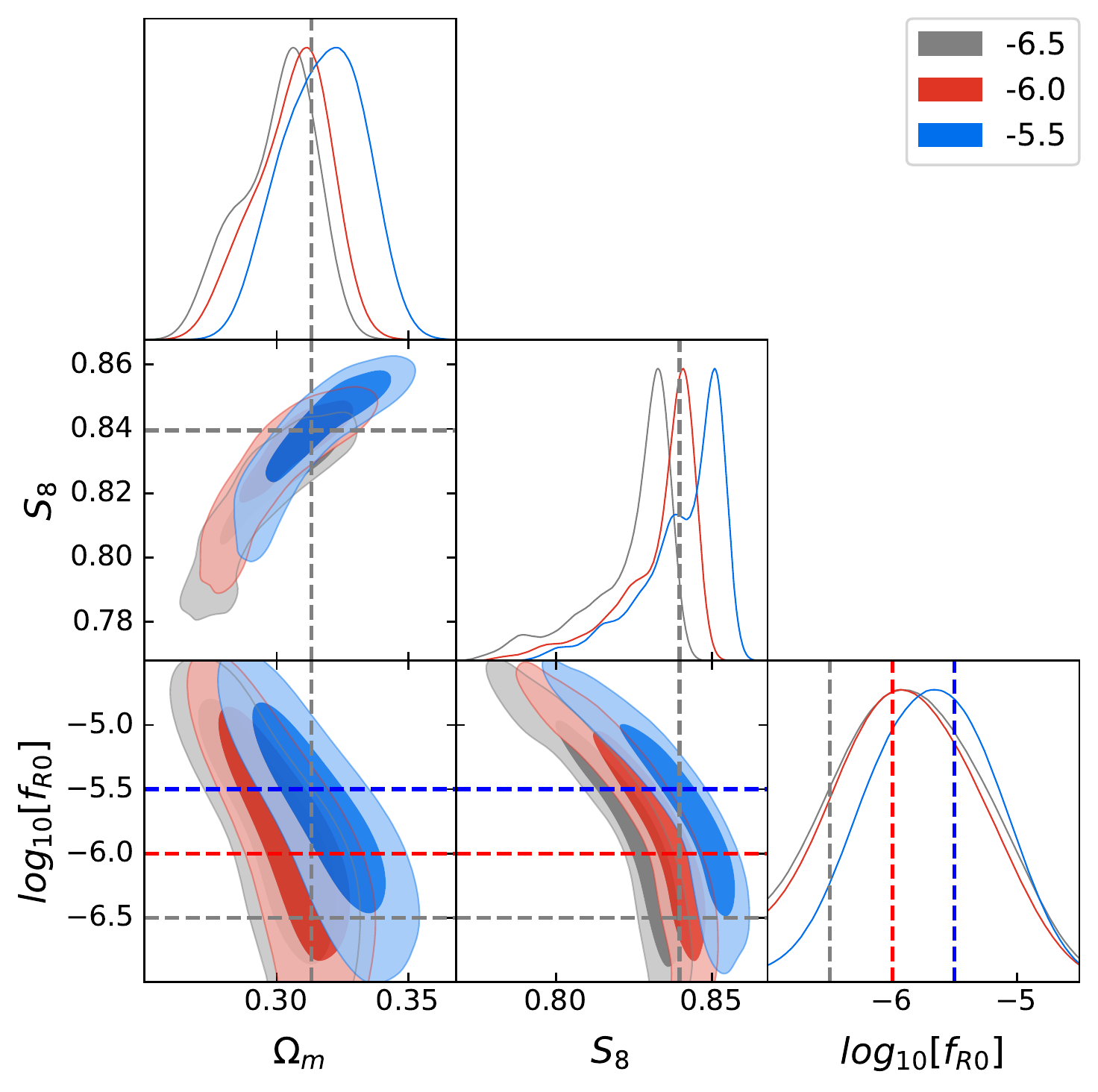}
\includegraphics[width=3.4in]{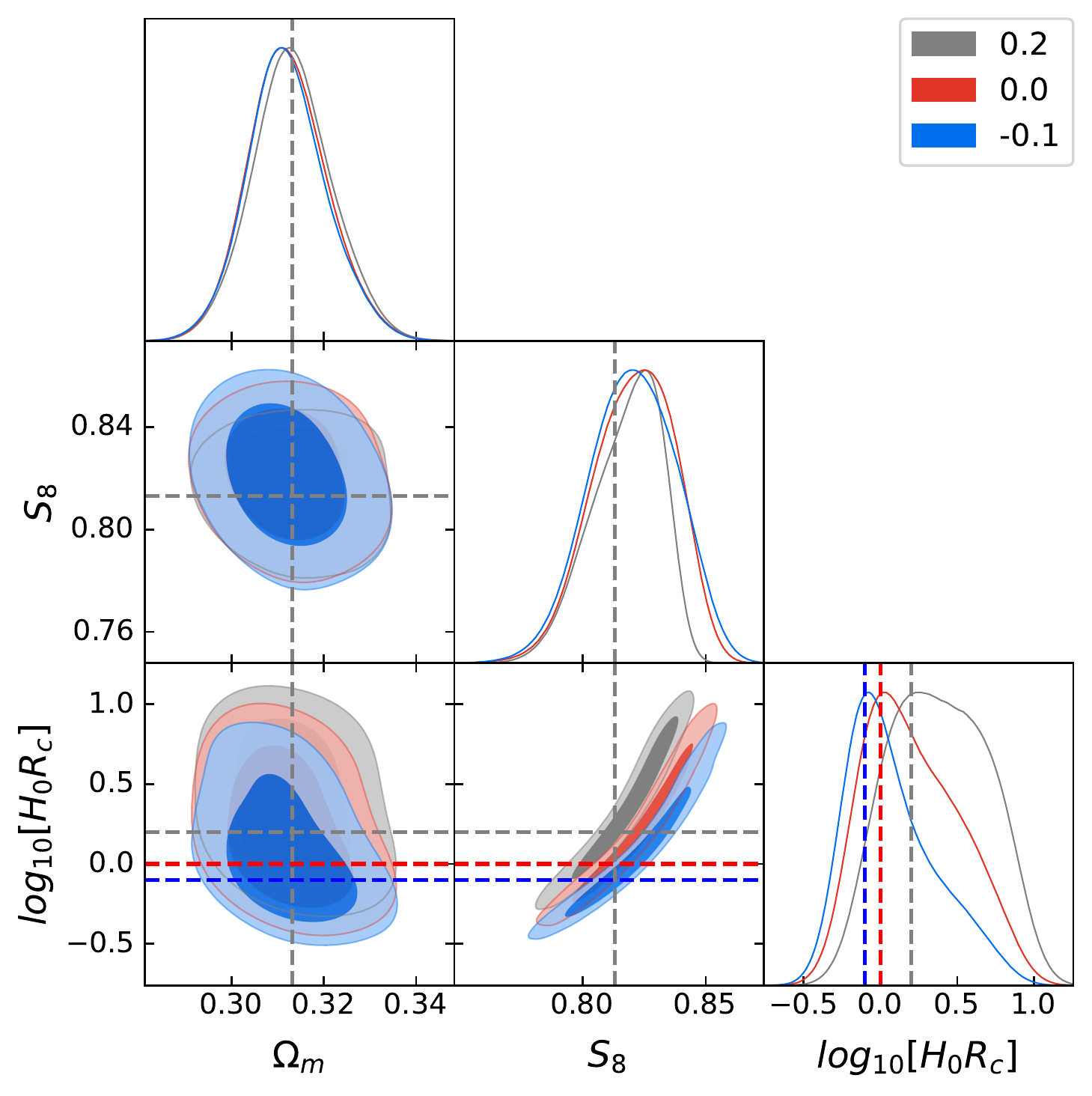}
\caption{Marginalised constraints on the \textsc{forge} (upper panel) and \textsc{bridge} (lower) parameters when analysing cosmology-00 in which the input data is taken directly from theory predictions,  for three values of log$_{10}\left[ f_{R_0} \right]$ and log$_{10}\left[ H_0 r_c \right]$  indicated in the legend.}
\label{fig:MCMC_model00_vary_fr}
\end{center}
\end{figure}

Forecasts on weak lensing $f_{R_0}$  and  $H_0 r_c$ constraints found in the literature need to be revisited, mostly due to recent improvements in modelling the deep non-linear matter power spectrum in presence of a screened fifth force. For example,  \citet{Pratten2016} forecast that with a full-sky 3D weak lensing analysis based on spectroscopic data, and assuming that the cosmological background is fixed by CMB data, one could constrain $f_{R_0}<5\times10^{-6}$. Their $\chi^2$ analysis is simpler than our full MCMC approach, they used a hybrid one-loop perturbation theory and halo-model to compute the $P(k)$ in presence of MG, and unlike us they do not include WL systematics. Other examples include the {\it Euclid} forecast of \citet{2009MNRAS.395..197T} that predicts from a Fisher analysis that the nDGP signal will be clearly detectable from lensing alone\footnote{In their work, \citet{2009MNRAS.395..197T} use a different DGP parametrisation, replacing  $H_0 r_c$ by a derived $\alpha$ parameter.}. \citet{2011PhRvD..83b3012M}  and \citet{2017PDU....18...73C} also predicts clear detection of MG signal from {\it Euclid}, this time using {\sc MGcamb} \citep{MGCAMB} for the $P(k)$ modelling, including $\ell$-modes up to 5000, and assuming the commonly used $(\mu, \Sigma)$  phenomenological parameterisation.  None of these adequately investigate the sensitivity of modern cosmic shear surveys, which we here begin to address.

Fig. \ref{fig:MCMC_model00_vary_fr} (top panel) presents the posterior distributions  from three likelihood samplings, in which the data is taken directly from the {\sc forge} emulator predictions, at cosmology-00 and for log$_{10}\left[ f_{R_0}\right]= -6.5, -6.0$ and $-5.5$. We observe a strong degeneracy between $f_{R_0}$ and $S_8$, expected from the fact that these two parameters both  modulate the overall amplitude of the lensing signal. This degrades the constraining performance with respect to our previous Fisher calculation (Sec. \ref{sec:fisher}). If $S_8$ was fixed, we could indeed detect with high significance these three models (imagine slicing through the $S_8-f_{R_0}$ contours along the vertical dashed line at the input $S_8$ value), however the two weakest models are hitting the  GR-limit when $S_8$ becomes large. The $f_{R_0}=10^{-5.5}$ model, on the other hand, would be detected  at the $\sim\!3\sigma$ level. This is an order of magnitude less constraining that what was found by our one-dimensional Fisher forecast, but is more realistic as we are now fully including gravity-cosmology degeneracies.

The lower panel of Fig. \ref{fig:MCMC_model00_vary_fr} shows a similar exercise carried out on nDGP data taken directly from the {\sc bridge} emulator. We observe that in all cases the three parameters are correctly inferred, and that the $[S_8 - H_0 r_c]$ degeneracy direction is inverted compared to $f_{R_0}$ due to the fact that in this model strongest deviations occur for smaller  $H_0 r_c$ values. Finally, whereas the posterior from weakest nDGP model in this figure (grey contours, corresponding to log$_{10}\left[H_0 r_c \right]=0.2$) is prior-dominated towards the higher $H_0 r_c=$ bound, the other two models are not:   $H_0 r_c < 1.0$ could be detected beyond  $3\sigma$ in this forecast. Once again this error is less constraining than our Fisher forecast, as expected from the added realism. Fixing cosmology would significantly help in this measurement as well, as the posteriors are narrow along a fixed $S_8$ value. 

\subsection{Recovering the GR-$\Lambda$CDM simulation}

\begin{figure}
\begin{center}
\includegraphics[width=3.4in]{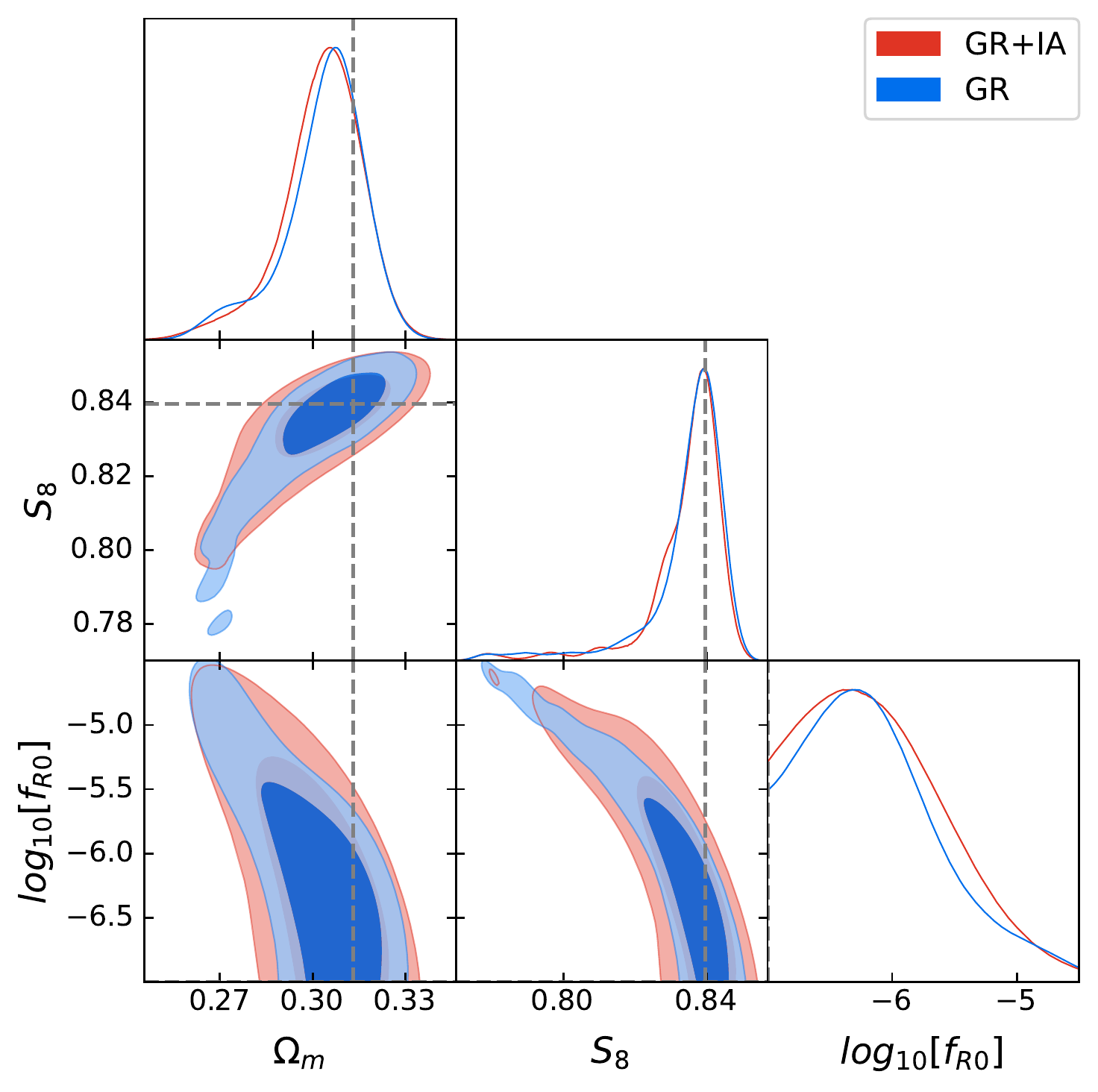}
\includegraphics[width=3.4in]{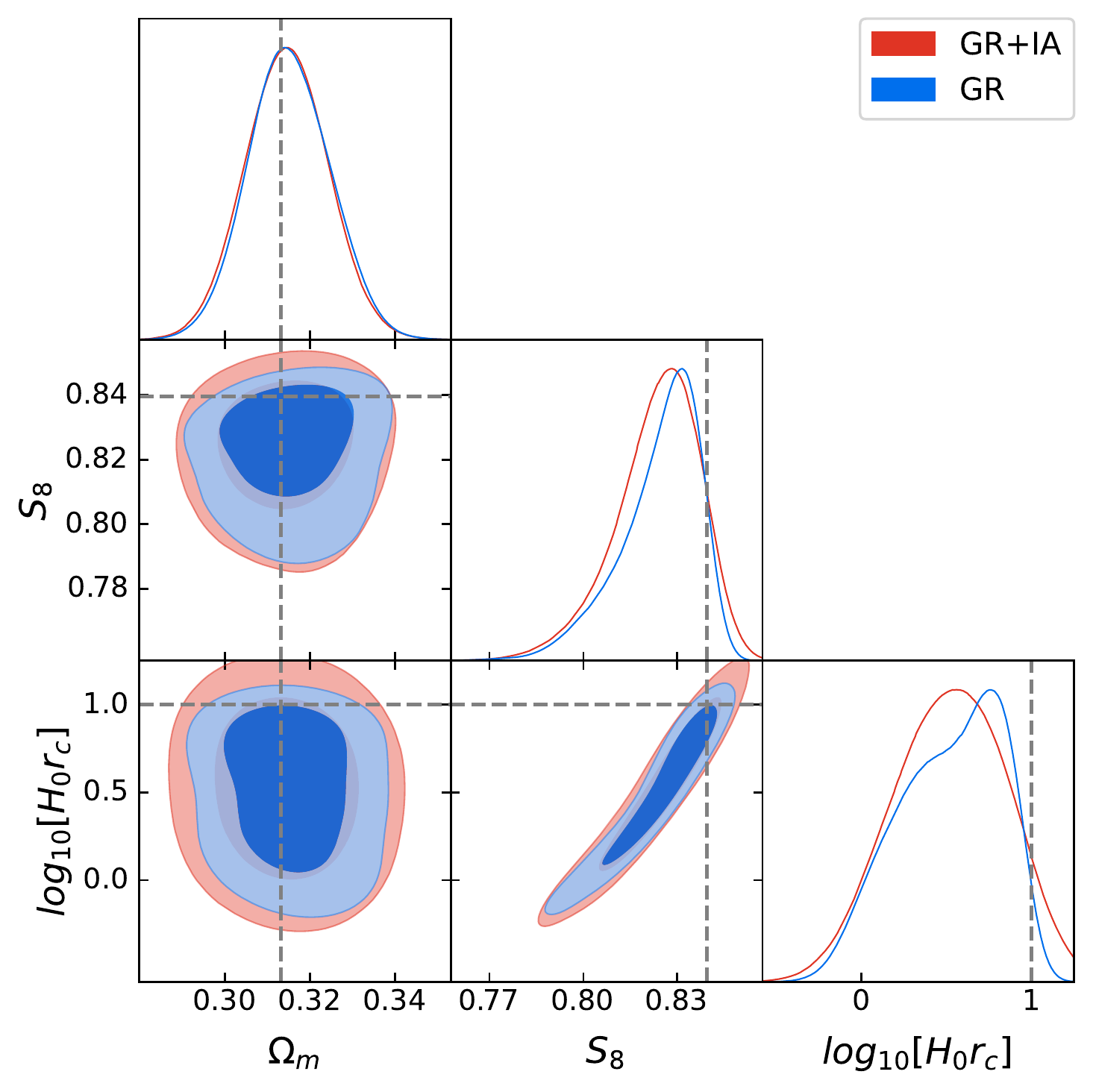}
\caption{Marginalised constraints on the \textsc{forge} (upper) and \textsc{bridge} (lower) parameters when analysing the GR simulations. Given our prior limits and the important degeneracy between $S_8$ and the MG parameters, we recover the expectation that the input truth is well inside the $1\sigma$ contours, but not necessarily at the center.}
\label{fig:MCMC_model00}
\end{center}
\end{figure}

Fig. \ref{fig:MCMC_model00} presents our first inference validation test on the \textsc{MGLenS} simulations, where we run  our analysis pipeline on the GR-only model, assuming consecutively a {\sc forge} and {\sc bridge} gravity model (top and bottom panels, respectively).  It is important to note here that our noise-free data has been measured from 5000 deg$^2$, and our analytical  covariance matrix assumes the same area and includes shape noise. We therefore expect the input truth to lie close to the center of the $1\sigma$  regions, but offset can be caused by residual sampling variance in the mocks and interpolation errors from the emulators. This is indeed consistent with what we observe in Fig. \ref{fig:MCMC_model00}, establishing that we correctly infer the input cosmological parameters, and  prefer modified gravity models that are beyond detection, with:
\begin{eqnarray}
 {\rm log}_{10}\left[f_{R0} \right] < -5.08\, , \nonumber
\end{eqnarray}
and
\begin{eqnarray}
 {\rm log}_{10}\left[H_0 r_c \right] > -0.051\, , \nonumber
\end{eqnarray}
in absence of systematics (both upper limits are reported with 95\%CL). 
Note that these one-sided limits depend on the prior range we adopt: larger sampled volumes (on the weak MG side) down-weight the tails and hence artificially increase the constraining power. For example, truncating the MCMC chains at ${\rm log}_{10}\left[f_{R0} \right]$=[$-7.0$,$-7.5$,$-8.0$] yield upper limits of [$-4.96$, $-5.08$ and $-5.24$], respectively. We selected the middle value in this work, but care must be taken when comparing these results with others found in the literature. Note that the results obtained here seems to contradict Fig. \ref{fig:model_00_fr}, in which models with  $f_{R_0}>10^{-6.0}$ clearly stand out of the 2$\sigma$ region at high-redshift, but again, this observation is ignoring the $[f_{R_0} - S_8]$ degeneracy.

An important feature of this figure is that the degeneracy between $f_{R_0}$ and  $S_8$ vanishes  when sampling lower $f_{R_0}$ values, as seen in the lower part of the contours which are close to vertical; this is also seen in Fig. \ref{fig:MCMC_model00_vary_fr}. That is likely due to the fact that a small $f_{R_0}$ tends to have little modification to the clustering in the linear regime on large scales, where the amplitude of clustering is influenced by $S_8$ more directly;
instead, it tends to cause stronger deviations to its GR counterpart only at the very small scales, where there is also a stronger nonlinearity, thus a weaker connection to the amplitude parameter $S_8$. Put together,  these two factors, the relatively stronger effect of $f_{R_0}$ on small scales and stronger nonlinearity,  naturally break the degeneracy between $f_{R_0}$ and $S_8$ when $f_{R_0}$ is small.
This is not the case for other {\sc forge} models with a stronger MG sector, as we will see in the following section.   

For nDGP, shown on the bottom panel of Fig. \ref{fig:MCMC_model00}, the degeneracy with $S_8$ is  present at every value of $H_0 r_c$, even for weak deviations from GR, but the input cosmology is well recovered, even though this model is at the edge of the Latin Hypercube. 

\begin{figure*}
\begin{center}
\includegraphics[width=3.0in]{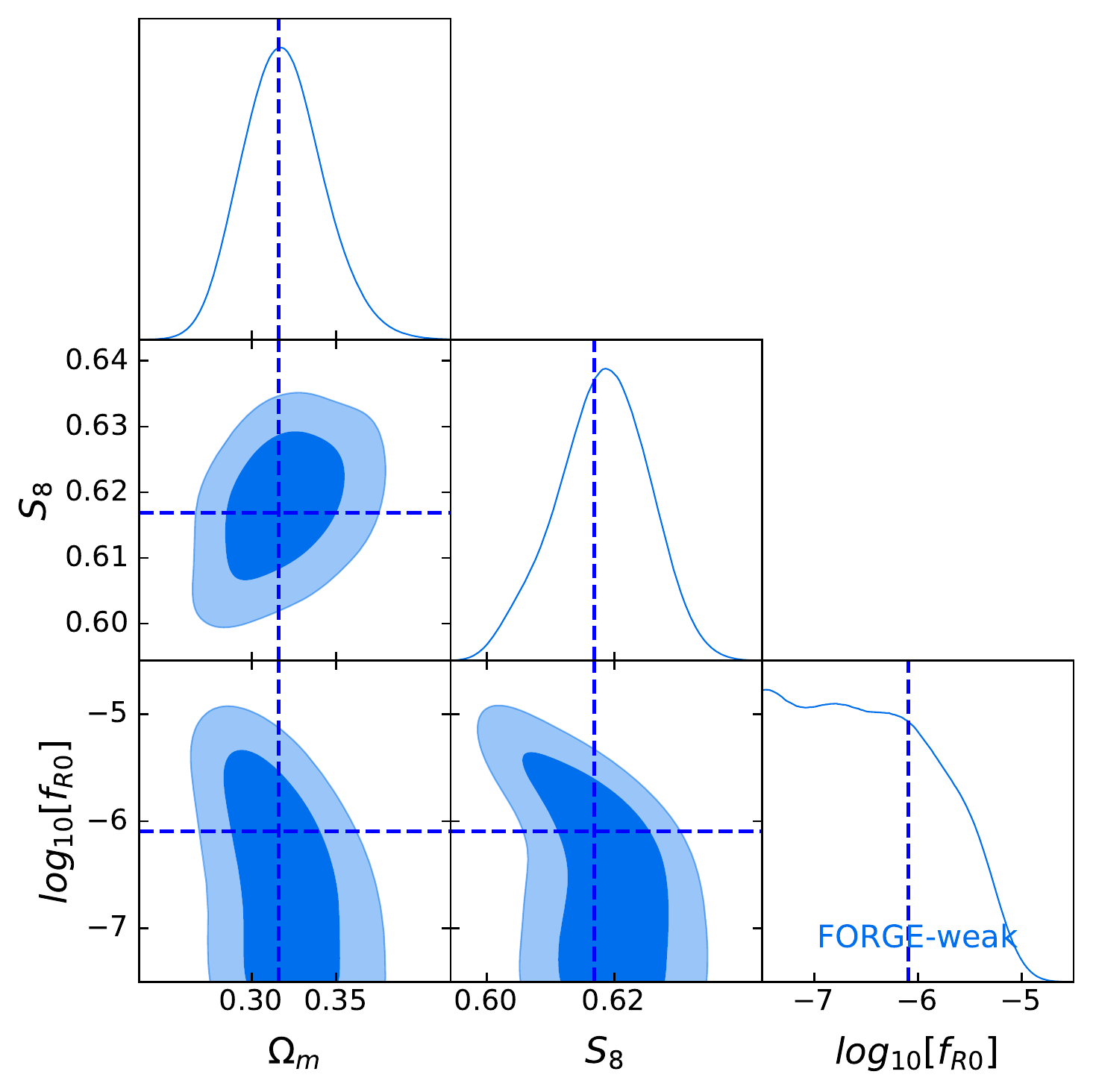}
\includegraphics[width=3.0in]{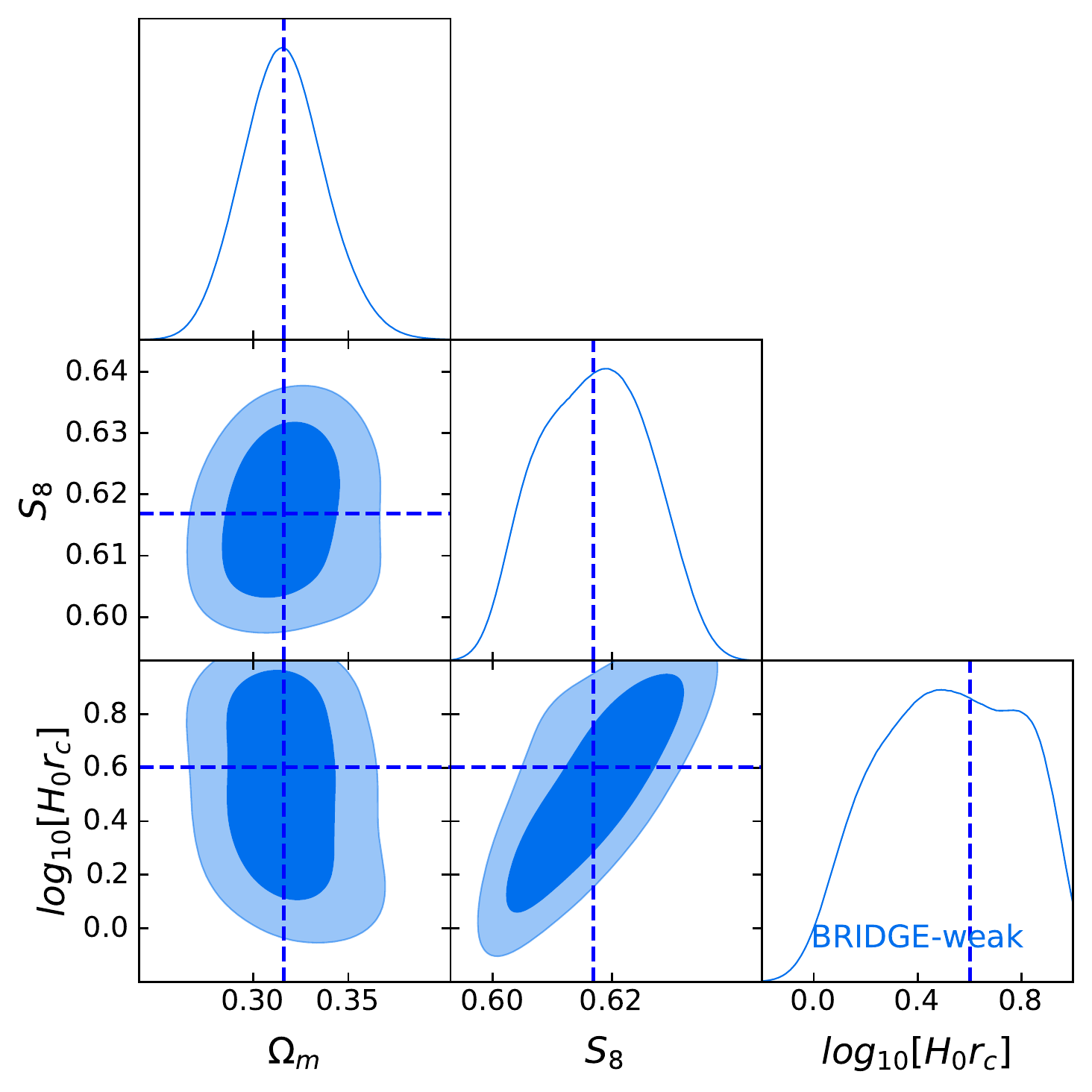}
\includegraphics[width=3.0in]{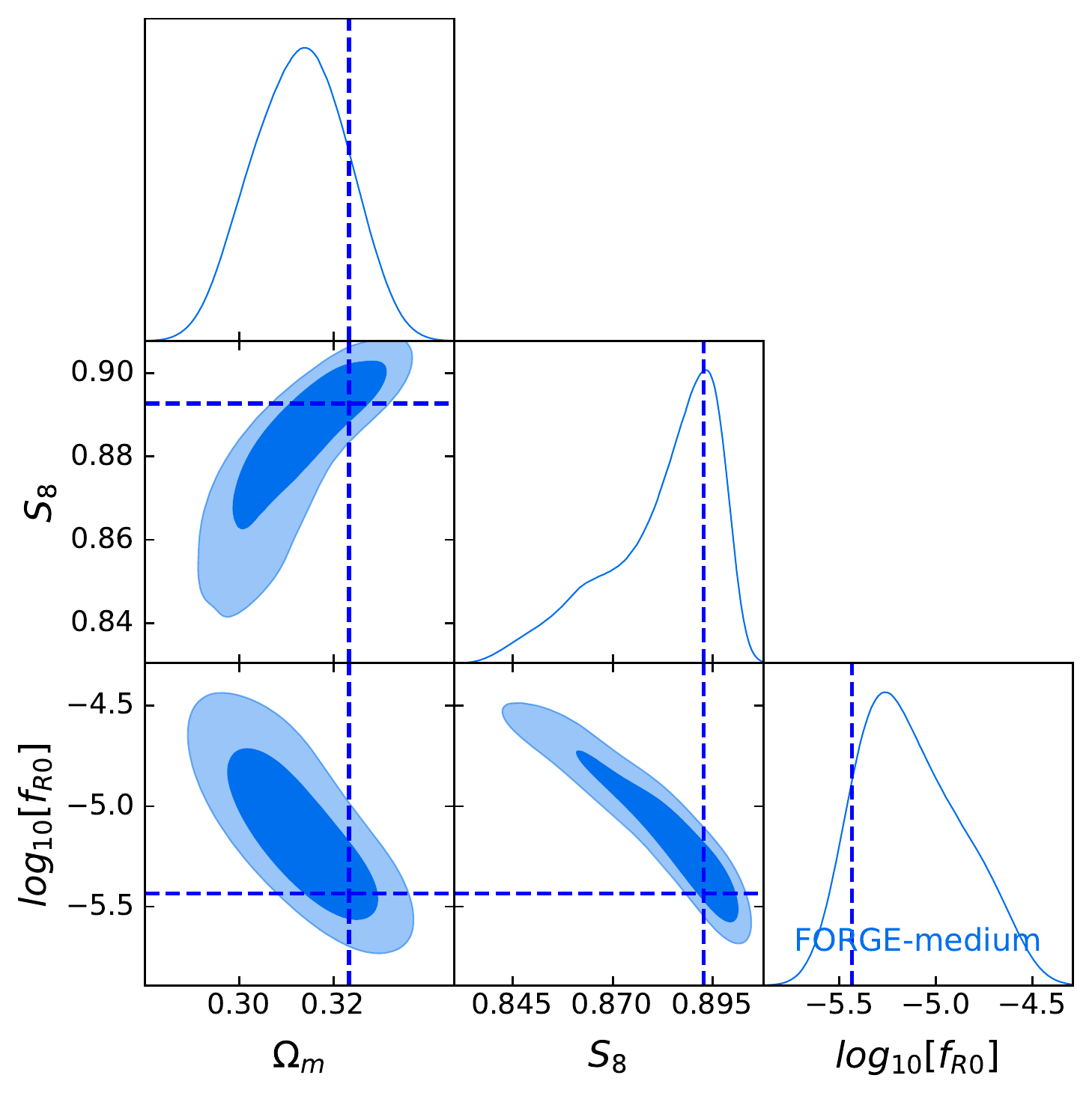}
\includegraphics[width=3.0in]{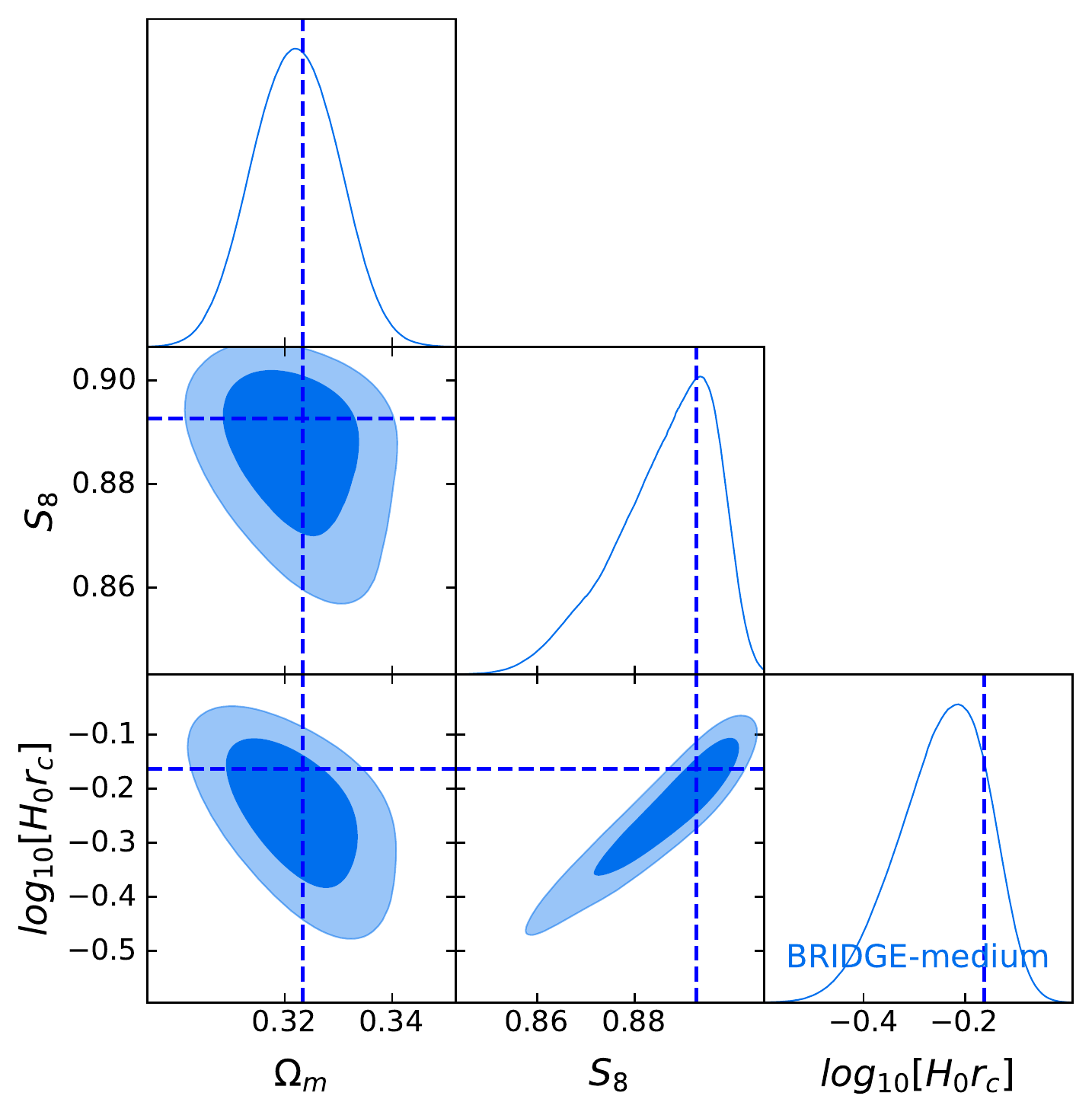}
\includegraphics[width=3.0in]{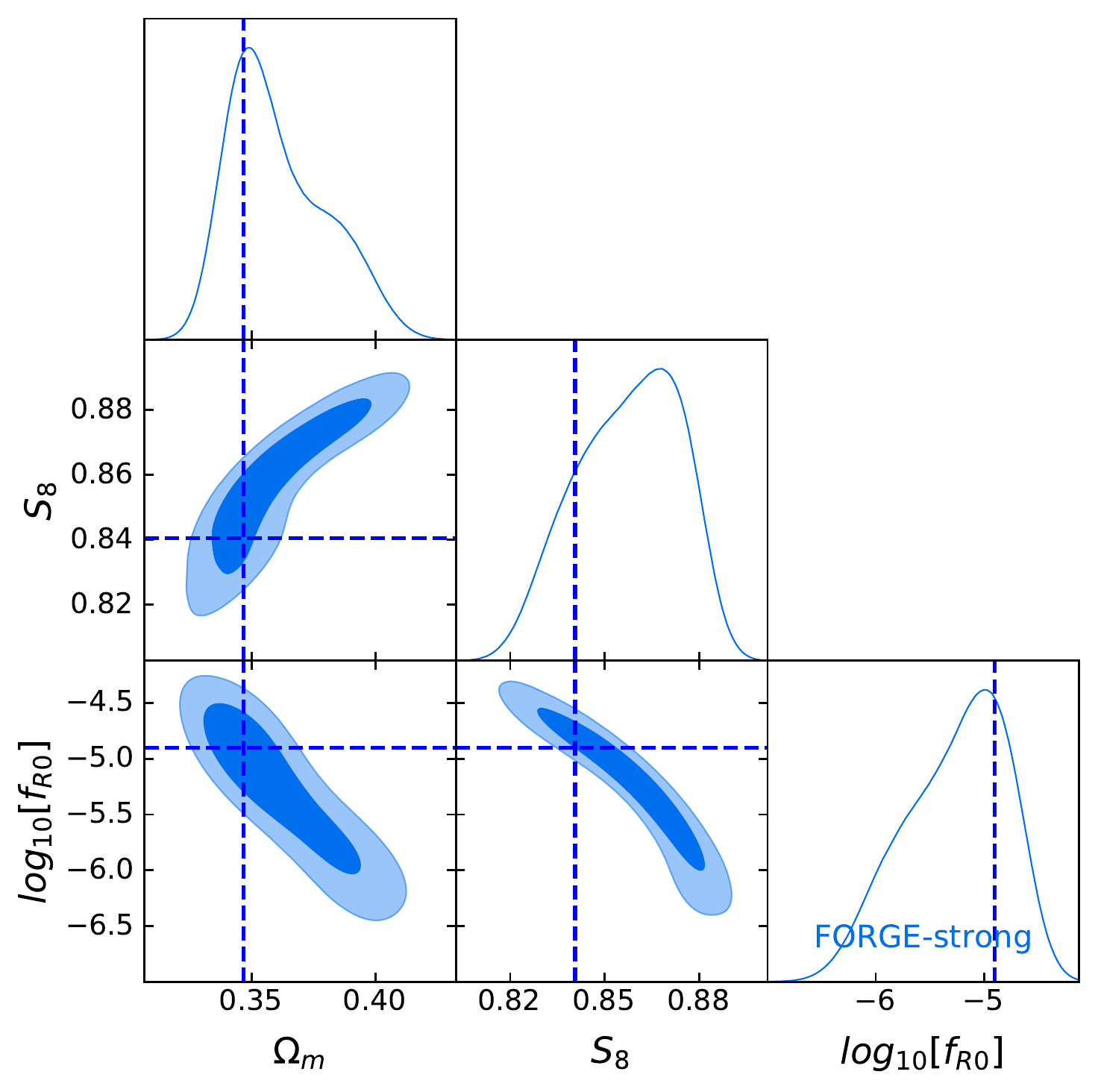}
\includegraphics[width=3.0in]{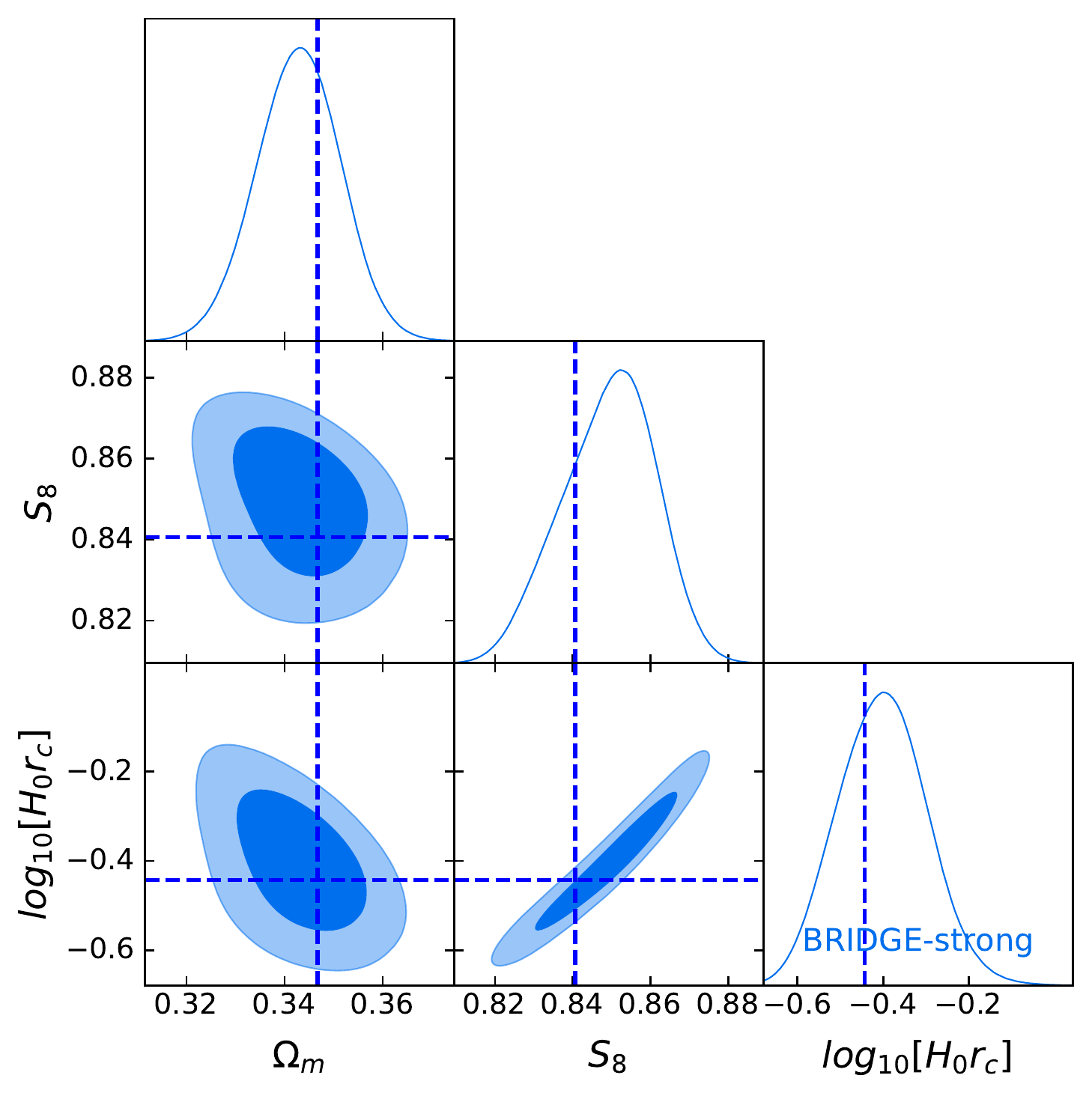}
\caption{Marginalised constraints on the \textsc{forge} (left) and \textsc{bridge} (right) parameters, for models-04 (upper panels, weak MG), -18 (middle panels, medium MG) and -13 (lower panels, strong MG). No systematics are included here.}
\label{fig:MCMC_forge}
\end{center}
\end{figure*}

\begin{figure*}
\begin{center}
\includegraphics[width=3.0in]{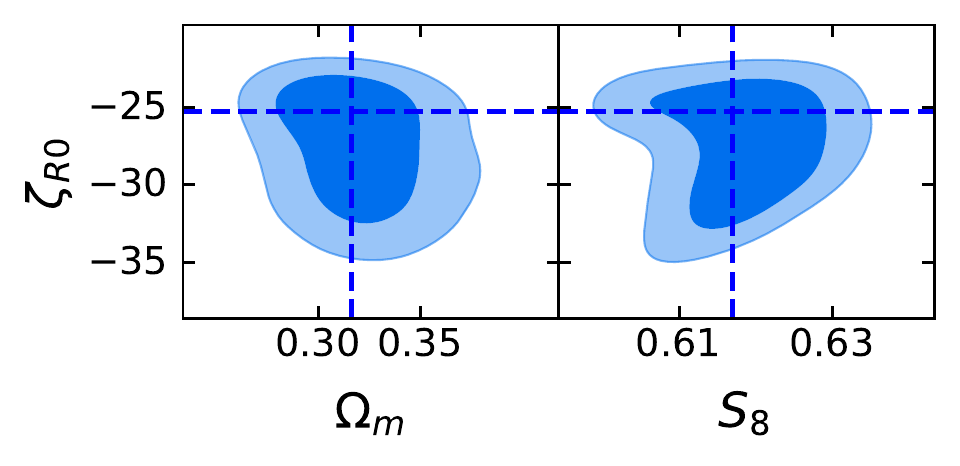}
\includegraphics[width=3.0in]{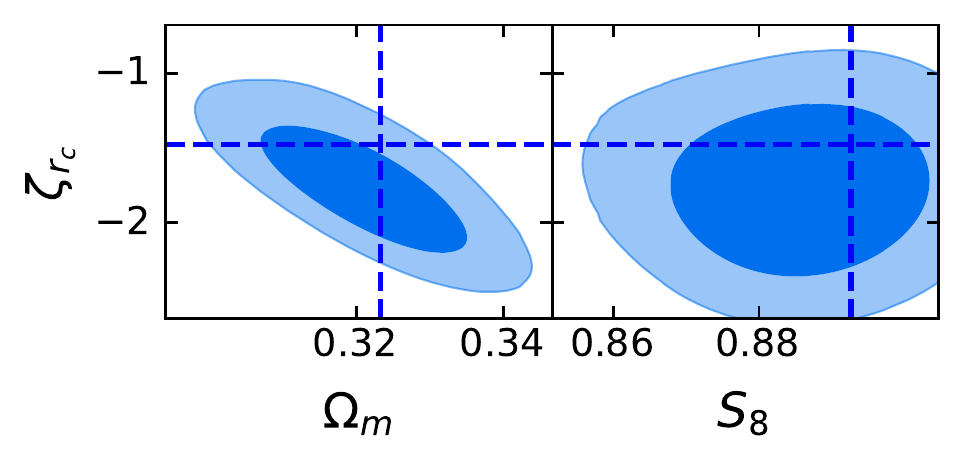}
\caption{Marginalised constraints on the two composite parameters introduced in this paper, $\zeta_{R_0}^{\alpha} $ and $\zeta_{r_c}^{\alpha} $ (see Eqs. \ref{eq:zeta} and \ref{eq:zeta_Rc}), which are best measured by cosmic shear data when  cosmology and modified gravity parameters are jointly varied. These are extracted at the weak $f(R)$ (left) and medium nDGP (right) gravity models. }
\label{fig:zeta}
\end{center}
\end{figure*}

\subsection{Recovering the {\sc forge} and {\sc bridge} simulations}

\begin{table}
   \centering
   \caption{Measurements of the modified gravity parameters inferred from the tomographic weak lensing power spectrum analysis of the {\sc forge} and \textsc{bridge} simulations. We show the results for a selection of models (top to bottom show GR, $f(R)$ and nDGP gravity). The last column shows the impact of marginalising over the $A_{\rm IA}$ nuisance parameter.  In our {\sc forge} and \textsc{bridge} emulators, the GR node is taken at log$_{10}\left[f_{R0} \right] =-7.0$ and log$_{10}\left[H_0 r_c \right] =1.0$, respectively. Upper and lower limits are reported at 95\%CL.}
   \tabcolsep=0.11cm
      \begin{tabular}{@{} lcccccc @{}} 
       \hline 
       \hline
 model& parameter &  truth & no-syst &  IA\\
\hline
\multirow{4}{*}{GR}	& $\Omega_{\rm m} $&  $0.313$& $0.315^{+0.010}_{-0.010}         $  & $0.314^{+0.009}_{-0.009}$ \\
	&$S_8                     $ & $0.840$& $0.834^{+0.011}_{-0.003}$            & $0.834^{+0.011}_{-0.003}  $\\
	&log$_{10}\left[f_{R0} \right] $  & $-\infty$ & $<-5.08$   & $<-5.24$ \\
	& log$_{10}\left[H_0 r_c \right]$  & $\infty$ & $>-0.05$  & $>-0.05$                      \\
\hline
\hline
\multirow{4}{*}{\textsc{forge}-weak}	&$\Omega_{\rm m} $ & $0.316$ & $0.319^{+0.022}_{-0.025}$ & $0.318^{+0.023}_{-0.027}   $ \\
&$S_8            $ & $0.617$ & $0.618^{+0.008}_{-0.007}$ & $0.618^{+0.008}_{-0.007}$\\
	&log$_{10}\left[f_{R0} \right]    $& $-6.09$& $-6.69^{+0.67}_{-1.00} $ & $-6.74^{+0.64}_{-1.00}            $\\
	& $\zeta_{R_0}$  & -25.3& $-27.5^{+3.9}_{-2.3}      $ &$-27.7^{+3.9}_{-2.5}     $\\
	\hline
\multirow{4}{*}{\textsc{forge}-medium}	&$\Omega_{\rm m} $ & $0.323$&  $0.313^{+0.010}_{-0.010}  $&  $0.310^{+0.010}_{-0.010}            $\\
&$S_8            $ &$0.893$ & $0.883^{+0.018}_{-0.007} $ & $0.879^{+0.018}_{-0.016}   $\\
	&log$_{10}\left[f_{R0} \right]   $ & $-5.43$& $-5.13^{+0.22}_{-0.33}     $&$-5.05^{+0.36}_{-0.32}    $ \\
	& $\zeta_{R_0}$& -3.55 & $-3.55^{+0.14}_{-0.06}   $ &  $ -3.57^{+0.13}_{-0.11}            $\\
	\hline
\multirow{4}{*}{\textsc{forge}-strong}	&$\Omega_{\rm m} $ & 0.347&  $0.360^{+0.014}_{-0.026}$& $0.361^{+0.021}_{-0.021}$  \\
&$S_8            $ &0.841 & $0.858^{+0.020}_{-0.014} $ & $0.858^{+0.020}_{-0.013} $\\
	&log$_{10}\left[f_{R0} \right]  $ & $-4.90$& $-5.25^{+0.59}_{-0.39}    $& $-5.26^{+0.58}_{-0.46}    $\\
	& $\zeta_{R_0}$ & -4.33&$-4.18^{+0.15}_{-0.10}   $& $-4.19^{+0.20}_{-0.15}            $\\
\hline
\hline
\multirow{4}{*}{\textsc{bridge}-weak}&$\Omega_{\rm m} $ & $0.316$ & $0.316^{+0.022}_{-0.022}   $ & $0.314^{+0.020}_{-0.023}            $ \\
&$S_8            $ & $0.617$ & $0.617^{+0.009}_{-0.009}$ & $0.617^{+0.009}_{-0.009}        $\\
&log$_{10}\left[H_0 r_c \right]    $& $0.602$&$0.53^{+0.34}_{-0.25}      $ & $0.53^{+0.30}_{-0.27}      $\\
	& $\zeta_{r_c}\times10^3$ & 0.36 &$0.40^{+0.14}_{-0.38}$ &$0.39^{+0.16}_{-0.37}           $\\
	\hline
\multirow{4}{*}{\textsc{bridge}-medium}&$\Omega_{\rm m} $ & $0.323$&  $0.322^{+0.008}_{-0.008}          $&  $0.321^{+0.007}_{-0.007}$\\
&$S_8            $ &$0.893$ & $0.886^{+0.013}_{-0.007}$ & $0.885^{+0.012}_{-0.009}$\\
&log$_{10}\left[H_0 r_c \right]   $ & $-0.163$& $-0.24^{+0.10}_{-0.07}  $&$-0.26^{+0.08}_{-0.08}  $ \\
	& $\zeta_{r_c}$ & -1.478 &$-1.75^{+0.26}_{-0.26}  $&  $-1.76^{+0.23}_{-0.23}   $ \\
	\hline
\multirow{4}{*}{\textsc{bridge}-strong}&$\Omega_{\rm m} $ & 0.347&  $0.343^{+0.008}_{-0.008}$& $0.339^{+0.006}_{-0.006}         $\\
&$S_8            $ &0.841 & $0.849^{+0.013}_{-0.011}            $ &$   0.846^{+0.012}_{-0.012}        $\\
&log$_{10}\left[H_0 r_c \right]  $ & $-0.443$& $-0.40^{+0.10}_{-0.10}      $& $-0.42^{+0.12}_{-0.14} $\\
	& $\zeta_{r_c}$ & -0.845& $-0.98^{+0.11}_{-0.16}         $ & $-0.92^{+0.12}_{-0.12}         $ \\
	\hline
\hline

\end{tabular}
    \label{table:inference}
\end{table}

 We now turn our attention to other \textsc{MGLenS} nodes, with Fig. \ref{fig:MCMC_forge} showing the inferred parameters  when analysing a series of  \textsc{forge} and {\sc bridge} data vectors (left and right panels, respectively), specifying the correct gravity framework ($f(R)$ or DGP) at the moment; we investigate later the result of specifying the wrong framework. We present, from top to bottom, models with increasing  deviations from GR. Once again the input cosmologies are recovered within $1\sigma$, which validates both the \textsc{MGLenS} simulations and the {\sc cosmoSIS} implementation of the \textsc{forge} and \textsc{bridge} emulators in our end-to-end cosmological inference.  
One of the most important features seen here is the strong degeneracy between the MG parameters ($f_{R_0},H_0 r_c$) and  $S_8$. 
Looking now at the posteriors, according to these results, if the gravitational physics of our Universe matched the medium or strong models in these survey conditions, we could strongly rule out GR and constrain the MG sector with our survey. The marginalised posteriors on the parameters of interests are summarised in Table \ref{table:inference}, where, for example, our measurement for the weak \textsc{forge}  yields 
log$_{10}\left[f_{R0} \right]  = -6.69^{+0.67}_{-1.00}$, which is fully consistent with the input truth (-6.09). 
Similar results can be seen for the nDGP inference analyses, where the large values of $H_0 r_c$ are heavily disfavoured, while successfully recovering the input simulation values.

The observed [$f_{R_0}, S_8$] degeneracy limits the precision we can achieve on these two parameters separately, which incites us to define a combination that is better measured. Inspired by the $\Sigma_8\equiv \sigma_8 (\Omega_{\rm m}/0.3)^\alpha$ composite lensing parameter,  we introduce a new variable which runs across the minor axis of the degeneracy ellipse:
\begin{eqnarray}
\zeta_{R_0}^{\alpha} \equiv \mbox{ log}_{10}\left[f_{R_0}\right] \left(\frac{S_8}{0.82} \right)^{\alpha} \, ,
\label{eq:zeta}
\end{eqnarray}
where $\alpha$ is a free parameter to be optimised. For small values of $f_{R_0}$, $\alpha=5.0$ returns a $\zeta_{R_0}^{\alpha}$ that is mostly orthogonal to both $S_8$ and $\Omega_{\rm m}$, making this an attractive target measurement for future cosmic shear experiments. 
We post-pone to future work the impact of letting $\alpha$ free in a likelihood analysis. 

The equivalent degeneracy-breaking parameter for nDGP models can be constructed as 
\begin{eqnarray}
\zeta_{r_c}^{\alpha} \equiv \mbox{ log}_{10}\left[H_0 r_c \right] \left(\frac{S_8}{0.82} \right)^{\alpha} \, ,
\label{eq:zeta_Rc}
\end{eqnarray}
where $\alpha= 26$ works better for the nDGP models. We show in Fig. \ref{fig:zeta} the marginalised constraints on these two new parameters, $\zeta_{R_0}^{\alpha} $ and $\zeta_{r_c}^{\alpha} $, where the degeneracy with respect to $S_8$ and $\Omega_{\rm m}$ is highly suppressed. The accuracy on these composite parameters is increased, where for example a 9\% measurement\footnote{ The precision is defined here as the ratio between the error and the best-fit value for a given parameter. } of log$_{10}\left[f_{R_0} \right]$ results in a 3\% precision on $\zeta_{R_0}^{\alpha} $ in the strong  \textsc{forge}  model. Similar improvements are seen on nDGP parameters, where a 25\% measurement of  log$_{10}\left[H_0 r_c \right]$ becomes a 14\% measurement of $\zeta_{r_c}^{\alpha} $ in the strong  \textsc{bridge}  model. The measurements reported in Table \ref{table:inference} indicate a net gain in precision for all models.

By construction, the variables  $\zeta_{R_0}^{\alpha} $ and  $\zeta_{r_c}^{\alpha}$ down-weight parameter regions of weak modified gravity, which therefore interacts with prior limits. 
These parameters are therefore mostly useful for medium and strong modified gravity models, but we advise against using them for one-sided limits.

\begin{figure*}
\begin{center}
\includegraphics[width=2.3in]{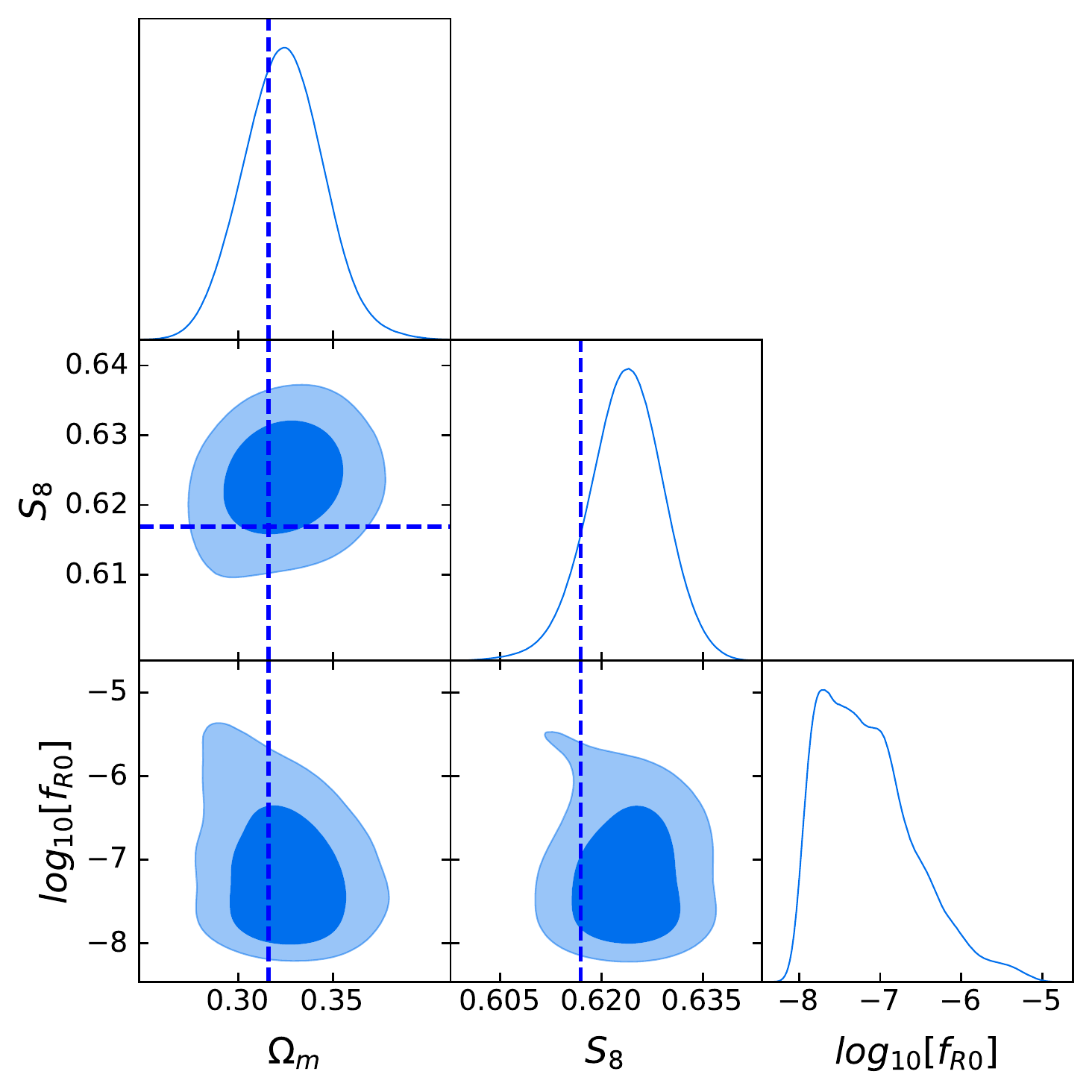}
\includegraphics[width=2.3in]{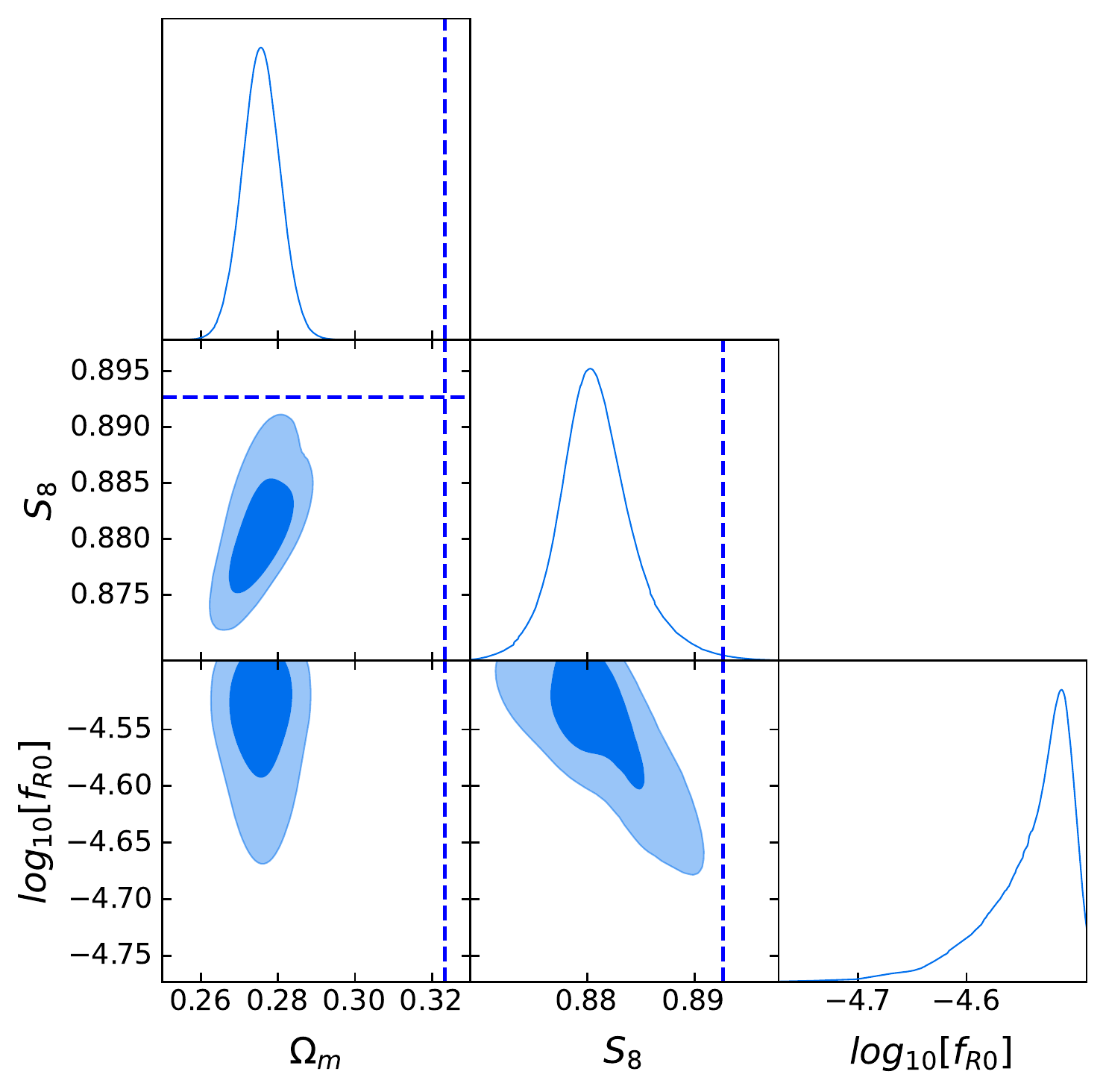}
\includegraphics[width=2.3in]{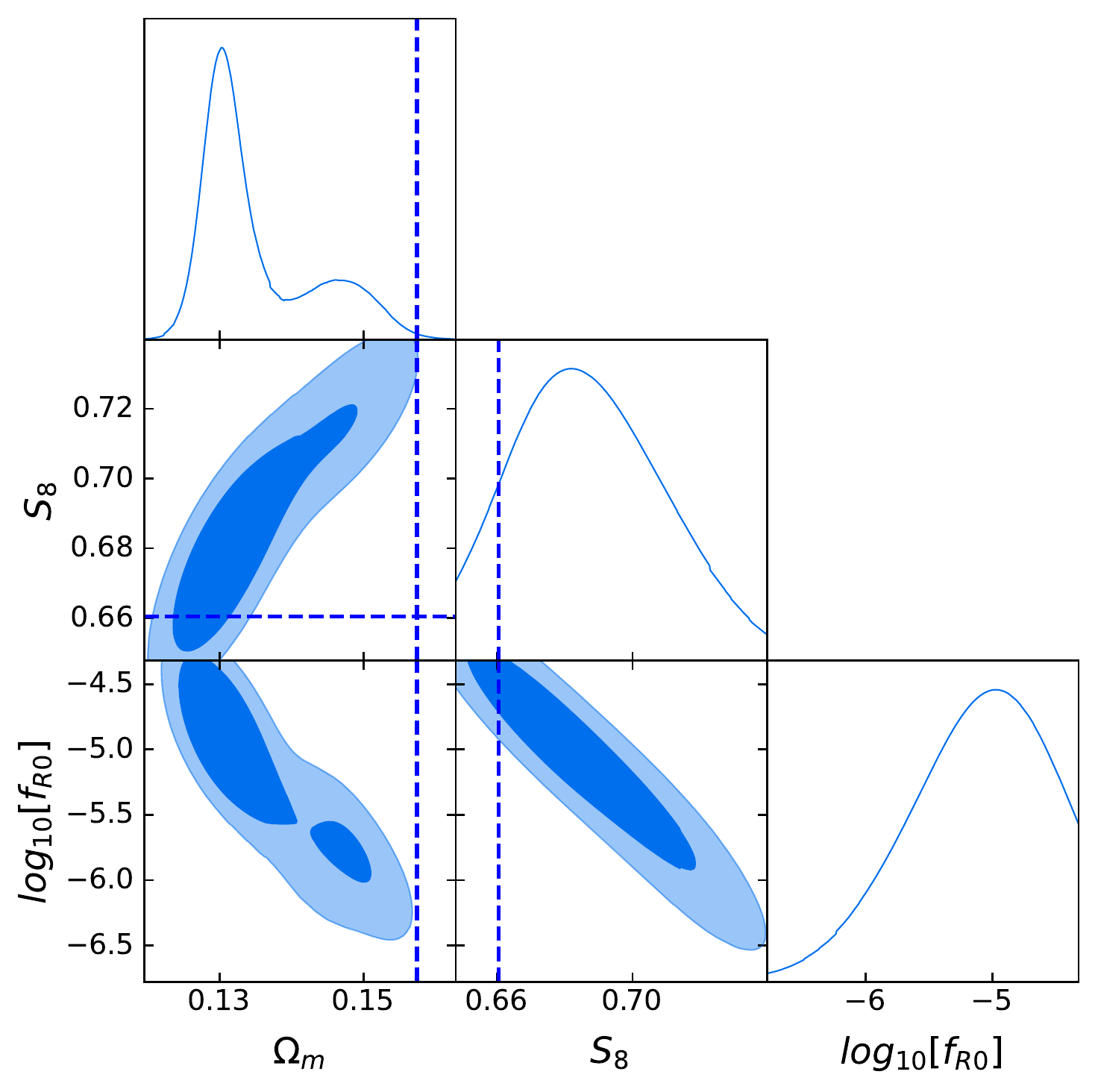}
\includegraphics[width=2.3in]{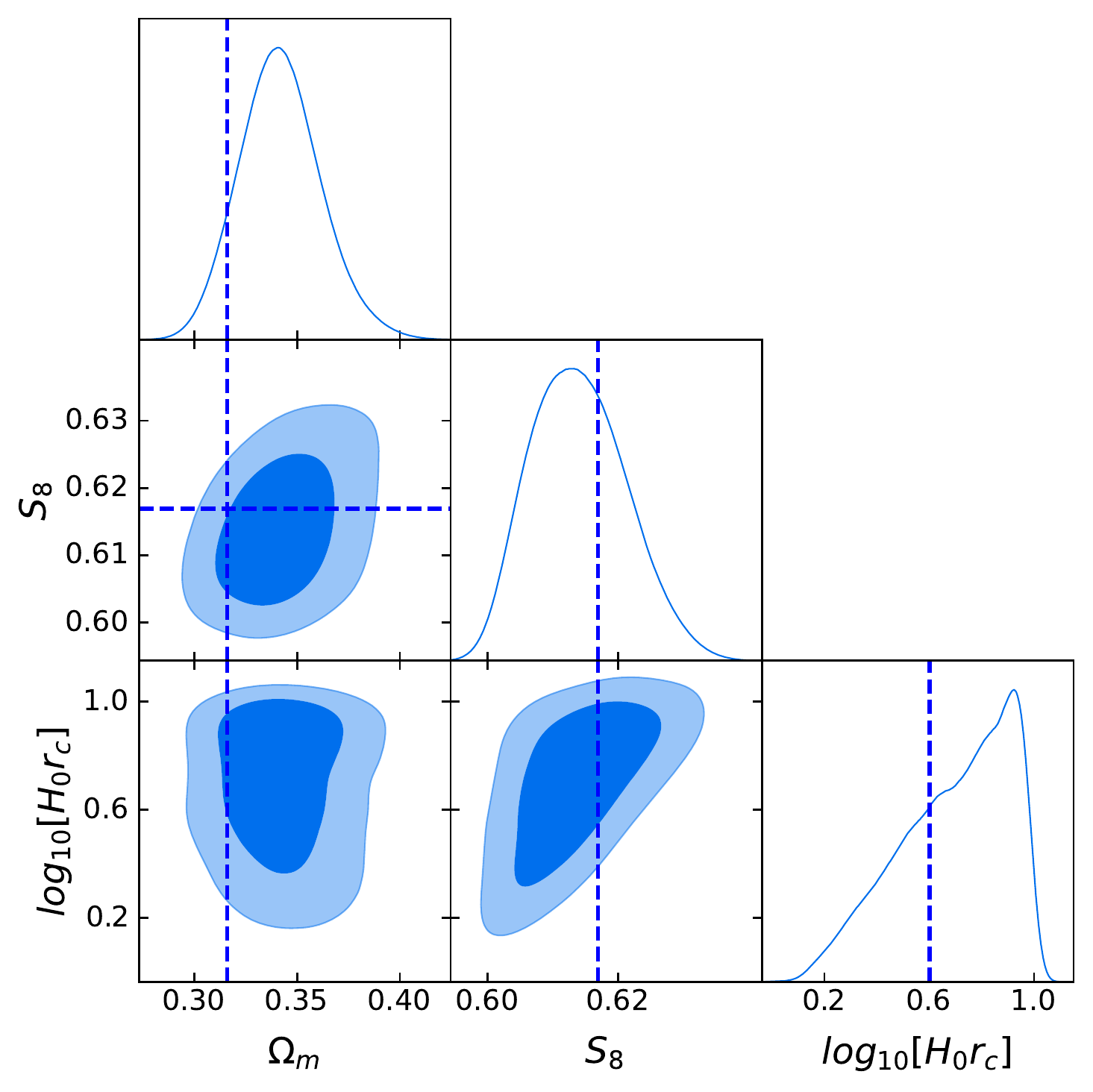}
\includegraphics[width=2.3in]{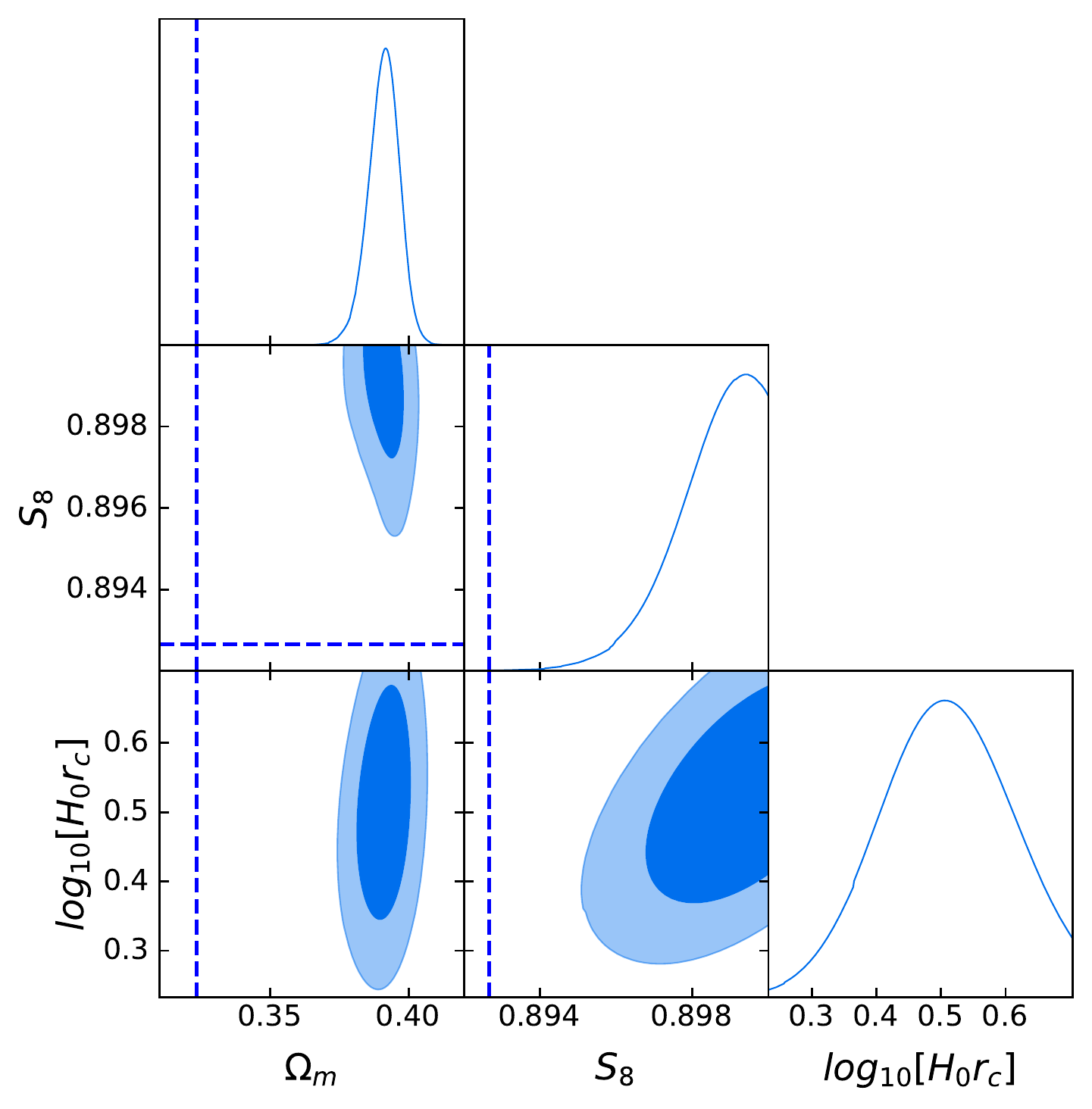}
\includegraphics[width=2.3in]{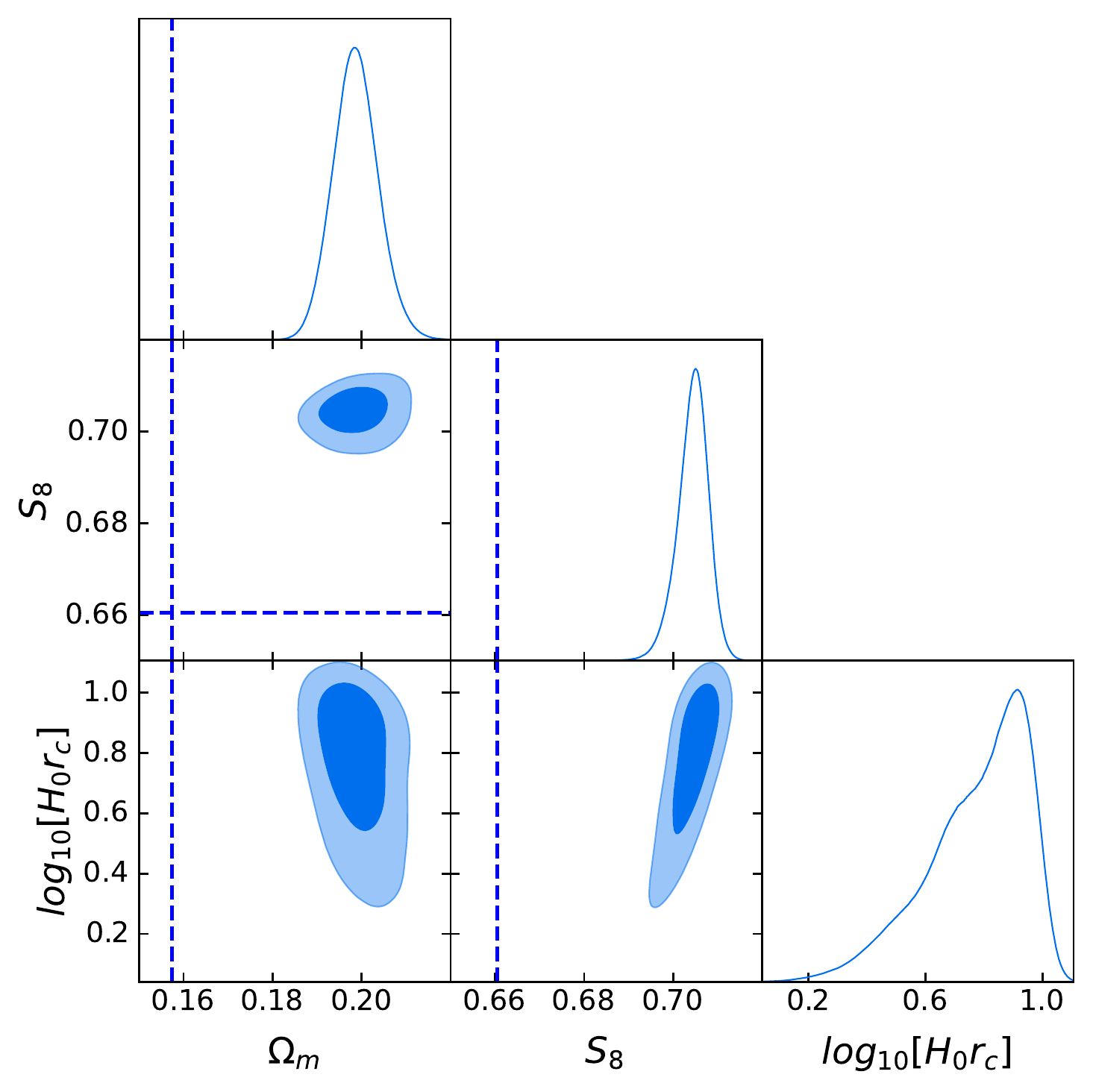}
\caption{({\it Upper}:) 
Marginalised parameter constraints when analysing {\sc forge} simulations (left is weak model, center is medium, right is  the strong model-05) with the {\sc bridge} emulator, yielding to catastrophic biases. 
({\it Lower}:) Counterpart of the upper panels, now analysing \textsc{bridge} simulations with the \textsc{forge} emulator.}
\label{fig:mix}
\end{center}
\end{figure*}

\subsection{Degeneracies between gravity models and cosmology}

\label{subsec:mixmodels}
\begin{table}
   \centering
   \caption{Impact of analysing MG simulated data with the wrong gravity model. Column $\Lambda$CDM+GR shows the results of analysing \textsc{MGLenS} simulations with a GR model (i.e. \textsc{Halofit}), while the `Wrong MG' columns consider {\sc forge} data analysed with the {\sc bridge} emulator and vice versa. The parameter shift are computed as $|$bestfit - true$|/\sigma$, and the $p$-values assume four free parameters. }
   \tabcolsep=0.11cm
\begin{tabular} {ccccccccc}
\hline
\hline
 True & & \multicolumn{2}{c}{True model}& \multicolumn{2}{c}{$\Lambda$CDM+GR} & \multicolumn{2}{c}{Wrong MG}\\
 gravity model & param &  shift &  $p$-value & shift & $p$-value&  shift &  $p$-value  \\
\hline
\multirow{2}{*}{{\sc forge}-weak} & $\Omega_{\rm m}       $ &0.7$\sigma$&\multirow{2}{*}{1.0} & $1.1\sigma$& \multirow{2}{*}{1.0}  &$1.4\sigma$&\multirow{2}{*}{1.0} \\
& $S_8            $ &0.4$\sigma$&& $0.3\sigma$ &  $   $&$0.4\sigma$&\\
\hline
\multirow{2}{*}{{\sc forge}-medium} & $\Omega_{\rm m}       $ &0.4$\sigma$&\multirow{2}{*}{0.77}& $23.7\sigma$& \multirow{2}{*}{0.0}&$11.3 \sigma$&\multirow{2}{*}{0.0}\\
& $S_8            $ &0.8$\sigma$&& $12.8\sigma$ &  $     $&$6.9\sigma$&\\
\hline
\multirow{2}{*}{{\sc forge}-strong} & $\Omega_{\rm m}       $ &1.8$\sigma$&\multirow{2}{*}{1.0}& $4.3\sigma$& \multirow{2}{*}{1.0}&$8.2\sigma$&\multirow{2}{*}{0.997}\\
& $S_8            $ &0.7$\sigma$&& $17.6\sigma$ &  $     $&$13.4\sigma$&\\
\hline
\hline
\multirow{2}{*}{{\sc bridge}-weak} &$\Omega_{\rm m}       $ &0.7$\sigma$&\multirow{2}{*}{1.0} &$0.6\sigma$& \multirow{2}{*}{1.0} &$0.4\sigma$&\multirow{2}{*}{1.0}\\
&$S_8            $ &0.2$\sigma$&& $1.2\sigma$& $ $&$1.2\sigma$&\\
\hline
\multirow{2}{*}{{\sc bridge}-medium} &$\Omega_{\rm m}       $ &0.2$\sigma$&\multirow{2}{*}{0.97}&$8.5\sigma$& \multirow{2}{*}{0.0} &$10.8\sigma$&\multirow{2}{*}{0.0}\\
&$S_8            $ &2.0$\sigma$&& -- & &$3.7\sigma$&\\
\hline
\multirow{2}{*}{{\sc bridge}-strong} &$\Omega_{\rm m}       $ &1.0$\sigma$&\multirow{2}{*}{1.0} &$5.3\sigma$& \multirow{2}{*}{0.002}&$2.2\sigma$&\multirow{2}{*}{0.003}\\
&$S_8            $ &2.1$\sigma$&& $20.7\sigma$& $ $&$1.2\sigma$&\\
\hline
\hline
\end{tabular}
\label{table:MGdata_GRmodel}
\end{table}

One of the main difficulties in detecting deviations from GR comes from the abundance of models to be tested, which each affect the growth of structures in different ways. A key question to be answered is whether one can confuse a clear detection of gravity model `A' at some cosmology with a different gravity model `B' at a different cosmology. The first part of the answer is already provided in the GR-only validation test, where both the \textsc{forge} and \textsc{bridge} emulators recognise negligible deviations from GR in model-00, both inferring the right cosmology. This is encouraging since it suggests that GR can be recognised as such.

Complications arise when analysing truly non-GR data with the wrong gravity model. The upper panels of Fig. \ref{fig:mix} shows such examples, where three \textsc{forge} data vectors are analysed with the \textsc{bridge} emulator.   For the weak model (left), this results in a minor bias in $\Omega_{\rm m}$ and $S_8$, and a wide posterior on $H_0 r_c$ that hits the upper edge prior, leading to inconclusive detection of MG. The central and right panels, however, reveal catastrophic biases on the cosmological parameters for the medium and strong models.  The two cosmological parameters are shifted towards higher values,  while the posteriors  indicate an apparent $H_0 r_c$  detection. Similarly catastrophic results are observed when, on the contrary, we analyse  nDGP data with the \textsc{forge} emulator (see the lower panels of Fig.~\ref{fig:mix}); in this case most inferred cosmological parameters are also far from the truth, and the $f_{R_0}$ parameter is falsely detected with high significance for both medium and strong nDGP models.

Biases also occur if data from a modified gravity universe is analysed with a GR-only model, in which case the additional structure formation caused by the fifth force is interpreted as  a higher $S_8$ value, as expected from the degeneracy between these quantities. Table \ref{table:MGdata_GRmodel} summarises these parameter shifts, normalised by the precision $\sigma$. We see again that the weak models has almost no impact on the inferred cosmology (shift $\sim1\sigma$), whereas the stronger model can offset $\Omega_{\rm m}$ by up to $23.7\sigma$ and $S_8$ by up to $17.6\sigma$. 

This inevitably raises the question of whether we could know that we are analysing the data with the wrong gravity model. One of the  standard approaches is to examine the {\it goodness-of-fit}, which informs us on the quality of the data-model match. This can be computed with the $p$-value measured at the best-fit parameters for different gravity models, from which one can test different model hypotheses\footnote{The $p$-value is computed from the $\chi^2$ conditional distribution function and the number of degrees of freedom; it is routinely used for rejection of null-hypotheses.}. A $p$-value below 0.01 generally indicates that the hypothesis should be rejected. Table \ref{table:MGdata_GRmodel} presents the measured $p$-values for different combinations of models and data. In our case this test is done with noise-free data, so we expect the distributions of  $p$-values to be sharply bimodal between 0.0 and 1.0. It turns out that some data (e.g. {\sc forge}-04) and {\sc bridge}-04 can be well fit by all three models, due to weakness of the departure from GR. GR and nDGP gravity can also provide a good fit to the {\sc forge}-05 data, which is achieved at the cost of significantly increasing both $\Omega_{\rm m}$ and $S_8$. This bias is clearly seen in the lower-right panel of Fig. \ref{fig:mix}. One would have problems, in such a case, to distinguish between gravity models, unless augmenting the analysis with prior knowledge of the cosmological parameters from e.g. the CMB. Other test cases are easier to reject based on the bad goodness-of-fit, such as {\sc forge}-18 and {\sc bridge}-18, which can only be well fit with the correct gravity model.

Other metrics exist to quantify tensions between data and models such as {\it Bayesian Evidence ratios} \citep{Evidence_1, Evidence_2} or {\it Suspiciousness} \citep{Suspiciousness}, the latter being more robust to prior-effects \citep[see also][for a recent discussion on the application of such metrics to real cosmological data]{2021A&A...647L...5J}. These metrics would be useful to explore as well, but have come with their own challenges and sets of requirements and hence we leave such investigations for future work.


\subsection{Impact of systematics}
\label{subsec:inference_syst}

\begin{figure*}
\begin{center}
\includegraphics[width=3.1in]{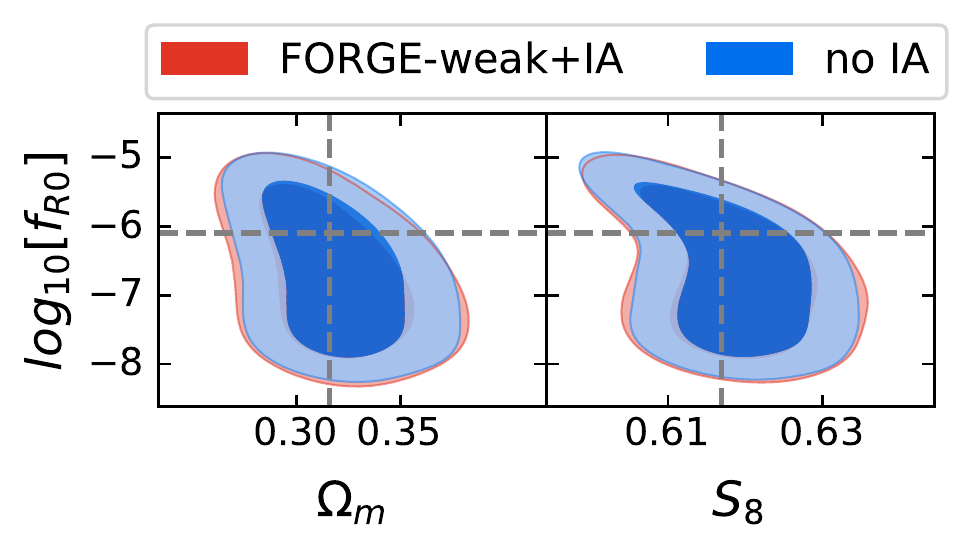}
\includegraphics[width=3.1in]{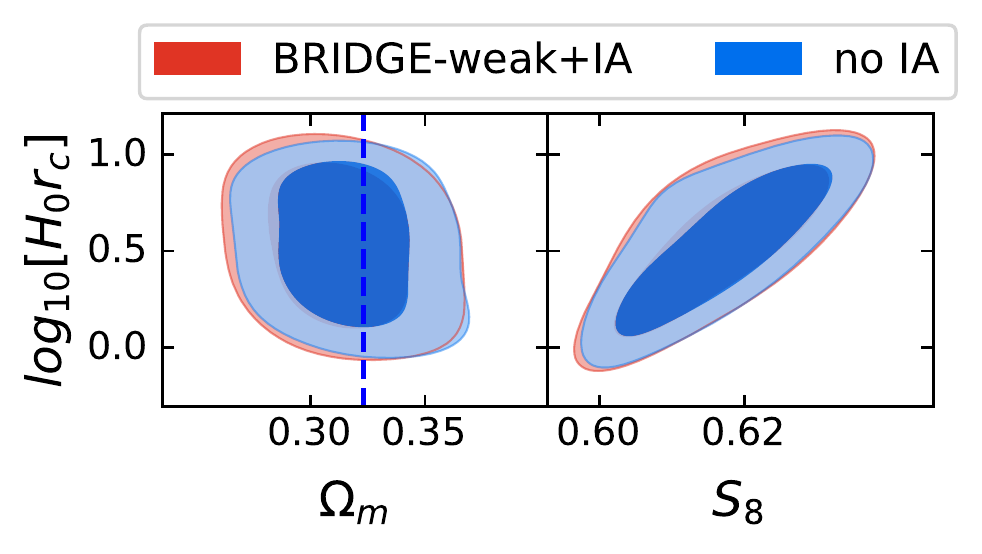}
\includegraphics[width=3.1in]{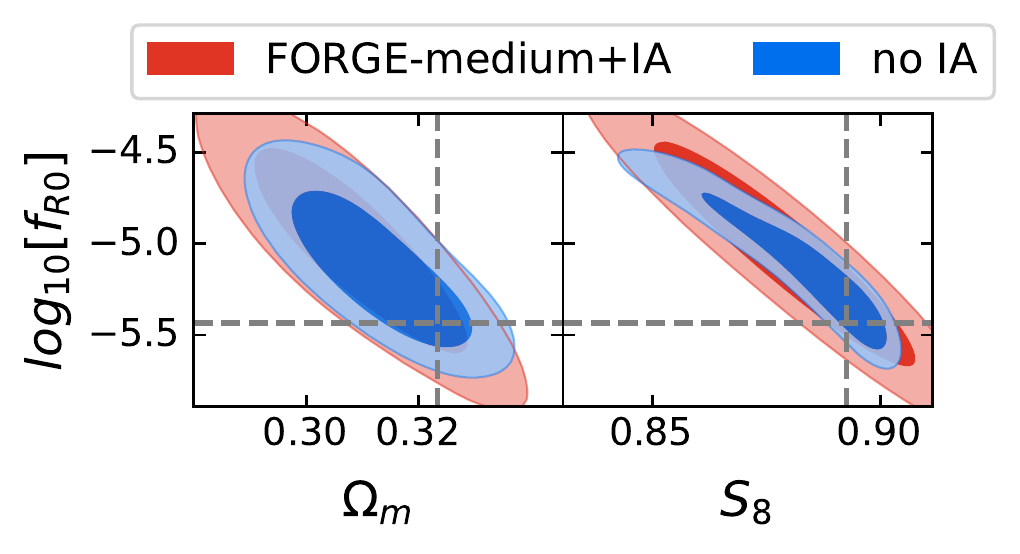}
\includegraphics[width=3.1in]{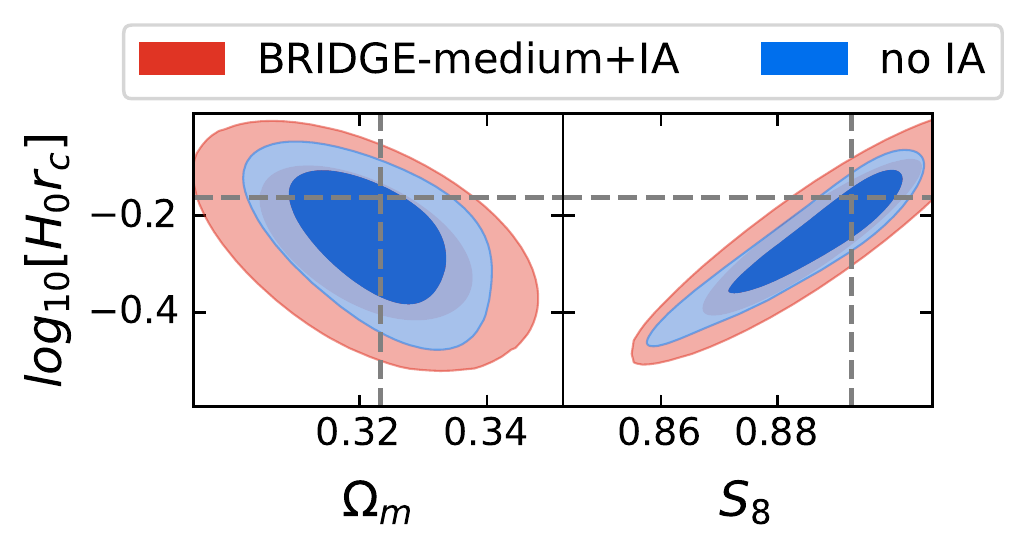}
\includegraphics[width=3.1in]{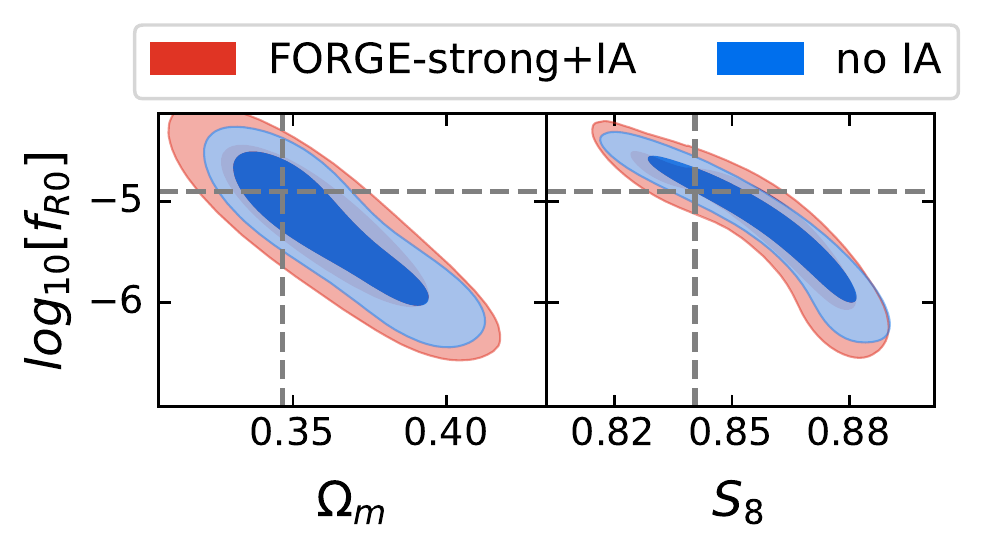}
\includegraphics[width=3.1in]{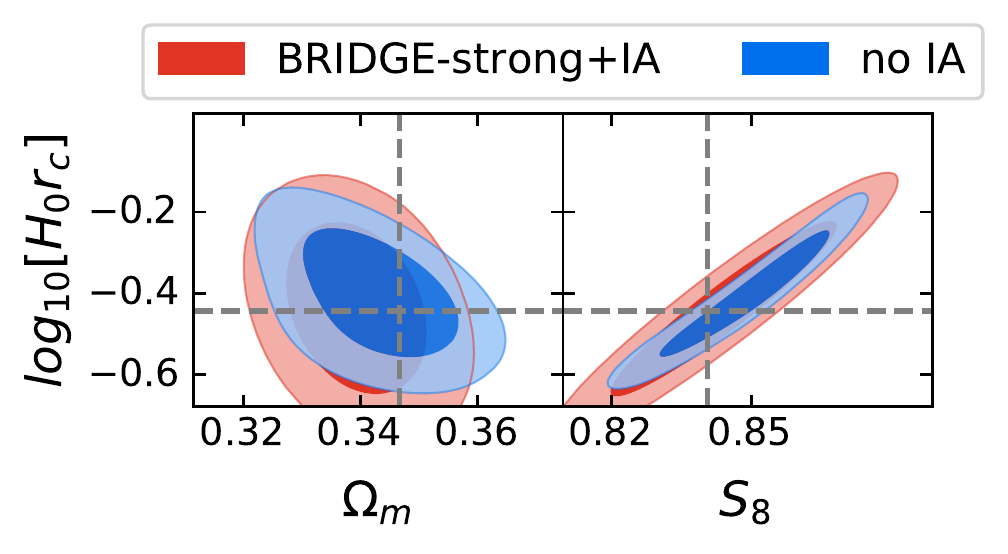}
\caption{Marginalised constraints on the parameters best probed by lensing, with and without including contamination from intrinsic alignment in the modelling. }
\label{fig:MCMC_forge_syst}
\end{center}
\end{figure*}

\begin{figure}
\begin{center}
\includegraphics[width=3.0in]{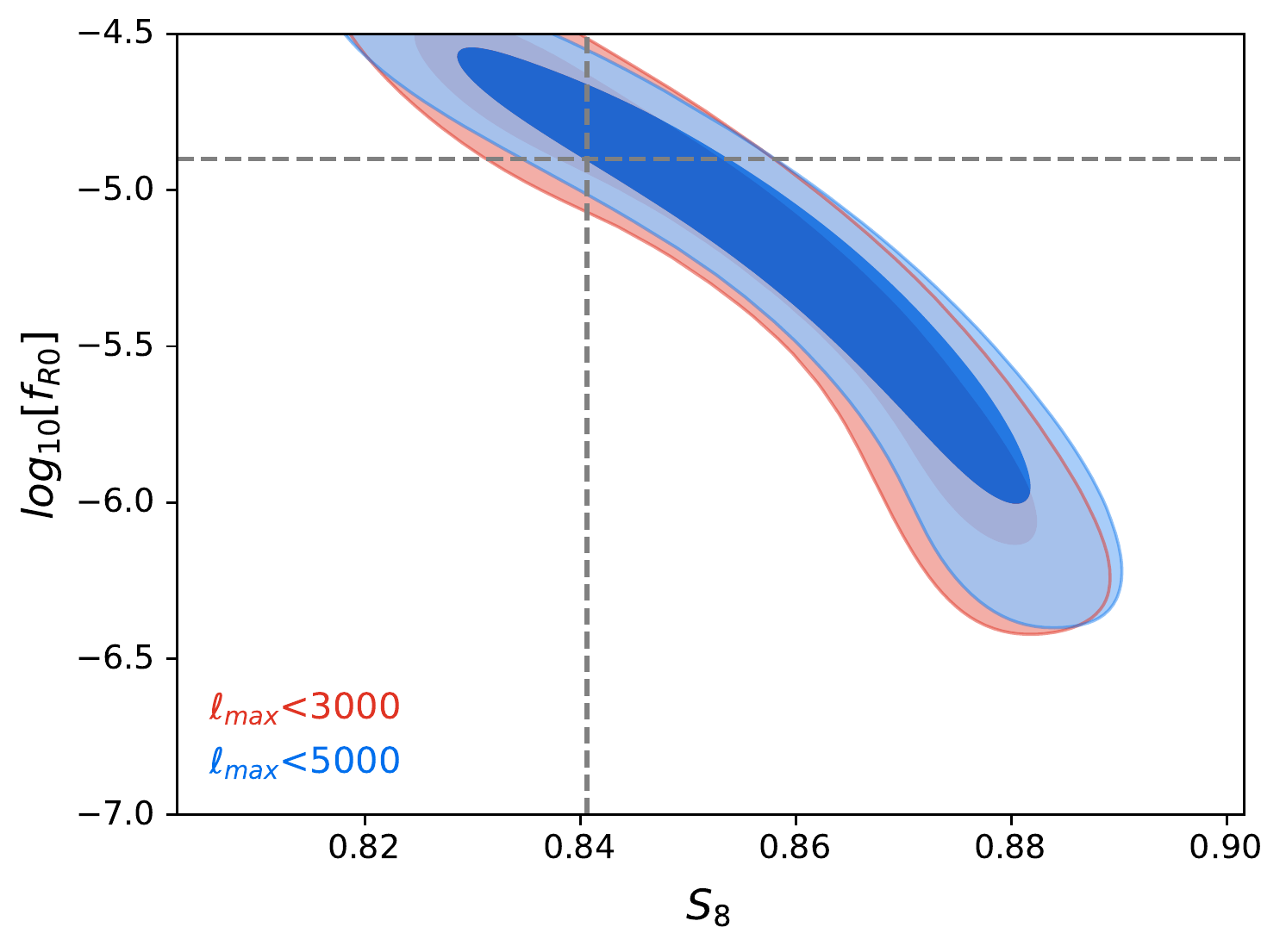}
\caption{Impact of scale cuts on the marginalised constraints, here shown for the {\sc forge} strong  model. }
\label{fig:ell_max}
\end{center}
\end{figure}

The  results from the beginning of Sec. \ref{sec:MCMC} are obtained in unrealistically clean conditions; as discussed previously, cosmic shear surveys are in fact affected by poorly constrained intrinsic alignments (IA), by uncertainty on the photometric redshift (photo-$z$) distributions and shape calibration, as well as by largely unconstrained baryonic feedback. Additionally the weak lensing signal is mildly sensitive to some of the other cosmological parameters such as the baryon density $\Omega_{\rm b}$, the  sum of neutrino masses $\Sigma m_{\nu}$ or the  tilt in the primordial power spectrum, $n_{\rm s}$, such that our constraints are likely slightly over-precise. Here we focus on two of these, namely the photo-$z$ and the IA, leaving a more comprehensive study of the others for future work. To some extent the impact of baryon can be reduced by removing some of the non-linear scales, which we also touch upon below. 

Using {\sc cosmoSIS} for the calculation of the theoretical cosmic shear predictions has key advantages  when it comes to modelling and marginalising over  the known weak lensing systematics. First, the public version  includes an implementation of the widely used non-linear alignment model \citep{BridleKing}, which describes the IA contamination from a linear coupling between the intrinsic galaxy orientations and the local tidal field. This model has been shown to accurately capture the IA signal in many cosmic shear analyses  \citep[see e.g. ][]{KiDS1000_Asgari, DESY1_Troxel}, with weak signs of potential limitations in the most recent DES-Y3 analysis by \citet{DESY3_Secco}. In all cases, IA  significantly bias the inferred cosmology if left unmodelled. Assuming no IA redshift evolution, the uncertainty is captured by a single scaling parameter, $A_{\rm IA}$, which we allow to vary over the range [-5.0, 5.0] in line with these previous analyses. 

Second,  {\sc cosmoSIS}  deals with the uncertainty on the redshift distribution by shifting the  tomographic $n^i(z)$ by a constant quantity $\delta_z^i$, which we treat independently for each tomographic bin $i$: $n^i(z) = n^i(z + \delta_z^i)$. It has been shown that in some cases these shift parameters are correlated \citep{Wright_KV450_SOM}, however we ignore this here.  Our five $\delta_z^i$ parameters are sampled assuming a Gaussian prior of width 0.01, similar to the accuracy achieved by current weak lensing surveys \citep[for example, an accuracy between 0.0084 and 0.0116  on these $\delta_z$ parameters is achieved with the KiDS-1000 data, see][]{KiDS1000_redshifts}. 
We do not include the uncertainty on shape calibration \citep[i.e. the $m$-bias, see][]{KiDS1000_Giblin} as it is currently subdominant compared to the effect of IA and photometric redshift \citep{KiDS1000_Asgari, DESY3_Secco}.  
Importantly, we neglect the impact of baryon feedback, which is arguably the largest approximation in our analysis. Indeed, baryons significantly redistribute the matter distribution and suppress the lensing signal by tens of percent depending on the scales and baryonic physics \citep{Semboloni11,HWVH15}. We could extend our results by using for instance the matter power spectrum provided by HMCode \citep{HMcode2020} in which the impact of baryons is 
modelled, but we leave this for future work. We  finally assume a constant total neutrino mass set to $\Sigma m_{\nu}=0.0$ eV, in order to be consistent with the \textsc{forge} and {\sc bridge} simulations. All of these analysis choices have an impact on the inference and will need to be revisited in order to make robust constraints on the MG parameters from cosmic shear data,  however our simplified  likelihood evaluations represent an important first step in this direction.

We show in Fig. \ref{fig:MCMC_forge_syst} (and summarise the results in Table \ref{table:inference}) the impact of IA on the marginalised constraints for some of the \textsc{forge} and \textsc{bridge} models. As expected, the presence of IA  degrades the constraints on most parameters, where for example the 1.4\% measurement of $S_8$ value in the \textsc{forge} medium model becomes a 1.9\% measurement. The same model sees the constraints on log$_{10}\left[ f_{R_0}\right]$ degrade from a 5.4\% to a 6.7\% measurement. We also note that for some models (e.g. \textsc{forge} medium, \textsc{bridge} strong), the IA contamination acts mostly along the $\left[  f_{R_0} - S_8 \right]$ or $\left[H_0 r_c - S_8 \right]$  degeneracy directions, whereas for other models the posterior is inflated in all dimensions (e.g. \textsc{forge}-strong). Finally low-$S_8$ models appear to be less affected (e.g. \textsc{forge} weak), which is expected since the IA signal also scales with $S_8$, causing them to be harder to distinguish from the cosmological signal given our fixed covariance matrix. Also worth repeating here is that our data vector does not include any of the cross-tomographic terms, which are more affected by IA as they are highly sensitive to the  `GI' alignment term, i.e. the coupling between the background shearing and the intrinsic alignment of foreground galaxies \citep{Hirata2004}. Adding this increases the contamination, but at the same time further help in constraining the IA sector and therefore self-calibrate.
Indeed, $A_{\rm IA}$ is one of parameters that is best measured by cosmic shear data \citep{KiDS1000_Asgari, DESY3_Secco, Homology2}, even though it is an `effective' model that depends on a number of physical selection effects  such as galaxy types, colours and bias \citep{Blazek2019}. Interestingly, there is a mild degeneracy between the $A_{\rm IA}$ and the MG parameters, such using the wrong gravity model can therefore lead to an apparent IA signal. The effect is generally small, but can lead to a false detection larger than $1\sigma$, as it is the case for the GR analysis of the strong \textsc{bridge} model.

The redshift error are in comparison very  small due to the narrow informative Gaussian prior that we are able to use. We have tested a few chains with the photo-$z$ nuisance turned on (plotted for example on Fig. \ref{fig:MCMC_model00}) and found almost no visible effect on the marginalised contours. Since this is the case for all models analysed we conclude that under these circumstances photo-$z$ errors are completely subdominant to IA and we do not investigate this any further.

Regarding baryons,  a common approach to protect analyses against their uncertain impact consists in excluding the deeply non-linear scales from the data vector \citep[as in, e.g.][]{DESY1_Troxel, DESY3_Amon}, which in our case are the high-$\ell$ modes. Lowering the highest $\ell$  from 5000 to 3000 typically results in a degraded constraint on the modified gravity parameters, largely due to an increased degeneracy with $S_8$, but this degradation is not catastrophic, as shown in Fig. \ref{fig:ell_max}. This is consistent with our Fisher calculations, according to which the information partly saturates by $\ell=3000$. 
Therefore, while we expect the impact of varying $\ell_{\rm max}$ to lower the precision, the amount by which it does is not easily predictable due to the highly non-trivial degeneracies that exists in the high-dimensional likelihood space. 

\begin{figure}
\begin{center}
\includegraphics[width=3.1in]{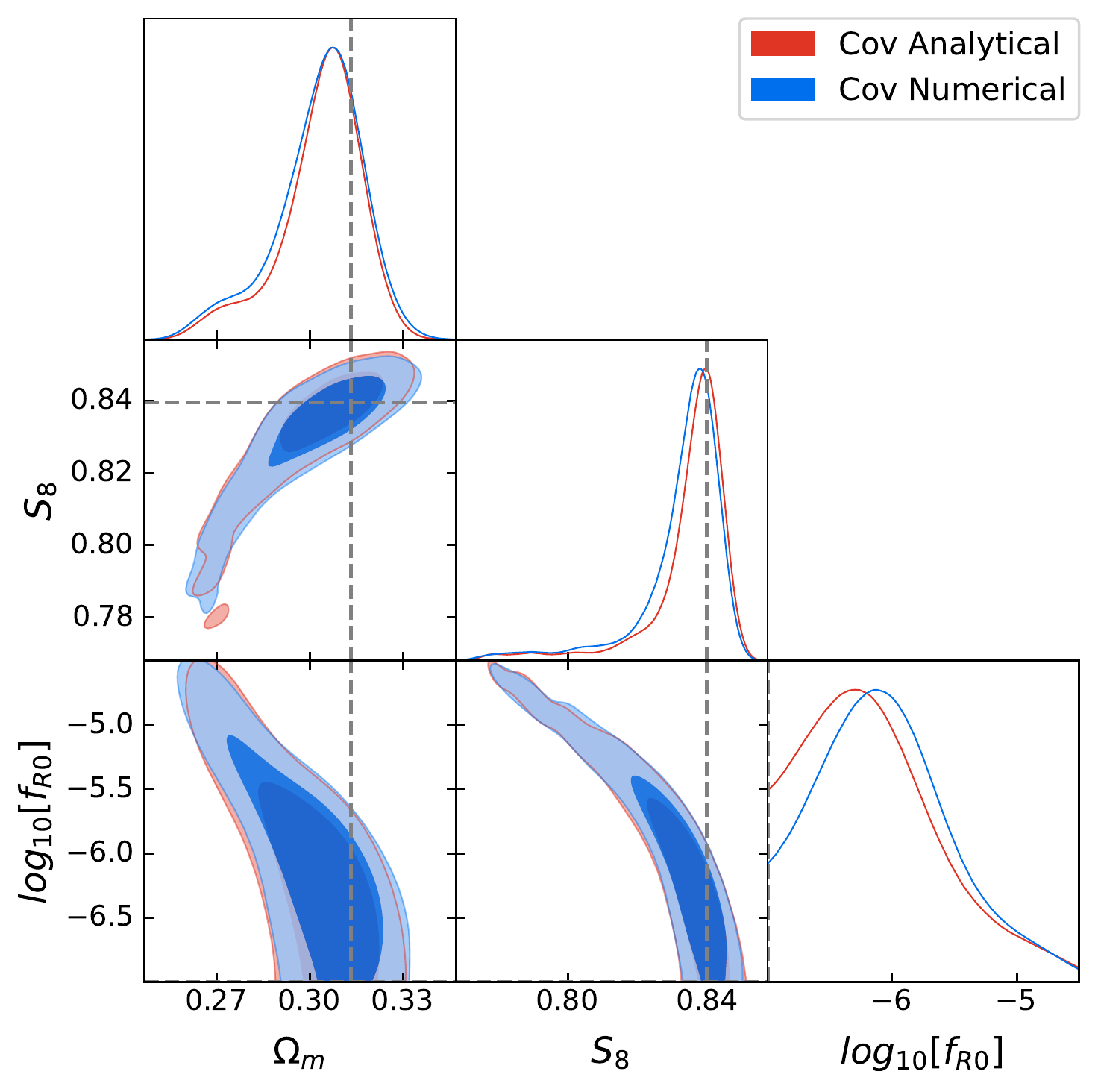}
\caption{Comparison between the cosmological inference resulting from using the analytical or the numerical covariance matrix when analysing the GR-$\Lambda$CDM simulated  data.}
\label{fig:CovComp}
\end{center}
\end{figure}

Finally, as mentioned earlier, an  ingredient central to cosmological inference is the covariance matrix, which in the case of two-point statistics can be either modelled analytically or estimated numerically. This choice is not guaranteed to exists for all probes, and in fact many other weak lensing statistics  must rely on an ensemble of mock data such as the SLICS to estimate the matrix. The validation process of these multi-purpose mocks generally includes a comparison  with the analytical predictions for covariance matrix about two-point statistics. A first step of this comparison is shown already in Fig. \ref{fig:CorrMatrix}, which visually demonstrate that the cross-correlation coefficient matrices are consistent with one another. A full quantitative validation must go beyond this, and we show in Fig. \ref{fig:CovComp} the cosmological inference resulting from using the two matrices. We observe that both posteriors fully overlap, providing identical best-fit values on $\Omega_{\rm m}$, and differences on $S_8$ that vary by less than $0.3\sigma$. The upper limits of log$_{10}\left[ f_{R_0} \right]$ shift by under 1\%, from $-5.08$ to $-5.05$. Note that the differences observed here are not exclusively caused by inaccuracies in the mocks, as  many other factors can source important deviations, such as choices in the implementation of shape noise or masking \citep{KiDS1000_Joachimi}. In particular, the total survey areas match in both cases, however the analytical calculations assume a spherical survey whereas the mocks are square-shaped. Thus the small observed shifts in the cosmological inferences should be viewed as systematic uncertainties, not as biases, which thereby establishes the precision on the covariance one can expect from these SLICS mocks for any alternative weak lensing probes.

Also note that in an actual data analysis, the accuracy of the $B_{\delta}(k,z)$ emulator itself should be propagated into the covariance matrix in order to capture the modelling uncertainty.

\section{Discussions and Conclusions}
\label{sec:conclusion}

This paper introduces the \textsc{MGLenS} simulations, a large set of lensing maps sampling five cosmological and MG parameters within a volume that is wide and dense enough to analyse current Stage-III cosmic shear surveys. 
We demonstrate that the lensing power spectra measured from these simulations match well with the theoretical predictions obtained by the \textsc{bridge} and \textsc{forge} emulators, both at the level of the data vector and in a series of full gravito-cosmological inference analyses based on two-point statistics. Our results are robust to systematics such as multipole scale cuts, photometric redshift errors and intrinsic alignments of galaxies. We test our inference framework on theoretical data vectors using either an analytical or simulation-based covariance matrix, finding an excellent recovery of the input data vector.

Notably, we find that next generation lensing surveys will be powerful at constraining the gravity sector: in our simplified systematics-free analysis, we forecast that 5000 deg$^2$ of upcoming data could lead to $3\sigma$ detection of a value of $f_{R_0}$ as weak as $3.2\times10^{-6}$, and $H_0 r_c$ as low as 1.0. For weaker MG models, we find we are able to place an upper limit of  $2.0 \times 10^{-7}$ on $f_{R_0}$, and a lower limit of $2.6$ on $H_0 r_c$, both at 95\% CL.
We acknowledge a number of caveats, including the absence of marginalisation over baryon feedback, or fixing the values of other cosmological parameters that have a secondary impact on the cosmic shear signal. These will inevitably translate into a slightly larger uncertainty budget in an more complete data analysis, however the statistical power displayed in our survey should remain relatively unchanged. Moreover, these forecasts are for auto-tomographic lensing data alone; adding cross-redshift bins, clustering and galaxy-galaxy lensing data could improve the constraints further. Another gain of precision could be achieved by analysing the data with non-Gaussian statistics.

When inferring cosmology from different input model vectors, we identify in many cases a strong degeneracy between the  input $S_8$ value (related to the primordial power spectrum amplitude $A_{\rm s}$) and the modified gravity parameters; we propose new composite parameters that are better measured by lensing, namely $\zeta_{R_0}^\alpha$ and $\zeta_{r_c}^\alpha$, on which the precision is increased  by up to a factor of two. 

We lastly explored the impact of analysing data with the wrong gravity model, typically finding a catastrophic impact on the inferred cosmology with biases exceeding at times 20$\sigma$ in some cases, as well as an unphysical detection of MG features. The goodness-of-fit is generally best when using the correct gravity models, but some data are well fitted by GR, $f(R)$ and nDGP. This means that other analysis methods will need to be developed in order to better differentiate the gravity sector, such as {\it Bayesian evidence ratios}, the {\it  Suspiciousness} metric,  the recent empirical approach of \citet{Campos}, or by looking at probes different from the lensing power spectrum.

The \textsc{MGLenS} simulations are organised as a series of flat-sky and curved-sky convergence maps, which can be analysed with any weak lensing statistics. Combined with the large SLICS ensemble produced for the evaluation of covariance matrix, the \textsc{MGLenS} suite are ideally suited to explore the sensitivity of novel statistics to cosmological and gravitational parameters, and can be used directly to analyse current Stage-III lensing surveys.

\section*{Acknowledgements}

JHD acknowledges support from an STFC Ernest Rutherford Fellowship (project reference ST/S004858/1). CH-A acknowledges support from the Excellence Cluster ORIGINS which is funded by the Deutsche Forschungsgemeinschaft (DFG, German Research Foundation) under Germany's Excellence Strategy -- EXC-2094 -- 390783311. CA and BL are supported by the European Research Council (ERC) through a starting Grant (ERC-StG-716532 PUNCA). BL is also supported by the UK Science and Technology Funding Council (STFC) Consolidated Grant No.~ST/I00162X/1 and ST/P000541/1. CTD is funded by the Deutsche Forschungsgemeinschaft (DFG, German Research Foundation) under Germany´s Excellence Strategy – EXC-2094 – 390783311. YC acknowledges the support of the Royal Society through a University Research Fellowship and an Enhancement Award. This work used the DiRAC@Durham facility managed by the Institute for Computational Cosmology on behalf of the STFC DiRAC HPC Facility (www.dirac.ac.uk). The equipment was funded by BEIS capital funding via STFC capital grants ST/K00042X/1, ST/P002293/1, ST/R002371/1 and ST/S002502/1, Durham University and STFC operations grant ST/R000832/1. DiRAC is part of the National e-Infrastructure.
The SLICS simulations were enabled by the Digital Research Alliance of Canada (alliancecan.ca).\\
\\
\\
{\footnotesize All authors contributed to the development and writing of this paper. JHD created the \textsc{MGLenS} suites, the {\sc cosmoSIS} interface module and led the inference analysis; CA and CHA respectively led the $N$-body computations for the \textsc{forge} and \textsc{bridge} simulations; CC developed the ML emulators, BL assisted in the design and in the data analysis while CTD and YC  helped in the interpretation of the data. }

\section*{Data Availability}

The \textsc{MGLenS} numerical simulations, the $B_{\delta}(k,z)$ CNN emulators and the {\sc cosmoSIS} interface module will be made available upon acceptance to the Journal, while the SLICS simulations are already public.



\bibliographystyle{hapj}
\bibliography{MGLenS} 




\appendix

\section{Curved-sky weak lensing light-cones} 
\label{sec:WL_sims_curved}

\begin{table}
   \centering
   \caption{Cosmological and gravity  parameters of the \textsc{forge} and \textsc{bridge} simulations. The listed values of the structure growth parameters $\sigma_8$ and $S_8$ correspond to the input truth in the corresponding  GR+$\Lambda$CDM simulations; the actual values in \textsc{MGLenS} are larger than these. Note that the emulators are specifically trained on $\Omega_{\rm m}$, $S_8$,  $h$ and either log$_{10}\left[ f_{R_0} \right]$ or log$_{10}\left[H_0 r_c \right]$. In this paper we focus on weak, medium and strong models, which are respectively models-04, -18 and -13.}
   \tabcolsep=0.11cm
      \begin{tabular}{@{} lcccccc @{}} 
       \hline 
       \hline
model & $\Omega_{\rm m}$ & $\sigma_8$ &  $S_8$ & $h$ &  $f_{R_0}$ & $H_0 r_c$\\
 \hline
 00 & 0.31315 & 0.82172 & 0.83954 & 0.6737 & 0 & Inf \\
01 & 0.54725 & 0.49342 & 0.66642 & 0.78699 & 3.5502e-06 & 0.72533 \\
02 & 0.53961 & 0.63783 & 0.85542 & 0.68393 & 3.0776e-06 & 0.81161 \\
03 & 0.10721 & 1.2297 & 0.73513 & 0.6109 & 3.3107e-06 & 0.76647 \\
04 & 0.31592 & 0.60111 & 0.61685 & 0.68845 & 8.0706e-07 & 3.9962 \\
05 & 0.15741 & 0.91175 & 0.66044 & 0.71067 & 1.2093e-05 & 0.37375 \\
06 & 0.35339 & 0.71886 & 0.78021 & 0.78052 & 5.2037e-06 & 0.56467 \\
07 & 0.1124 & 1.2341 & 0.75539 & 0.79318 & 3.1185e-05 & 0.25000 \\
08 & 0.39303 & 0.72152 & 0.82585 & 0.752 & 7.1372e-07 & 6.7113 \\
09 & 0.18096 & 1.0378 & 0.80599 & 0.76132 & 9.1585e-07 & 3.3057 \\
10 & 0.42927 & 0.5035 & 0.60228 & 0.77667 & 4.5479e-06 & 0.62132 \\
11 & 0.40249 & 0.55523 & 0.64312 & 0.6912 & 1.3401e-06 & 1.7208 \\
12 & 0.21286 & 1.0669 & 0.89867 & 0.70661 & 7.1154e-06 & 0.47331 \\
13 & 0.34671 & 0.78191 & 0.84059 & 0.70056 & 1.2573e-05 & 0.36029 \\
14 & 0.15464 & 0.9339 & 0.6705 & 0.77273 & 4.0961e-06 & 0.65314 \\
15 & 0.28172 & 0.71367 & 0.69158 & 0.64968 & 4.9744e-06 & 0.59191 \\
16 & 0.37032 & 0.61264 & 0.68066 & 0.76204 & 2.7753e-06 & 0.86134 \\
17 & 0.41627 & 0.74242 & 0.87454 & 0.63427 & 1.4375e-05 & 0.33547 \\
18 & 0.32331 & 0.85987 & 0.89266 & 0.81749 & 3.6751e-06 & 0.6877 \\
19 & 0.47784 & 0.56403 & 0.71183 & 0.66724 & 6.7404e-06 & 0.49385 \\
20 & 0.20509 & 0.75641 & 0.62541 & 0.64437 & 5.8109e-06 & 0.53938 \\
21 & 0.44103 & 0.50237 & 0.60912 & 0.62046 & 6.2281e-06 & 0.51583 \\
22 & 0.46403 & 0.5862 & 0.72906 & 0.80296 & 1.4121e-06 & 1.5615 \\
23 & 0.13644 & 1.2584 & 0.84862 & 0.62473 & 1.0481e-06 & 2.4364 \\
24 & 0.18832 & 0.85396 & 0.67659 & 0.80174 & 1.668e-05 & 0.32401 \\
25 & 0.12066 & 1.3159 & 0.83454 & 0.69563 & 2.4559e-06 & 0.91639 \\
26 & 0.28854 & 0.65331 & 0.6407 & 0.73943 & 8.7041e-06 & 0.43601 \\
27 & 0.45016 & 0.72241 & 0.88492 & 0.71954 & 2.174e-05 & 0.2835 \\
28 & 0.17155 & 1.1394 & 0.86159 & 0.62768 & 1.5757e-06 & 1.4266 \\
29 & 0.51949 & 0.59577 & 0.78399 & 0.74473 & 9.6963e-06 & 0.40305 \\
30 & 0.43909 & 0.61327 & 0.74195 & 0.67856 & 1.7774e-06 & 1.3111 \\
31 & 0.49786 & 0.58288 & 0.75088 & 0.80806 & 1.8337e-06 & 1.2109 \\
32 & 0.40909 & 0.54179 & 0.63268 & 0.73799 & 1.211e-06 & 1.9119 \\
33 & 0.23227 & 0.86433 & 0.76052 & 0.60028 & 1.9037e-05 & 0.30276 \\
34 & 0.3839 & 0.61174 & 0.69201 & 0.6557 & 2.2527e-06 & 1.0462 \\
35 & 0.26234 & 0.88665 & 0.82914 & 0.76998 & 1.0089e-06 & 2.8097 \\
36 & 0.25453 & 0.76212 & 0.702 & 0.66918 & 1.7789e-05 & 0.31312 \\
37 & 0.29762 & 0.79347 & 0.79031 & 0.673 & 2.3584e-06 & 0.97764 \\
38 & 0.22423 & 0.88911 & 0.76866 & 0.64603 & 1.3881e-05 & 0.34755 \\
39 & 0.30799 & 0.71046 & 0.71985 & 0.66001 & 1.1732e-06 & 2.1452 \\
40 & 0.51288 & 0.61834 & 0.80849 & 0.79098 & 7.8299e-06 & 0.45407 \\
41 & 0.14061 & 1.1712 & 0.80186 & 0.73101 & 1.0743e-05 & 0.38798 \\
42 & 0.33782 & 0.66702 & 0.70781 & 0.72256 & 7.9806e-07 & 5.0232 \\
43 & 0.5252 & 0.66452 & 0.87924 & 0.81347 & 2.3279e-05 & 0.27454 \\
44 & 0.19435 & 1.0172 & 0.8187 & 0.63911 & 2.7347e-05 & 0.25781 \\
45 & 0.26963 & 0.91366 & 0.86618 & 0.75511 & 9.4886e-06 & 0.41903 \\
46 & 0.49135 & 0.50927 & 0.65176 & 0.60766 & 2.5865e-05 & 0.26599 \\
47 & 0.47207 & 0.58056 & 0.72827 & 0.61562 & 2.0816e-06 & 1.1234 \\
48 & 0.24424 & 0.85676 & 0.77304 & 0.71436 & 6.6853e-07 & 10.0000 \\
49 & 0.36187 & 0.56321 & 0.61856 & 0.72861 & 2.0258e-05 & 0.2929 \\
     \hline 
    \hline
    \end{tabular}
    \label{table:forge_nodes}
\end{table}

\begin{figure*}
\begin{center}
\includegraphics[width=6.1in]{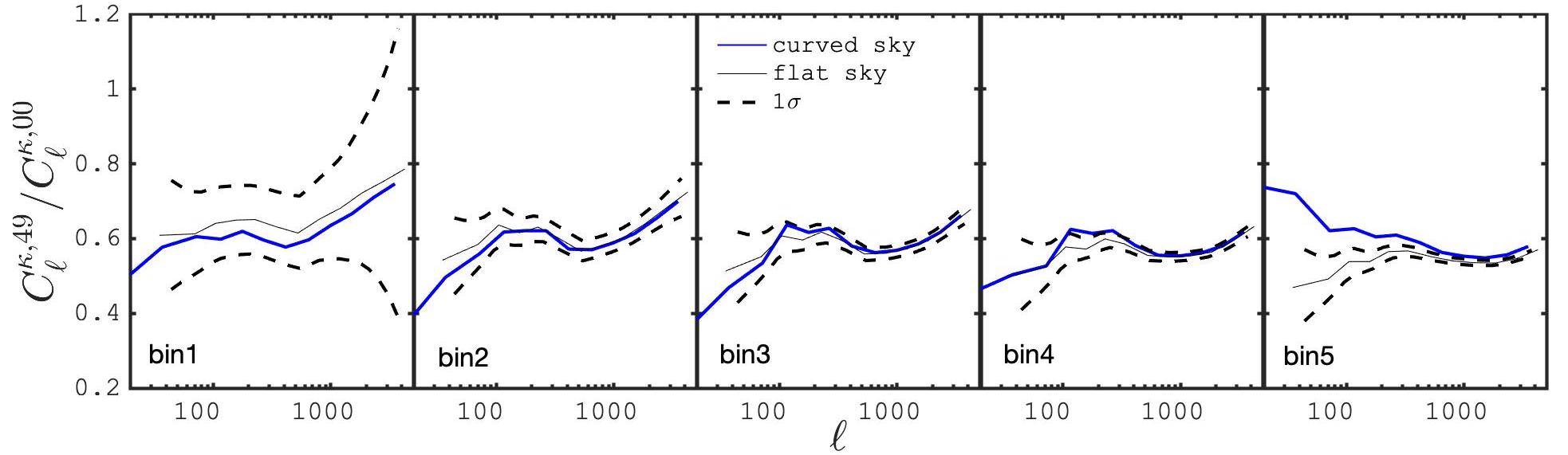}
\caption{Comparison between the curved- and flat-sky lensing power spectra. Plotted is the ratio between the measurements from the nodes 49 and 00, for all five tomographic redshift bins. The right-most plot exhibits large-scales systematics due to masking, which are increasingly important towards higher redshifts. Our flat-sky methods are mostly immune to this. }
\label{fig:Cell_kappa_flat_curve}
\end{center}
\end{figure*}

We develop a curved-sky ray-tracing algorithm adapted from {\sc UFalcon}\footnote{cosmo-docs.phys.ethz.ch/UFalcon}  \citep{Zuercher2020a}, in which the particle data falling into spherical mass shells are assigned onto a {\sc HealPIX} \citep{Healpix} maps with {\sc nside=4096}, instead of the Cartesian grids used in this paper.   
We again use periodic boundary conditions to fill the light-cone volume whenever it exits the simulation box, and repeat the procedure for 24 different observer's positions. We modified the original {\sc UFalcon} full-sky map making algorithm to implement instead a pencil-beam method, significantly reducing the memory load required to fill the high-redshift shells.  This is achieved by stacking  the simulation boxes along the [RA-Dec] = [0,0] direction only, and masking any pixel with RA/Dec $>$ 12 deg. {\it Pseudo}-independent light-cones are then extracted by selecting at random one of the 24 shells for each redshift, repeating the procedure 24 times per $N$-body simulation.
The curved-sky angular power spectrum measurements are obtained from the {standard \sc Healpy}\footnote{healpy.readthedocs.io/en/latest/} routine {\sc map2alm}, which performs Legendre transforms on the sphere and provides measurements for  $\ell \in [1 - 12288]$, which we rebin to match the flat-sky measurements for  an improved comparison. 

We show that both flat- and curved-sky lensing simulations produce similar $C_{\ell}^{\kappa}$ measurements. Figure \ref{fig:Cell_kappa_flat_curve} presents the ratio between the lensing spectra from  two models (the $f(R)$ model-49 and the GR model-00). The thin black lines present the mean over all  flat-sky measurements while the thin blue lines show the curved-sky equivalent. The agreement between these two methods is excellent in the first four tomographic bins, whereas the last tomographic bin exhibits strong discrepancies on large scales. This is caused by the mixing between the maps and the mask, and can be removed with {\it pseudo}-$C_{\ell}$ estimators such as {\sc NaMaster} \citep{NaMaster}.

\bsp	
\label{lastpage}
\end{document}